\documentclass[aps,prd,preprint,groupedaddress,longbibliography,nofootinbib,showkeys]{revtex4-1}

\usepackage{amsmath}
\usepackage{graphicx}
\newcommand{\nt}{\notag}
\newcommand{\lb}{\left\lbrace}
\newcommand{\rb}{\right\rbrace}
\newcommand{\vev}[1]{\left\langle #1 \right\rangle}
\newcommand{\vvev}[1]{\left\langle\kern-0.3em\vev{#1}\kern-0.3em\right\rangle}
\newcommand{\G}{\mathcal{G}}
\newcommand{\cont}{\mathrm{cont}}
\newcommand{\F}{\mathcal{F}}
\newcommand{\eff}{\mathrm{eff}}
\newcommand{\mph}{m_{\mathrm{ph}}}
\newcommand{\ep}{\epsilon}

\allowdisplaybreaks

\begin{document}

\preprint{KOBE-TH-23-02}

\title{Exact Renormalization Group in Large $N$}


\author{Hidenori SONODA}
\email[]{hsonoda@kobe-u.ac.jp}
%\thanks{}
\affiliation{Physics Department, Kobe University, Kobe 657-8501, Japan}


\date{22 February 2023}

\begin{abstract}
  We apply the exact renormalization group formalism to compute the
  effective action and potential of the four dimensional O$(N)$ linear
  sigma model in large $N$.  With a finite momentum cutoff in place,
  the model is well defined.  In the naive continuum limit where the
  cutoff is taken to infinity, the effective action suffers from a
  tachyon, and the effective potential is unbounded from below and
  develops a negative imaginary part.  These problems disappear once a
  small enough cutoff is restored.  The effective potential of the
  naive continuum limit, obtained by analytic continuation, is given
  explicitly in terms of the Lambert $W$ function.
\end{abstract}


\maketitle

\section{Introduction\label{sec-introduction}}

The large $N$ approximation was first introduced for the spherical
model \cite{Berlin-Kac:1952} on a lattice by Berlin and Kac.  The
model that imposes a global constraint on the size of the spins turned
out to be equivalent to the O$(N)$ non-linear
sigma model in large $N$ \cite{Stanley:1968}.  The large $N$ approximation, being
consistent with scaling, has been an essential tool for our
understanding of critical phenomena. (See Ma \cite{Ma:1973zu} for
review.)

The approximation was first introduced to quantum field theory in
\cite{Wilson:1972cf}.  With the hope that physics, at least
qualitatively, does not depend much on $N$, it has become a popular
exercise to study the large $N$ limit.  (See \cite{Moshe:2003xn} for a
review of the early works in large $N$.)  The subject of this paper is
the four dimensional O$(N)$ linear sigma model in large $N$, which was
first studied in \cite{Dolan:1973qd, Schnitzer:1974ji,
  Schnitzer:1974ue, Coleman:1974jh}.

We have set two purposes for the paper:
\begin{enumerate}
\item The malaise of the large $N$ limit in four dimensions found in
  \cite{Coleman:1974jh} is pursued further.
\item The technique of the exact renormalization group (ERG, for
  short) is used for the calculations of the effective action and
  potential.
\end{enumerate}

In \cite{Coleman:1974jh} two main problems were discovered: (a) the
effective potential $V_\eff (\varphi)$ ($\varphi$ will be defined
precisely in Sec.~\ref{sec-overview}) can be defined only up to a
maximum $\varphi_{\max}$ beyond which the potential becomes complex,
(b) the effective action suffers from tachyon poles.  We will confirm
both problems.  As for (a), we give an analytic expression for
$V_\eff (\varphi)$ in terms of the Lambert $W$ function
\cite{Corless:1996zz, wiki:Lambert_W_function}.  This has actually
been done already in \cite{Sonoda:2013jia}, but we go one step further
here.  We extend $V_\eff (\varphi)$ beyond $\varphi_{\max}$ by
analytic continuation.  As argued in \cite{Coleman:1974jh}, the
potential is complex for $\varphi > \varphi_{\max}$.  But our finding
is contrary to the expectation of \cite{Coleman:1974jh} that there may
be a true minimum beyond $\varphi_{\max}$.  We find that the real part
of the potential is unbounded from below.  We will show that both (a)
and (b) go away if the momentum cutoff $\Lambda_0$ is kept finite,
agreeing with the observations made in Sec.~III of
\cite{Sonoda:2013jia}.

The second purpose of the paper is to demonstrate that both the
effective action and the effective potential can be calculated by
integrating the ERG differential equation for the Wilson action.  ERG
was first introduced in sect.~11 of \cite{Wilson:1973jj}.  Given a
theory with momentum cutoff $\Lambda_0$ we integrate over the fields
with momenta below $\Lambda_0$ all the way to zero.  In the ERG
formalism we do the integration incrementally scale by scale by
solving the ERG differential equation.  The ERG formalism has been
actively pursued from the 1990's, and by now many reviews are
available.  (See \cite{Morris:1993qb, Becchi:1996an, Berges:2000ew,
  Pawlowski:2005xe, Gies:2006wv, Igarashi:2009tj, Rosten:2010vm,
  Dupuis:2020fhh} among others.)  Yet, to introduce our notation, we
will start our paper with a short overview of the formalism.

The paper is organized as follows.  In Sec.~\ref{sec-overview} we
overview the ERG formalism, and explain the simplification obtained in
large $N$.  In Sec.~\ref{sec-solution}, we give a general solution to
the ERG differential equation derived in the previous section.  In
Sec.~\ref{sec-absence}, we explain that the ERG differential equation
has no nontrivial fixed-point solution.  This implies that there is no
interacting continuum limit, but as an alternative we introduce a
``naive continuum limit'' to be studied in the following sections.  In
Sec.~\ref{sec-effective action}, we compute the effective action of
the naive continuum limit, and confirm the presence of a tachyon which
was first observed in \cite{Coleman:1974jh}.  Then, in
Sec.~\ref{sec-effective potential}, we compute the corresponding
effective potential $V_\eff (\varphi)$.  The effective potential,
defined only up to a maximum field $\varphi_{\max}$, can be given an
analytic form in terms of the Lambert $W$ function
\cite{Corless:1996zz, wiki:Lambert_W_function}, as was shown already
in \cite{Sonoda:2013jia}.  We can continue $V_\eff (\varphi)$
analytically beyond $\varphi_{\max}$ but find that the potential
develops a negative imaginary part, and that its real part is
unbounded from below.  In Sec.~\ref{sec-cutoff}, we replace the naive
continuum limit by a theory with a bare cutoff $\Lambda_0$.  By using
the general solution of the ERG differential equation given in
Sec.~\ref{sec-solution}, we construct both the effective action and
the effective potential.  The effective action is free from tachyons.
The effective potential agrees with the naive continuum limit for
small fields, but approaches the bare potential for large fields.  In
Sec.~\ref{sec-comparison}, we compare the naive continuum limit with
the bare theory.  We observe how a bound state is converted to a
tachyon as we raise the cutoff beyond the allowed maximum, the Landau
pole.  In \ref{sec-conclusions} we give concluding remarks.  We have
prepared five appendices to give details omitted in the main text.

All through the paper we work in the momentum space and use the
shorthand notation
\[
  \int_p \equiv \int \frac{d^D p}{(2 \pi)^D},\quad
  \delta (p) \equiv (2 \pi)^D \delta^{(D)} (p)\,.
\]

\section{Short overview of the ERG formalism\label{sec-overview}}

We consider a Wilson action $S_t [\phi]$ for the O$(N)$ linear sigma
model consisting of a real scalar field $\phi^I$ ($I = 1,\cdots, N$)
with $N$ components (we use Einstein's convention for repeated
indices).  It satisfies the differential equation
\begin{align}
  \partial_t e^{S_t [\phi]}
  &= \int_p \Bigg[ \left( - p \cdot \partial_p \ln K(p) + \frac{D+2}{2} -
    \gamma_t + p \cdot \partial_p \right) \phi^I (p) \cdot
    \frac{\delta}{\delta \phi^I (p)}\notag\\
  &\quad + \left(- p \cdot \partial_p \ln R(p) + 2 - 2 \gamma_t
    \right) \frac{K(p)^2}{R(p)} \frac{1}{2} \frac{\delta^2}{\delta
    \phi^I (p) \phi^I (-p)} \Bigg] e^{S_t [\phi]}\,,
    \label{II-ERG-S}
\end{align}
where $K$ and $R$ are two independent momentum cutoff functions, both
of which are decreasing positive functions of $p^2$, are $1$ at $p=0$,
and vanish rapidly as $p^2 \to \infty$ \cite{Sonoda:2015bla}.  We
denote the usual correlation functions using single
brackets:\footnote{Please note our sign convention for the Wilson
  action.  The Boltzmann weight is $e^{S_t}$, not $e^{- S_t}$.}
\begin{equation}
  \vev{ \cdots }_{S_t} \equiv \int [d\phi] \, e^{S_t [\phi]} \,
  \cdots\,.
\end{equation}
The $t$-dependence (\ref{II-ERG-S}) of the Wilson action is determined so that
the correlation functions, modified as \cite{Sonoda:2015bla}
\begin{align}
&  \vvev{\phi^{I_1} (p_1) \cdots \phi^{I_n} (p_n)}_t\notag\\
& \equiv \prod_{i=1}^n \frac{1}{K(p_i)}\cdot \vev{\exp \left( - \int_p
      \frac{K(p)^2}{R(p)} \frac{1}{2} \frac{\delta^2}{\delta \phi^I
        (p) \delta \phi^I (-p)} \right) \phi^{I_1} (p_1) \cdots
    \phi^{I_n} (p_n)}_{S_t}\,,
\end{align}
satisfy the scaling relations
\begin{align}
&  \vvev{\phi^{I_1} (p_1 e^{t-t'}) \cdots \phi^{I_n} (p_n e^{t-t'})}_t\notag\\
&  = \exp \left[ n \left(- \frac{D+2}{2} (t-t') + \int_{t'}^t ds\,
      \gamma_s \right)\right] \cdot  \vvev{\phi^{I_1} (p_1) \cdots
        \phi^{I_n} (p_n)}_{t'}\,.
\end{align}
Clearly, $\gamma_t$ gives the scale dependent anomalous dimension of $\phi^I$.

To define the one-particle-irreducible (1PI) part of the Wilson
action, it is convenient first to define a generating functional with
an IR cutoff by
\begin{equation}
  W_t [J] \equiv S_t [\phi] + \frac{1}{2} \int_p \frac{J^I (p) J^I
    (-p)}{R (p)}\,,
\end{equation}
where
\begin{equation}
  J^I (p) \equiv \phi^I (p) \frac{R(p)}{K(p)}\,.
\end{equation}
The cutoff function $R(p)$ provides an IR cutoff of order $1$ because
the momenta below the cutoff has not been integrated yet.
Eq.~(\ref{II-ERG-S}) for $S_t$ implies
\begin{align}
  \partial_t e^{W_t [J]}
  &= \int_p \Bigg[ \left( p \cdot \partial_p + \frac{D-2}{2} + \gamma_t
    \right) J^I (p) \cdot \frac{\delta}{\delta J^I (p)}\notag\\
  &\quad + \left( - p \cdot \partial_p +2 - 2 \gamma_t \right) R(p)
    \cdot \frac{1}{2} \frac{\delta^2}{\delta J^I (p) \delta J^I (-p)}
    \Bigg]\, e^{W_t [J]}\,,
    \label{II-ERG-W}
\end{align}
which is simpler than Eq.~(\ref{II-ERG-S}) in that it depends only on
the cutoff function $R (p)$, but not on $K (p)$.

We can now define the 1PI Wilson action $\Gamma_t [\Phi]$ as
the Legendre transform of $W_t [J]$ by
\begin{equation}
  \Gamma_t [\Phi] - \frac{1}{2} \int_p R(p) \Phi^I (p) \Phi^I (-p)
  \equiv W_t [J] - \int_p J^I (-p) \Phi^I (p)\,,
  \label{II-def-Gamma}
\end{equation}
where we define
\begin{equation}
  \Phi^I (p) \equiv \frac{\delta W [J]}{\delta J^I (-p)}\,.
\end{equation}
Differentiating (\ref{II-def-Gamma}) with respect to $\Phi^I$, we obtain
\begin{equation}
  J^I (p) = R (p) \Phi^I (p) - \frac{\delta \Gamma_t [\Phi]}{\delta
    J^I (-p)}\,.
\end{equation}
From Eq.~(\ref{II-ERG-W}) we obtain the ERG equation for
$\Gamma_t [\Phi]$ as\footnote{This equation is known as the Wetterich
  equation.  It is introduced and popularized in \cite{Wetterich:1992yh}.}
\begin{align}
  \partial_t \Gamma_t [\Phi]
  &= \int_p \left( p \cdot \partial_p + \frac{D+2}{2} - \gamma_t
    \right) \Phi^I (p) \cdot \frac{\delta \Gamma_t [\Phi]}{\delta
    \Phi^I (p)}\notag\\
  &\quad + \int_p \left( - p \cdot \partial_p + 2 - 2 \gamma_t \right)
    R(p) \cdot \frac{1}{2} \G_{t; p,-p}^{II} [\Phi]\,,
\end{align}
where 
\begin{align}
  \G_{t; p,-q}^{IJ} [\Phi] \equiv
  \frac{\delta^2 W_t [J]}{\delta J^I (p) \delta J^J (-q)}
\end{align}
is determined by the differential equation
\begin{align}
  \int_q \G_{t; p, -q}^{IJ} [\Phi]
  \left( R (q) \delta^{JK} \delta (q-r) - \frac{\delta^2 \Gamma_t
  [\Phi]}{\delta \Phi^J (q) \delta \Phi^K (-r)} \right) = \delta^{IK} \delta
  (p-r)\,.
  \label{II-def-G}
\end{align}

Let us now introduce the large $N$ approximation following
\cite{DAttanasio:1997yph}.  We first split $\Gamma_t$ into the
Gaussian and interaction parts
\begin{equation}
  \Gamma_t [\Phi] = - \frac{1}{2} \int_p p^2 \Phi^I (p) \Phi^I (-p) +
  N \Gamma_{I, t} [\varphi]\,,
  \label{II-split}
\end{equation}
and then \textbf{assume} that the interaction part is a functional of
\begin{equation}
  \varphi (p) \equiv \frac{1}{2 N} \int_q \Phi^I (q+p) \Phi^I (-q)\,.
  \label{II-def-varphi}
\end{equation}
Since the interaction part has no kinetic term, the assumption
(\ref{II-split}) implies the vanishing anomalous dimension
\begin{equation}
  \gamma_t = 0\,.
\end{equation}

Let us verify the consistency of this approximation in the large $N$
limit.  Substitution of (\ref{II-def-varphi}) gives
\begin{equation}
\int_p \left( p \cdot \partial_p + \frac{D+2}{2} \right) \Phi^I (p)
    \cdot \frac{\delta \Gamma_{I,t} [\varphi]}{\delta 
    \Phi^I (p)}\notag
= \int_p \left(2 + p \cdot \partial_p \right) \varphi
    (p) \cdot \frac{\delta \Gamma_{I,t}[\varphi]}{\delta \varphi
      (p)}\,.
\end{equation}
Since we can approximate
\begin{align}
N  \frac{\delta^2 \Gamma_{I,t} [\varphi]}{\delta \Phi^J (q) \delta
  \Phi^K  (-r)}
  &= \frac{\delta}{\delta \Phi^J (q)} \int_s \Phi^K (s+r)
    \frac{\delta \Gamma_{I,t} [\varphi]}{\delta \varphi (s)}\notag\\
  &= \delta^{JK} \frac{\delta \Gamma_{I,t} [\varphi]}{\delta \varphi
    (q-r)} + \int_{s,t} \frac{1}{N} \Phi^K (s+r) \Phi^J (t-q)
    \frac{\delta^2 \Gamma_{I,t} [\varphi]}{\delta \varphi (s) \delta
    \varphi (t)}\notag\\
  &\overset{N \gg 1}{\longrightarrow}  \delta^{JK} \frac{\delta
    \Gamma_{I,t} [\varphi]}{\delta \varphi  (q-r)} 
\end{align}
for large $N$, we obtain
\begin{equation}
  \G^{IJ}_{t; p,-q} [\Phi] \overset{N \gg 1}{\longrightarrow}
  \delta^{IJ} \G_{t; p,-q} [\varphi]\,,
\end{equation}
where $\G_{t; p,-q} [\varphi]$ is a functional of $\varphi$, defined by
\begin{equation}
\int_q \G_{t; p,-q} [\varphi] \left[ \left( q^2 + R(q) \right) \delta
(q-r) - \frac{\delta \Gamma_{I,t} [\varphi]}{\delta \varphi (q-r)}
\right] = \delta (p-r)\,.
\end{equation}
Hence, for large $N$, it is consistent to assume the structure (\ref{II-split}),
and we obtain the ERG equation for $\Gamma_{I, t} [\varphi]$ as
\begin{equation}
  \partial_t \Gamma_{I,t} [\varphi]
  = \int_p (2 + p \cdot \partial_p ) \varphi (p) \cdot
    \frac{\delta \Gamma_{I,t} [\varphi]}{\delta \varphi (p)} + \int_p (2 - p
    \cdot \partial_p ) R(p) \cdot \frac{1}{2} \G_{t; p,-p}
    [\varphi]\,.
\label{II-ERG-N}    
\end{equation}
We will solve this in the next section.

Before proceeding to the next section, we would like to explain
briefly the relation between $\Gamma_{I, t} [\varphi]$ and the
physical effective action $\Gamma_{I, \eff}[\varphi]$.
$\Gamma_{I, t} [\varphi]$ is constructed in the dimensionless
convention; all the physical quantities are dimensionless, measured in
units of appropriate powers of the momentum cutoff.  To restore
physical dimensionful quantities, we introduce $\Lambda = \mu e^{-t}$
that corresponds to the physical momentum cutoff.  This gives us an
effective action $\Gamma_{I, \Lambda} [\varphi]$ in the dimensionful
convention.  The cutoff $\Lambda$ is an IR cutoff, since the fields
with momenta less than $\Lambda$ have not been integrated out for the
construction of $\Gamma_{I, \Lambda} [\varphi]$.  By taking
$\Lambda \to 0+$, we obtain the physical effective action
$\Gamma_{I, \mathrm{eff}} [\varphi]$ which gives the 1PI correlation
functions in the dimensionful convention \cite{Sonoda:2017rro}. In
general the limit depends on the finite momentum scale $\mu$ which
plays the role of a renormalization scale.  We give some details in
Appendix \ref{Appendix-dimension}.

What about renormalizability of the theory?  The theory is
renormalizable if $\Gamma_{I, t} [\varphi]$ has a well defined limit
as we take $t \to - \infty$; the limit is a UV fixed point of the ERG
transformation. In four dimensions there is no fixed point, and the
theory is not renormalizable in the strict sense.  We discuss this in
Sec.~\ref{sec-absence}.

\section{Solving the ERG equation in $D=4$\label{sec-solution}}

To solve Eq.~(\ref{II-ERG-N}), we use a well known trick of the
Legendre transformation, which has been used extensively for the
studies of the effective potential in the large $N$
\cite{DAttanasio:1997yph,Morris:1997xj,Litim:2018pxe}. We apply the
same trick to the 1PI Wilson action.\footnote{The same Legendre
  transformation was used in \cite{Coleman:1974jh}.  The variable
  $\sigma$ in this paper corresponds to the auxiliary field $- \chi$
  in \cite{Coleman:1974jh}.  Eq.~(2.6a) of \cite{Coleman:1974jh}
  corresponds to our (\ref{III-varphi-sigma}).}

We introduce the Legendre transform $F_t [\sigma]$ by
\begin{equation}
  \Gamma_{I, t} [\varphi] = F_t [\sigma] + \int_p \sigma (p) \varphi
  (-p)\,,\label{III-Gamma-F}
\end{equation}
where
\begin{equation}
  \sigma (p) = \frac{\delta \Gamma_{I,t} [\varphi]}{\delta \varphi
    (-p)}
  \label{III-sigma-varphi}
\end{equation}
so that we obtain
\begin{equation}
  \varphi (p) = - \frac{\delta F_t [\sigma]}{\delta \sigma (-p)}\,.
  \label{III-varphi-sigma}
\end{equation}
Once we obtain $F_t [\sigma]$, we can construct the corresponding
$\Gamma_{I,t} [\varphi]$ by Eq.~(\ref{III-Gamma-F}), where $\varphi$
is given by Eq.~(\ref{III-varphi-sigma}).

Now, the ERG equation for $F_t [\sigma]$ is given by
\begin{align}
  \partial_t F_t [\sigma]
  &= \partial_t \Gamma_{I,t} [\varphi]\notag\\
  &=  \int_p \left(D-2+p \cdot \partial_p\right)
  \sigma (p) \cdot \frac{\delta F_t [\sigma]}{\delta \sigma (p)} +
  \int_p (2 - p \cdot \partial_p) R(p) \cdot \frac{1}{2} \G_{t; p,-p}
    [\varphi]\,,
    \label{III-diffeq-F}
\end{align}
where Eq.~(\ref{II-def-G}) defines $\G_{t; p,-q} [\varphi]$ by
\begin{equation}
  \int_q \G_{t; p,-q}[\varphi] \left[ (q^2 + R(q)) \delta (q-r) - \sigma (-q+r)
  \right] = \delta (p-r)\,.
\end{equation}
As a functional of $\sigma$, $\G_{t; p,-q} [\varphi]$ has no explicit
$t$-dependence, and we will write
\begin{equation}
  \G_{p,-q} [\sigma] = \G_{t; p,-q} [\varphi]
\end{equation}
from now on.

Solving
\begin{equation}
  \int_q \G_{p,-q} [\sigma] \left[ (q^2+R(q)) \delta (q-r) - \sigma
  (-q+r) \right] = \delta (p-r)\label{III-def-G}
\end{equation}
in powers of $\sigma$, we obtain
\begin{align}
&  \G_{p,-q} [\sigma]
  = h (p) \delta (p-q)\notag\\
  &\, + h (p) \left[ \sigma (-p+q) + \int_{p_1, p_2} \sigma (p_1)
    h(p+p_1) \sigma (p_2) \delta (p_1+p_2+p-q)\right.\notag\\
  &\quad\left. + \int_{p_1, p_2, p_3} \sigma (p_1) h (p+p_1) \sigma
    (p_2) h (p+p_1+p_2) \sigma (p_3) \delta (p_1+p_2+p_3+p-q) + \cdots
    \right] h (q)\,,
\end{align}
where we define the high momentum propagator by
\begin{equation}
  h (p) \equiv \frac{1}{p^2 + R(p)}\,.
  \label{III-def-h}
\end{equation}
($R(p)$ is of order $1$ for $p < 1$, but nearly zero for $p \gg 1$.)

\subsection{General solution for $D=4$}

The general solution of Eq.~(\ref{III-diffeq-F}) depends on $D$, and we
now specialize $D=4$.\footnote{For $2 < D < 4$, there is a
  $t$-independent particular solution $I [\sigma]$ that corresponds to
  the Wilson-Fisher fixed point.}  It is given by the sum of a
particular solution $I_t [\sigma]$ and an arbitrary homogeneous
solution:
\begin{equation}
  F_t [\sigma] = I_t [\sigma] + \tilde{F} [\sigma_{-t}]\,.\label{III-general}
\end{equation}
We first explain the particular solution, which is given by
\begin{equation}
  I_t [\sigma] = - \frac{1}{4} \int_p f(p) \, \sigma (0) + \frac{1}{2}
  \int_p \sigma (p) \sigma (-p) \left( \frac{t}{(4 \pi)^2} + \F(p)
  \right) + I [\sigma]\,.\label{III-It}
\end{equation}
The functional $I [\sigma]$, defined by
\begin{equation}
  I [\sigma] \equiv \sum_{n=3}^\infty \frac{1}{2n} \int_{p_1, \cdots, p_n}
    \sigma (p_1) \cdots \sigma (p_n)\, \delta \left( \sum_1^n p_i
    \right)\, I_n (p_1, \cdots, p_n)\label{III-def-I}
\end{equation}
and
\begin{equation}
  I_{n\ge 3} (p_1, \cdots, p_n) \equiv \int_q h(q) h(q+p_1) h(q+p_1+p_2)
  \cdots h (q+p_1+\cdots + p_{n-1})\,,\label{III-def-In}
\end{equation}
satisfies the differential equation
\begin{align}
  &  \int_p \left( p \cdot \partial_p + 2 \right) \sigma (p) \cdot
    \frac{\delta I[\sigma]}{\delta \sigma (p)}
    = \frac{1}{2} \int_p (p \cdot \partial_p - 2) R (p) \cdot \Big[
    \G_{p,-p} [\sigma] \nt\\
  &\qquad - h (p) \delta (0) - h(p)^2 \sigma (0) - h(p)^2
    \int_q \sigma (q) h (p+q) \sigma (-q)  \Big]\,.\label{III-diffeq-I}
\end{align}
Note $I_{n\ge 3}$ is a convergent integral for $D=4$, and it satisfies
the differential equation
\begin{align}
  & \left(\sum_i p_i \cdot \partial_{p_i} + 2 (n-2) \right) I_n (p_1,
    \cdots, p_n)\nt\\
  &= \int_p \Big( f(p) h(p+p_1) \cdots h(p+ \sum_1^{n-1} p_i)
   + h(p) f(p+p_1) \cdots h (p+ \sum_1^{n-1} p_i) \nt\\
  &\qquad+ \cdots
    + h(p) h(p+p_1) \cdots f (p+\sum_1^{n-1} p_i) \Bigg)\,,
    \label{III-diffeq-In}
\end{align}
where
\begin{equation}
  f(p) \equiv \left( p \cdot \partial_p + 2 \right) h (p) =
  \frac{(2-p\cdot \partial_p) R(p)}{\left(p^2+R(p)\right)^2}\,.\label{III-def-f}
\end{equation}
The function $\F(p)$ is determined by
\begin{equation}
  p \cdot \partial_p \F(p) = \int_q f(q) \left( h(q+p) - h(q)\right)\label{III-diffeq-calF}
\end{equation}
and
\begin{equation}
  \F (0) = 0\,.\label{III-zero-calF}
\end{equation}
It is given explicitly by
\begin{equation}
  \F (p) = \frac{1}{2} \int_p h(q) \left( h(q+p) - h (q)\right)\,.\label{III-calF}
\end{equation}
Since
\begin{align}
  \int_p f(p) h(p) &=  \int_p (p\cdot \partial_p - 2) R(p) \cdot
                     h(p)^3 = \frac{1}{2} \int_p
                (p \cdot \partial_p + 4 ) h(p)^2 \nt\\
              &= \frac{1}{(4\pi)^2} \int_0^\infty dp^2\, 
                \frac{d}{dp^2} \left( p^4 h (p)^2 \right) =
                \frac{1}{(4\pi)^2}\,,
                \label{III-fh-integral}
\end{align}
we can rewrite (\ref{III-diffeq-calF}) as
\begin{equation}
  p \cdot \partial_p \F (p) = \int_q f(q) h(q+p) - \frac{1}{(4
    \pi)^2}\,.\label{III-diffeq2-calF}
\end{equation}
So much for the particular solution $I_t [\sigma]$.  (Please see
Appendix \ref{Appendix-I} for more details on $I [\sigma]$.)

The homogeneous solution $\tilde{F} [\sigma_{-t}]$ is an arbitrary
functional of
\begin{equation}
  \sigma_{-t} (p) \equiv e^{2 t} \sigma (p e^t)\,.
\end{equation}
We find
\begin{equation}
  \partial_t \tilde{F} [\sigma_{-t}] = \int_p \left( p \cdot
    \partial_p + 2 \right) \sigma
  (p) \cdot \frac{\delta \tilde{F} [\sigma_{-t}]}{\delta \sigma (p)}\,.
\end{equation}
Expanding $\tilde{F} [\sigma]$ in powers of $\sigma$
\begin{equation}
  \tilde{F} [\sigma] = f_1  \sigma (0) + \sum_{n=2}^\infty \frac{1}{n!}
  \int_{p_1, \cdots, p_n} \sigma (p_1) \cdots \sigma (p_n)\, \delta
  \left(\sum_1^n p_i\right)\, f_n (p_1, \cdots, p_n)\,,
\end{equation}
we obtain
\begin{align}
  \tilde{F} [\sigma_{-t}]
  &= f_1 e^{2t} \sigma (0) + \frac{1}{2} \int_p \sigma (p) \sigma (-p) 
    f_2 (p e^{-t}, - p e^{-t})\nt\\
  &\quad + \sum_{n=3}^\infty \frac{1}{n!} \int_{p_1, \cdots, p_n}
    \sigma (p_1) \cdots \sigma (p_n)\, \delta 
    \left(\sum_1^n p_i\right)\, e^{2(2-n) t} f_n (p_1 e^{-t}, \cdots,
    p_n e^{-t})\\
  &\overset{t \to +\infty}{\longrightarrow} f_1 e^{2t} \sigma (0) +
   f_2 (0,0)  \frac{1}{2}  \int_p \sigma (p) \sigma (-p)\,.
\end{align}

We have thus found the most general solution as
\begin{align}
  F_t [\sigma]
  &= \left( - \frac{1}{4} \int_p f(p) + f_1 e^{2t}\right) \sigma (0)
    + \frac{1}{2} \int_p \sigma (p) \sigma (-p)
    \left[\frac{t}{(4\pi)^2} + \F (p) + f_2 ( p e^{-2t}, - p e^{-2t} )
    \right] \nt\\
  &\quad + I [\sigma] + \sum_{n=3}^\infty \frac{1}{n!} \int_{p_1, \cdots, p_n}
    \sigma (p_1) \cdots \sigma (p_n) \, \delta \left( \sum_1^n
    p_i\right)\, e^{- 2 (n-2)t} f_n \left( p_1 e^{-t}, \cdots, p_n
    e^{-t}\right)\,,\label{III-generalFt}
\end{align}
where $f_n$'s are all arbitrary.  The parameter $f_1$ is relevant, and
$f_2 (0, 0)$ is marginal.

\subsection{1PI potential}

For the constant field
\begin{equation}
  \varphi (p) = \varphi \delta (p)\,,
\end{equation}
we obtain
\begin{equation}
  \Gamma_{I, t} [\varphi] = G_t (\varphi)\, \delta (0)\,,
\end{equation}
where $\delta (0) = \int d^4 x$ is the four dimensional space volume.
We call $G_t (\varphi)$ a 1PI potential.  In an appropriate limit, to
be discussed in Sections \ref{sec-effective potential} and
\ref{sec-cutoff}, it gives \textbf{minus} the effective potential,
$- V_\eff (\varphi)$.  Eq.~(\ref{II-ERG-N}) for
$\Gamma_{I, t} [\varphi]$ amounts to the ERG equation
\begin{equation}
  \partial_t G_t (\varphi) = 4 G_t (\varphi) - 2 \varphi
  \frac{\partial}{\partial \varphi} G_t (\varphi) + \frac{1}{2}
  \frac{\partial}{\partial \varphi} G_t (\varphi) \cdot \int_p \frac{f(p)}{1
    - h(p) \frac{\partial}{\partial \varphi} G_t (\varphi)}\,.
  \label{III-diffeq-Gt}
\end{equation}

We define the Legendre transform $F_t (\sigma)$ by
\begin{equation}
  G_t (\varphi) = F_t (\sigma) + \sigma \varphi\,,
\end{equation}
where
\begin{equation}
  \sigma = \frac{\partial G_t (\varphi)}{\partial \varphi}\,.
\end{equation}
The inverse transformation gives
\begin{equation}
  \varphi = - \partial_\sigma F_t (\sigma)\,.
\end{equation}

The ERG equation for $F_t (\sigma)$ is obtained from
(\ref{III-diffeq-Gt}) as
\begin{align}
  \partial_t F_t (\sigma)
  &= \partial_t G_t (\varphi)\nt\\
  &= 4 \left(F_t (\sigma) - \sigma \partial_\sigma F_t (\sigma)\right) + 2
    \partial_\sigma F_t (\sigma) \cdot \sigma + \frac{1}{2} \sigma
    \int_p \frac{f(p)}{1 - h (p) \sigma}\nt\\
&= 4 F_t (\sigma) - 2 \sigma \partial_\sigma F_t (\sigma) +
                                             \frac{1}{2} \sigma \int_p
                                             f(p) \frac{1}{1 - h (p) \sigma}\,.
\end{align}
The most general solution is given by
\begin{align}
  F_t (\sigma) &= \sigma \left(- \frac{1}{4} \int_p f(p) + f_1 e^{2t}\right)
 + \frac{1}{2} \sigma^2 \left(\frac{t}{(4 \pi)^2} +
                 f_2\right)\nt\\
  &\quad + I (\sigma) + \sum_{n=3}^\infty \frac{1}{n!} \sigma^n
    e^{-2(n-2)t} f_n\,,
\end{align}
where the function $I (\sigma)$ is defined by
\begin{align}
  I (\sigma)
  &\equiv \frac{1}{2} \int_p \left( - \ln (1 - \sigma h(p))
    - \sigma h(p) - \frac{1}{2} (\sigma h(p))^2 \right)\nt\\
  &= \sum_{n=3}^\infty \frac{1}{2n} I_n\, \sigma^n\quad \left(
    I_n \equiv \int_p h(p)^n\right)\,,\label{III-function-I}
\end{align}
and the constants $f_n$ are given by
\begin{equation}
  f_{n \ge 2} = f_n (p_1=0, \cdots, p_n=0)\,.
\end{equation}

\subsection{Convexity of $F_t$}

In fact the constants $f_{n \ge 2}$ cannot be arbitrary; they are
constrained so that the Legendre transformation between $F_t (\sigma)$
and $G_t (\varphi)$ is well defined.  We demand $F_t (\sigma)$ be
convex:
\begin{equation}
  \partial_\sigma^2 F_t (\sigma) > 0\,.\label{III-F-convex}
\end{equation}
This is equivalent to
\begin{equation}
  \partial_\varphi^2 G_t (\varphi) < 0\,.\label{III-G-concave}
\end{equation}
Since $G_t (\varphi)$ becomes minus the effective potential
$- V_\eff (\varphi)$ in an appropriate limit, (\ref{III-G-concave})
would imply that the effective potential be convex:
\begin{equation}
  \frac{d^2}{d\varphi^2} V_\eff (\varphi) > 0\,,
\end{equation}
as required by the positivity of the squared fluctuation of $\varphi$.

Analogously, the coefficient functions $f_{n \ge 2} (p_1, \cdots,
p_n)$ are not arbitrary; they are constrained so that
\[
  \frac{\delta^2 F_t [\sigma]}{\delta \sigma (p) \delta \sigma (q)}
\]
be positive.  Equivalently,
\[
  \frac{\delta^2 \Gamma_{I, t} [\varphi]}{\delta \varphi (p) \delta
    \varphi (q)}
\]
are required to be negative.

\section{Absence of a continuum limit\label{sec-absence}}

To construct a continuum limit we introduce a solution of (\ref{III-diffeq-F})
\begin{align}
  F_t^{T} [\sigma]
  &\equiv  \left( - \frac{1}{4} \int_p f(p) + f_1 e^{2(t-T)}\right) \sigma (0)\nt\\
&\quad    + \frac{1}{2} \int_p \sigma (p) \sigma (-p)
    \left(\frac{t-T}{(4\pi)^2} + \F (p) + f_2 \right) + I [\sigma]\label{IV-FTt}
\end{align}
with two parameters $f_1, f_2$ that satisfies the boundary condition
\begin{equation}
  F_{t=T}^T [\sigma] = \left( - \frac{1}{4} \int_p f(p) + f_1 \right)
  \sigma (0) + \frac{1}{2} \int_p \sigma (p) \sigma (-p) \left(\F (p)
    + f_2 \right) + I [\sigma]\,.\label{IV-FTT}
\end{equation}
$F_{t=T}^T [\sigma]$ is independent of $T$, but we cannot interpret it
as a fixed point.  $F_t^T [\sigma]$ does not make sense as
$t \to - \infty$.

\begin{figure}[h]
  \centering
  \includegraphics[width=0.45\textwidth]{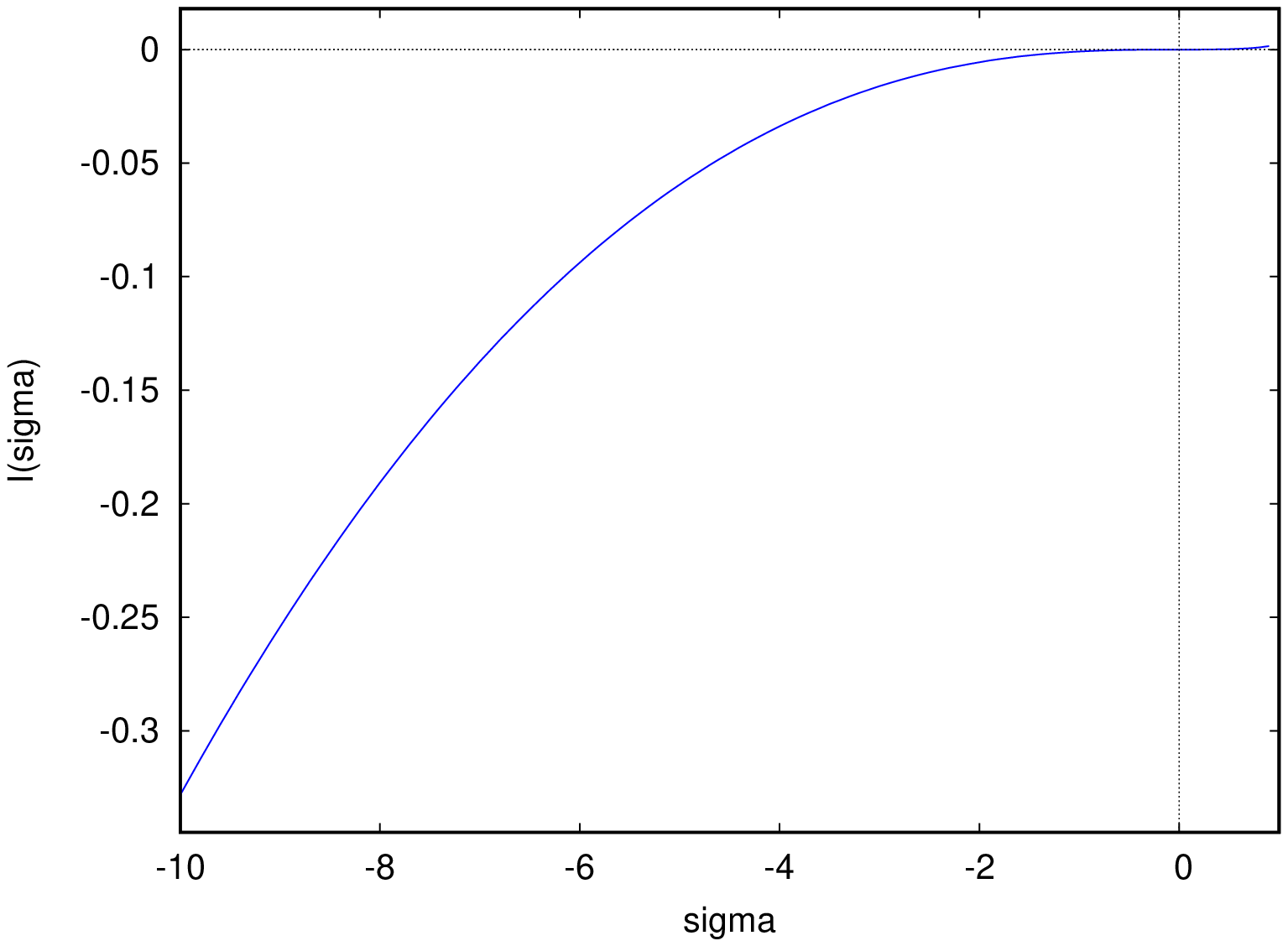}
  \hspace{0.5cm}
  \includegraphics[width=0.45\textwidth]{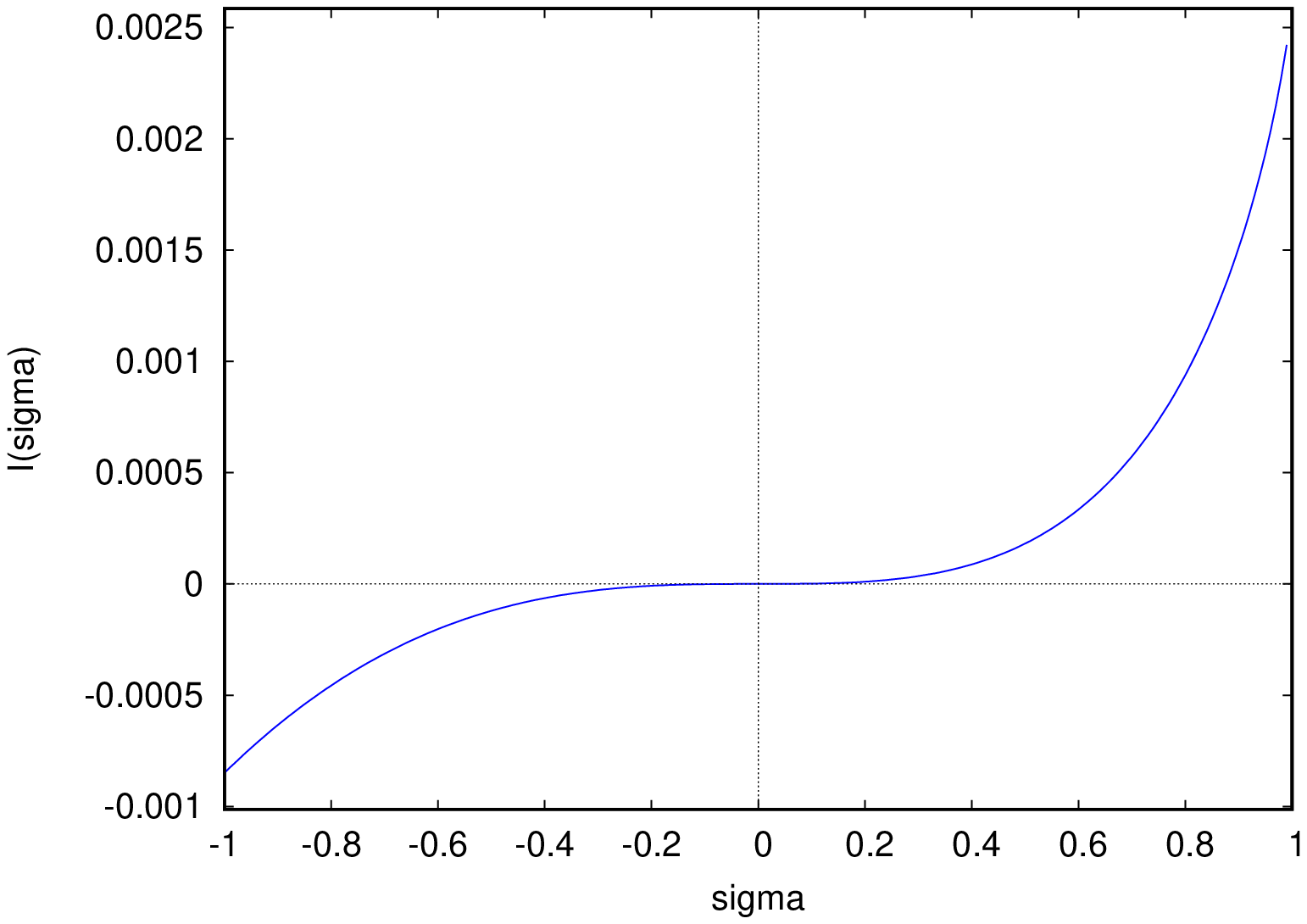}
  \caption{Plot of $I (\sigma)$ for the choice $R (p) =
    e^{-p^2}$.}
  \label{fig-I}
\end{figure}

To understand better what this means, let us consider a constant field
\begin{equation}
  \sigma (p) = \sigma \delta (p)\,.
\end{equation}
We then obtain the Legendre transform of the 1PI potential as
\begin{equation}
  F_t^T (\sigma) = \left( - \frac{1}{4} \int f + f_1 e^{2(t-T)}\right)
  \sigma
  + \frac{1}{2} \sigma^2 \cdot \left( \frac{t-T}{(4 \pi)^2} + f_2
    \right) + I (\sigma)\,,
  \end{equation}
where $I (\sigma)$ is defined by (\ref{III-function-I}).  For the
cutoff function
\begin{equation}
  R (p) = e^{- p^2}\,,
\end{equation}
$I (\sigma)$ is well defined for $\sigma < 1$.  We plot $I (\sigma)$
in Fig.~\ref{fig-I}.

$F_t^T (\sigma)$ is well defined in $\sigma < 1$.  For this to be the
Legendre transform of the corresponding 1PI potential
$G_t^T (\varphi)$, $F_t^T (\sigma)$ must be convex at least up to
negative enough $\sigma$.

The problem is that $I (\sigma)$ is concave in $\sigma < 0$.
Especially, for $- \sigma \gg 1$, we find the asymptotic behavior (see
Appendix \ref{Appendix-I} for derivation)
\begin{equation}
  I'' (\sigma) \sim - \frac{1}{2} \frac{1}{(4\pi)^2} \ln (-\sigma)\quad (- \sigma \gg 1)\,.
\end{equation}
Hence, $F_t^T (\sigma)$ is convex only in a finite range $
\sigma_{\min} < \sigma$, where $\sigma_{\min}$ is estimated as
\begin{equation}
  \frac{t-T}{(4 \pi)^2} + f_2 - \frac{1}{2} \frac{1}{(4 \pi)^2} \ln
  (-\sigma) > 0
  \Longleftrightarrow - \sigma < \exp \left( 2 (t-T) + 2 (4\pi)^2 f_2
  \right) \sim - \sigma_{\min}\,.
\end{equation}
To get a large $- \sigma_{\min} \gg 1$, we must have
\begin{equation}
  t-T + (4 \pi)^2 f_2 \gg 1\,.
\end{equation}
No matter how large we take $f_2$, we can take $t$ negative enough 
that $- \sigma_{\min}$ is small.  Hence, we cannot reach a well defined
fixed point as $t \to - \infty$.

We have shown that there is no fixed point, but for the time being let
us pretend that
\begin{equation}
 F_{t=T}^T [\sigma] = \left( f_1 - \frac{1}{4} \int_p f(p) \right)
  \sigma (0) + \frac{1}{2} \int_p \sigma (p) \sigma (-p) \left(\F (p)
    + f_2 \right) + I [\sigma]
\end{equation}
for
\begin{equation}
  f_2 \gg 1
\end{equation}
gives a ``continuum limit'' on a renormalized trajectory.  We call
this a \textbf{naive continuum limit}.  We discuss the general
properties of the 1PI potential $G_{t=T}^T$ in the following, and then
compute the 1PI action $\Gamma_{I, t=T}^T$ for the two cases of $f_1$.

\subsection*{General properties of $G_{t=T}^T (\varphi)$}

For a constant field $\sigma (p) = \sigma \delta (p)$ we obtain
\begin{equation}
  F_{t=T}^T (\sigma) = \left(f_1 - \frac{1}{4} \int_p f(p) \right)
  \sigma + \frac{1}{2} f_2 \sigma^2 + I (\sigma)\,.
\end{equation}
We then find
\begin{equation}
  \varphi = - \frac{d}{d\sigma} F_{t=T}^T (\sigma)
  = - \left(f_1 - \frac{1}{4} \int_p f(p) \right) - \sigma f_2 - I' (\sigma)\,,
\end{equation}
and
\begin{equation}
  \frac{d^2}{d\sigma^2} F_{t=T}^T (\sigma) = f_2 + I'' (\sigma)\,.
\end{equation}
$F_{t=T}^T (\sigma)$ is convex for $\sigma > \sigma_{\min}$ where
$\sigma_{\min}$ is determined by
\begin{equation}
  f_2 + I'' (\sigma_{\min}) = 0\,.
\end{equation}
$\varphi$ is a decreasing function of $\sigma > \sigma_{\min}$.  The
larger $f_2$ is, the smaller $\sigma_{\min}$ gets.  For
$- \sigma_{\min} \gg 1$, we can approximate
\begin{equation}
  I'' (\sigma_{\min}) \simeq - \frac{1}{2} \frac{1}{(4 \pi)^2} \ln
  \left[(-\sigma_{\min}) e^{A+1} \right] = - f_2\,,
\end{equation}
where $A$ is a cutoff dependent constant.  (See Appendix \ref{Appendix-I}.)
Hence, we obtain
\begin{equation}
  - \sigma_{\min} \simeq e^{2 (4 \pi)^2 f_2 - A - 1}
\end{equation}
which is very large compared with $1$ if $f_2 \gg 1$.  This gives the
maximum of $\varphi$:
\begin{align}
  \varphi_{\max}
  &= - \left( f_1 - \frac{1}{4} \int_p f(p)\right) -
    \sigma_{\min} f_2 - I' (\sigma_{\min})\nt\\
  &\simeq - f_1 + \frac{1}{2} \frac{1}{(4 \pi)^2} e^{2 (4 \pi)^2 f_2 - A
    - 1} \gg 1\,.
\end{align}
For $\varphi < \varphi_{\max}$, the relation between $\sigma$ and
$\varphi$ is one-to-one.

\begin{figure}[h]
  \centering
  \includegraphics[width=0.6\textwidth]{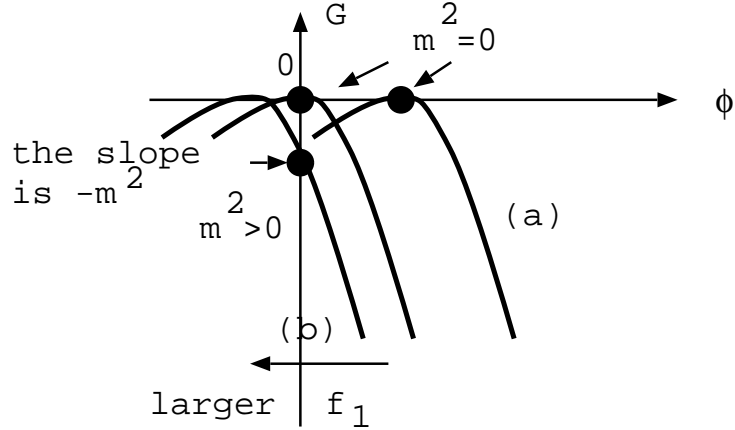}
  \caption{$f_1$ shifts $\varphi$ to the left by $f_1$}
  \label{fig-shift}
\end{figure}

Before going into more details, let us understand the role of $f_1$
for the 1PI potential
\begin{equation}
  G_{t=T}^T (\varphi) = F_{t=T}^T (\sigma) + \sigma \varphi\,,
\end{equation}
where
\begin{equation}
  \varphi = - \frac{d}{d\sigma} F_{t=T}^T (\sigma)
  = - \left( f_1 - \frac{1}{4} \int_p f(p)\right) - f_2 \sigma - I'
  (\sigma)\,.
\end{equation}
As a function of $\sigma$, 
\begin{equation}
  G_{t=T}^T (\varphi) = - \frac{1}{2} f_2 \sigma^2 + I (\sigma) -
  \sigma I' (\sigma)
\end{equation}
does not depend on $f_1$.  Hence, $f_1$ simply shifts $\varphi$
keeping the shape of $G_{t=T}^T (\varphi)$.  (See Fig.~\ref{fig-shift}.)

We have two cases to consider:
\begin{itemize}
\item[Case (a)] $f_1 < \frac{1}{4} \int_p f(p)$ (broken symmetry type)
  --- $\sigma=0$ corresponds to
  \begin{equation}
    \varphi_{\min} = - \left( f_1 - \frac{1}{4} \int_p f(p)\right) > 0\,.
  \end{equation}
\item[Case (b)]  $f_1 > \frac{1}{4} \int_p f(p)$ (symmetric type) ---
  $\varphi = 0$ corresponds to $\sigma_{\max} < 0$, which satisfies
  \begin{equation}
    \sigma_{\max} f_2 + I' (\sigma_{\max}) = -
   \left( f_1  - \frac{1}{4} \int_p f(p)\right)  < 0\,.\label{IV-sigmamax}
  \end{equation}
\end{itemize}

\subsection*{Case (a) $f_1 < \frac{1}{4} \int_p f(p)$}

Expanding in powers of $\sigma$, we obtain
\begin{equation}
  F_{t=T}^T (\sigma)
  = \underbrace{\left( f_1 - \frac{1}{4} \int_p f(p)
    \right)}_{- \varphi_{\min}} \sigma + \frac{1}{2}
  f_2 \sigma^2 + \underbrace{\sum_{n=3}^\infty \frac{1}{2n} I_n\, \sigma^n}_{I(\sigma)}\,.
\end{equation}
$\varphi > \varphi_{\min}$ is defined by
\begin{equation}
  \varphi = - \frac{d}{d\sigma} F_{t=T}^T (\sigma)
  = \varphi_{\min} - f_2 \sigma - I' (\sigma)\quad (\sigma_{\min} <
  \sigma < 0)\,.
\end{equation}
The 1PI potential can be expanded as
\begin{align}
  G_{t=T}^T (\varphi)
  &= F_{t=T}^T (\sigma) + \sigma \varphi\nt\\
  &= F_{t=T}^T (\sigma) + \sigma \varphi_{\min} + \sigma (\varphi -
    \varphi_{\min})\nt\\
  &= \sum_{n=2}^\infty \frac{1}{n!} c_n (\varphi - \varphi_{\min})^n\,,
\end{align}
where
\begin{subequations}
\begin{align}
  c_2
  &= - \frac{1}{f_2}\,,\\
  c_3
  &= c_2^3 \cdot I_3\,,\\
  c_4
  &= c_2^4 \cdot \left( 3 I_4 + 3 I_3^2 c_2 \right)\,,
\end{align}
\end{subequations}
and so on.  $f_1$ merely shifts $\varphi_{\min}$ as we have noted above.

\begin{figure}[h]
  \centering
  \includegraphics[width=0.45\textwidth]{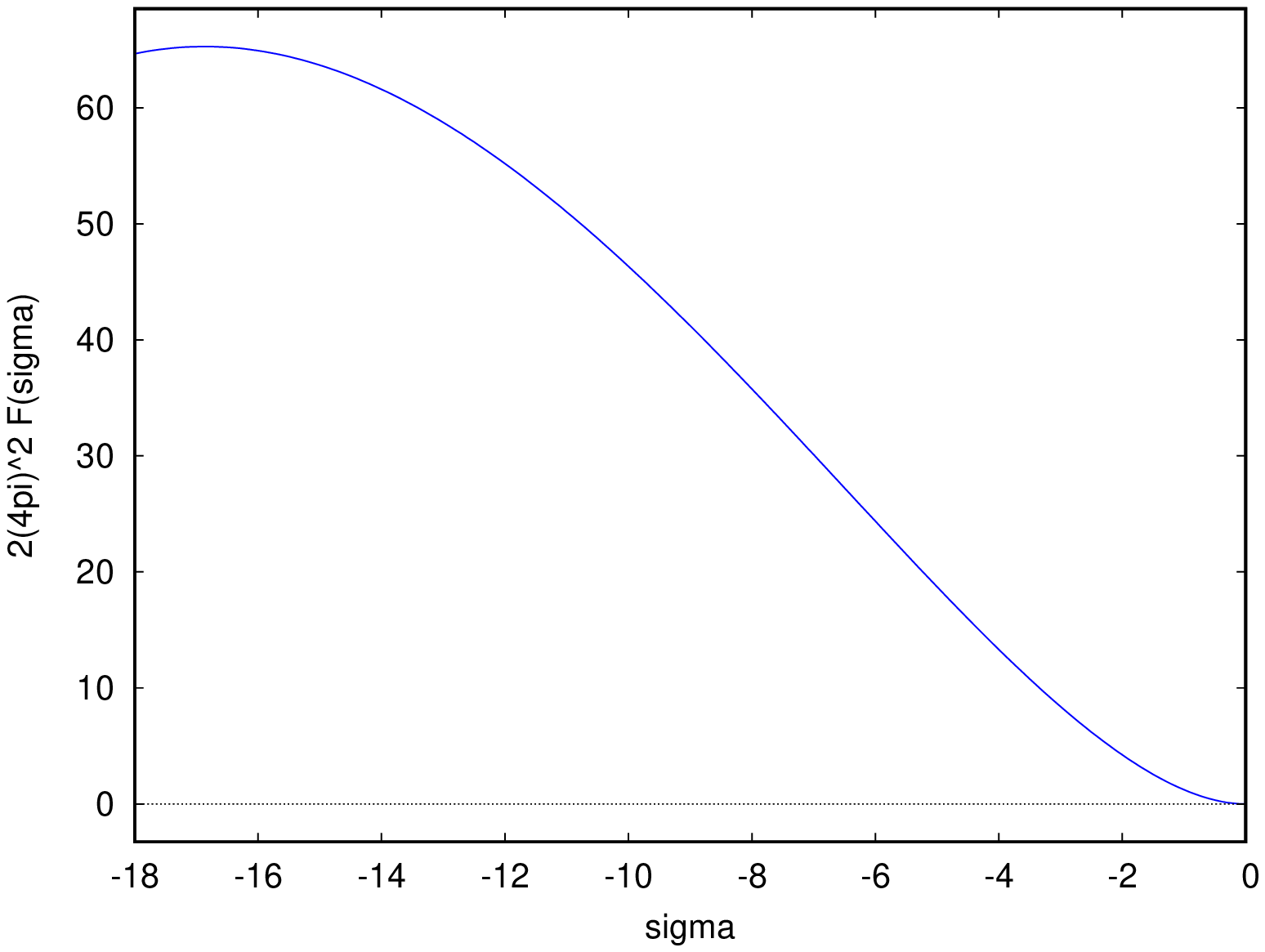}
  \hspace{0.5cm}
  \includegraphics[width=0.45\textwidth]{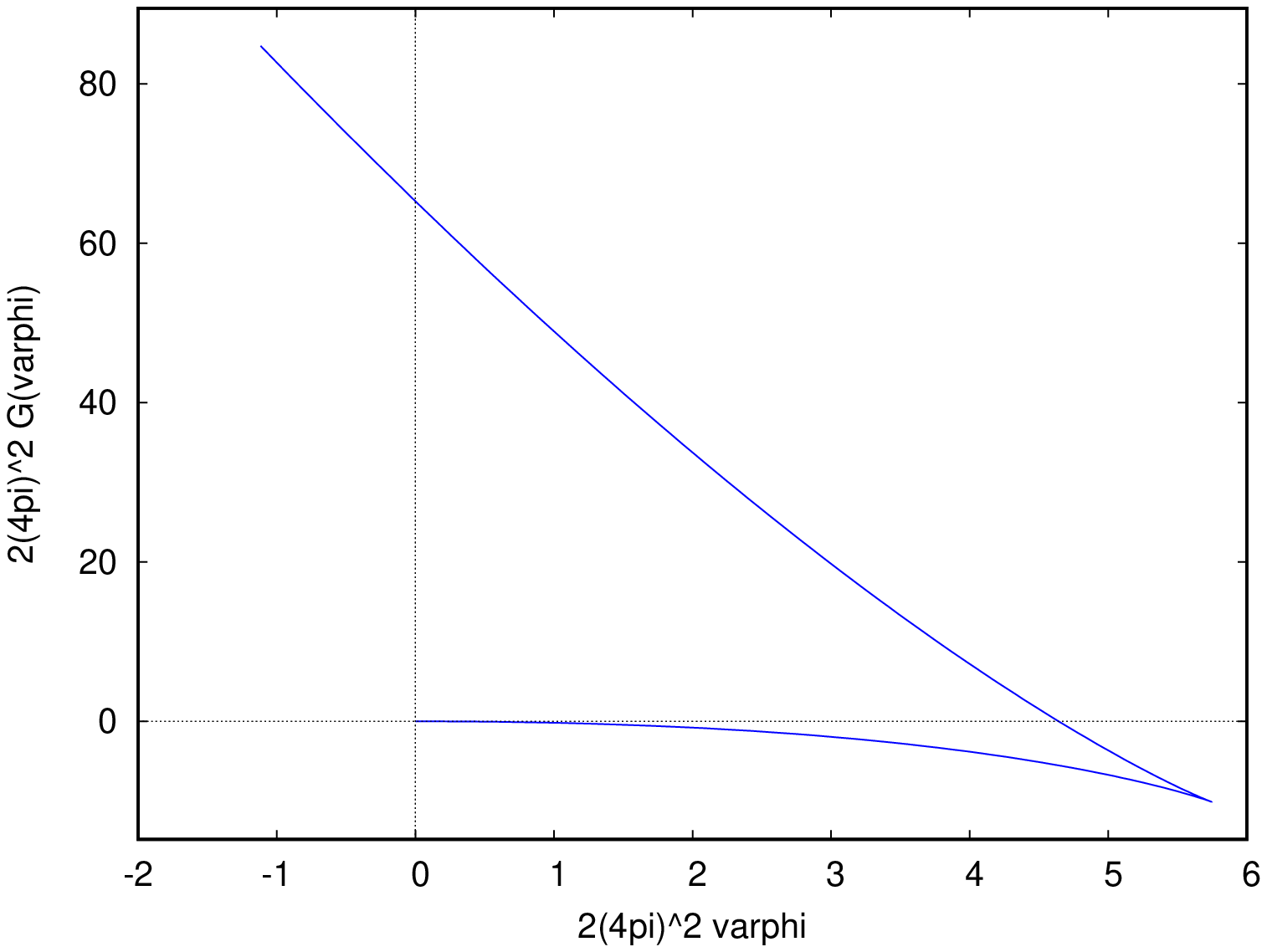}
  \caption{$2(4\pi)^2 F_{t=T}^T (\sigma), 2 (4\pi)^2 G_{t=T}^T (\varphi)$ for $2 (4 \pi)^2
    f_2 = 3$}
  \label{fig-FG}
\end{figure}

We plot $F_{t=T}^T (\sigma), G_{t=T}^T (\varphi)$ for $2(4\pi)^2 f_2 = 3,
\varphi_{\min} = 0$. (Fig.~\ref{fig-FG})
The value $2 (4\pi)^2 f_2 = 3$ (not so big) was chosen for the ease of
plotting.  $\varphi$ is well defined only up to
$\varphi_{\max} \simeq \frac{5.7}{2(4\pi)^2}$, which corresponds to
the inflection point $F_{t=T}^T\,'' (\sigma_{\min}) = 0$.  The convex
part of $G (\varphi)$ corresponds to the concave part of $F(\sigma)$
in $\sigma < \sigma_{\min}$.

We next consider the Legendre transform of
\begin{equation}
  F_{t=T}^T [\sigma] = \left( f_1 - \frac{1}{4} \int_p f(p)\right)
  \sigma (0) + \frac{1}{2} \int_p \sigma (p) \sigma (-p) \left( f_2 + \F (p)
    \right) + I [\sigma]\,.
\end{equation}
We obtain
\begin{align}
  \Gamma_{I, t=T}^T [\varphi]
  &=  F_{t=T}^T [\sigma] + \varphi_{\min} \sigma (0) + \int_p \sigma (-p) \left( \varphi (p) -
    \varphi_{\min} \delta (p) \right)\nt\\
  &=  \frac{1}{2} \int_p \sigma (p) \sigma (-p) \left( f_2 + \F (p)
    \right) + I [\sigma] + \int_p \sigma (-p) \left( \varphi (p) -
    \varphi_{\min} \delta (p) \right)\nt\\
  &= \sum_{n=2}^\infty \frac{1}{n!} \int_{p_1, \cdots, p_n} \left(
    \varphi (p_1) - \varphi_{\min} \delta (p_1) \right) \cdots  \left(
    \varphi (p_n) - \varphi_{\min} \delta (p_n) \right) \nt\\
  &\quad \times \delta
    \left(\sum_1^n p_i \right)\, c_n (p_1, \cdots, p_n)\,,
\end{align}
where
\begin{subequations}
\begin{align}
  c_2 (p, -p)
  &= - \frac{1}{f_2 + \F (p)}\,,\\
  c_3 (p_1, p_2, p_3)
  &= \prod_{i=1}^3 c_2 (p_i, - p_i) \cdot I_3 (p_1, p_2, p_3)\,,\\
  c_4 (p_1, \cdots, p_4)
  &= \prod_{i=1}^4 c_2 (p_i, - p_i) \cdot \left[
    I_4 (p_1, p_2, p_3, p_4) + I_4 (p_1, p_2, p_4, p_3) + I_4 (p_1,
    p_4, p_2, p_3)\right.\nt\\
  &\quad + I_3 (p_1, p_2, p_3+p_4) c_2 (p_1+p_2, p_3+p_4) I_3 (p_3,
    p_4, p_1+p_2)\nt\\
  &\quad\left. + (\textrm{t-, u-channels}) \right]\,,
\end{align}
\end{subequations}
and so on.

\subsection*{Tachyon pole}

We know that $F_{t=T}^T (\sigma)$ is convex only for $\sigma >
\sigma_{\min}$, and accordingly $G_{t=T}^T (\varphi)$ is concave only
for $\varphi < \varphi_{\max}$.  We are not surprised that the 1PI
action has a problem.  We can show that $c_2 (p,-p)$ has a tachyon
pole.  To see this, we use the asymptotic behavior of $\F
(p)$ (see Appendix \ref{Appendix-asymp}):
\begin{equation}
  \F (p) \overset{p^2 \gg 1}{\longrightarrow} - \frac{1}{2(4\pi)^2}
  \ln p^2 + B \,,\label{IV-calF-asymp}
\end{equation}
where the constant $B$ depends on the choice of the cutoff function $R$.
Since $\F (p=0) = 0$, there is some value of $p^2$ for which
\begin{equation}
  f_2 + \F (p) = 0
\end{equation}
is satisfied.  This corresponds to a tachyon pole.

\subsection*{Case (b) $f_1 > \frac{1}{4} \int_p f(p)$}

We denote $\sigma_{\max} < 0$ defined by Eq.~(\ref{IV-sigmamax}) as $-
m^2$ so that
\begin{equation}
   m^2 f_2 - I' (- m^2) = f_1 - \frac{1}{4} \int_p f(p) > 0\,.
\end{equation}
At $\sigma = -m^2$, we obtain $\varphi = 0$.  We then rewrite
\begin{equation}
  F_{t=T}^T (\sigma)
  = \frac{1}{2} f_2 (\sigma + m^2)^2 + I (\sigma) -
    I(-m^2) - I' (-m^2) (\sigma +m^2) + \underbrace{F_{t=T}^T (-m^2)}_{\mathrm{const}}\,.
\end{equation}
We use an equality proven in Appendix \ref{Appendix-I}:
\begin{equation}
  I (\sigma) - I (-m^2) - I' (-m^2)
  (\sigma+m^2)
  =  I(m^2; \sigma + m^2)\,,
\end{equation}
where $I (m^2; \sigma+m^2)$ is defined by
\begin{align}
  &  I(m^2; \sigma+m^2)  \equiv \frac{1}{2} \F_{m^2} \cdot (\sigma + m^2)^2
    + \frac{1}{2}  \int_p \Big[ - \ln \left(1 - (\sigma+m^2) h(m^2, p)\right) \nt\\
  &\qquad - (\sigma+m^2) h(m^2, p) - \frac{1}{2} (\sigma+m^2)^2 h(m^2, p)^2 \Big]\,.
\end{align}
Here,
\begin{equation}
  h(m^2, p) \equiv \frac{1}{p^2 + m^2 + R (p)}
\end{equation}
is a massive high-momentum propagator, and the constant $\F_{m^2}$ is
defined by
\begin{equation}
  \F_{m^2} \equiv \frac{1}{2} \int_p \left( h(m^2, p)^2 - h(p)^2
  \right)\,.
\end{equation}
Expanding in powers of $\sigma
+ m^2$, we obtain
\begin{equation}
   I (m^2; \sigma+m^2) - \frac{1}{2} \F_{m^2} \cdot
    (\sigma+m^2)^2 =  \sum_{n=3}^\infty \frac{1}{2n} I_n (m^2) \cdot
    (\sigma+m^2)^n\,,
\end{equation}
where
\begin{equation}
  I_{n \ge 3} (m^2) \equiv \int_p h (m^2, p)^n\,.
\end{equation}
Ignoring an additive constant, we obtain at last
\begin{align}
  F_{t=T}^T (\sigma)
  &= \frac{1}{2} f_2  \left(\sigma +
    m^2\right)^2 + I (m^2; \sigma+m^2)\nt\\
  &= \frac{1}{2} \left( f_2 + \F_{m^2}\right) \left(\sigma +
    m^2\right)^2 + \sum_{n=3}^\infty \frac{1}{2n} I_n (m^2) \cdot
    \left(\sigma + m^2\right)^n\,.
\end{align}

\begin{figure}[h]
\centering
\includegraphics[width=0.5\textwidth]{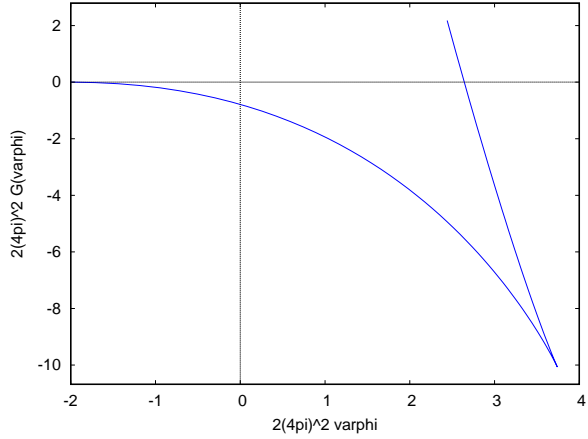}
\caption{$2 (4\pi)^2 G_{t=T}^T (\varphi)$ for $2 (4\pi)^2 f_2 = 3$ and
  $2 (4\pi)^2 (f_1 - \frac{1}{4} \int_p f(p))=2$}
\label{fig-Gshift}
\end{figure}

The Legendre transformation gives
\begin{align}
  G_{t=T}^T (\varphi)
  &= F_{t=T}^T (\sigma) + \sigma \varphi\nt\\
  &= - m^2 \varphi + F_{t=T}^T (\sigma) + \varphi (\sigma + m^2)\nt\\
  &= - m^2 \varphi + \sum_{n=2}^\infty \frac{1}{n!} c_n (m^2)
    \varphi^n\,,
\end{align}
where
\begin{subequations}
\begin{align}
  c_2 &= - \frac{1}{f_2 + \F_{m^2}}\,,\\
  c_3 &= c_2^3 \cdot I_3 (m^2)\,,\\
  c_4 &= c_2^4 \cdot \left( 3 I_4 (m^2) + 3 I_3 (m^2)^2 c_2 \right)\,,
\end{align}
\end{subequations}
and so on.  If $f_2 \gg \F_{m^2}$, we can expand
\begin{equation}
  c_2 = - \lambda + (-\lambda)^2 \F_{m^2} + (-\lambda)^3 \F_{m^2}^2 +
  \cdots\,,
\end{equation}
where the $\phi^4$ coupling is given by
\begin{equation}
  \lambda \equiv \frac{1}{f_2} \ll 1\,.
\end{equation}
We plot $G_{t=T}^T (\varphi)$ for $2(4\pi)^2 f_2 = 3$ and $2(4\pi)^2 \left(f_1 -
  \frac{1}{4} \int_p f(p) \right) = 2$.
(Fig.~\ref{fig-Gshift})
Note this is merely shifted to the left compared with the plot of
$G_{t=T}^T$ in Fig.~\ref{fig-FG}.  The slope at $\varphi = 0$ gives $- m^2$.

We next consider the functional
\begin{align}
  F_{t=T}^T [\sigma]
  &= \left( f_1 - \frac{1}{4} \int_p f(p) \right) \sigma (0) +
    \frac{1}{2} \int_p \sigma (p) \sigma (-p) \left(\F (p) +
    f_2\right) + I [\sigma]\nt\\
  &= \underbrace{\mathrm{const}}_{\textrm{to be ignored}} +
    \underbrace{\left( f_1 - \frac{1}{4} \int_p f(p) 
    + I' (-m^2) - m^2 f_2\right)}_{=0}  \sigma (0)\nt\\
  &\quad + \frac{1}{2} \int_p \left(\sigma (p) + m^2 \delta
    (p)\right)\left(\sigma (-p) + m^2 \delta (p)\right) \F (p) +
    I(m^2) [\sigma + m^2 \delta]\,,
\end{align}
where
\begin{align}
  I (m^2) [\sigma + m^2 \delta]
  &\equiv I [\sigma] - I [-m^2\delta] - \int_p \frac{\delta
    I[\sigma]}{\delta \sigma (p)}\Bigg|_{\sigma = -m^2 \delta}
    \left(\sigma (p) + m^2 \delta (p)\right)\nt\\
  &= \frac{1}{2} \int_p \left( \F (m^2; p) - \F (p)\right)
    \left(\sigma (p) + m^2 \delta (p)\right)\left(\sigma (-p) + m^2
    \delta (p)\right)\nt\\
  &\quad +
    \sum_{n=3}^\infty \frac{1}{2n} \int_{p_1, \cdots, p_n}
    \left(\sigma (p_1) + m^2 \delta (p_1)\right) \cdots \left( \sigma
    (p_n) + m^2 \delta (p_n)\right)\,\delta \left(\sum_1^n
    p_i\right)\nt\\
  &\qquad \times I_n (m^2; p_1, \cdots, p_n) \,,
\end{align}
and
\begin{align}
    \F (m^2; p)
    &\equiv \frac{1}{2} \int_q \left( h(m^2, q) h (m^2, q+p) - h(q)^2
      \right)\,,\\
    I_{n \ge 3} (m^2; p_1, \cdots, p_n)
    &\equiv \int_p h(m^2, p) h(m^2, p+p_1) \cdots h \left( m^2,
      p+\sum_{i=1}^{n-1} p_i \right)\,.
\end{align}
See Appendix \ref{Appendix-I} for derivation.  We thus obtain
\begin{align}
  F_{t=T}^T [\sigma]
  &= \frac{1}{2} \int_p \left(\sigma (p) + m^2 \delta
    (p)\right)\left(\sigma (-p) + m^2 \delta (p)\right) \left(f_2 + \F (m^2; p)\right)
    \nt\\
   &\quad +  \sum_{n=3}^\infty \frac{1}{2n} \int_{p_1, \cdots, p_n}
    \left(\sigma (p_1) + m^2 \delta (p_1)\right) \cdots \left( \sigma
    (p_n) + m^2 \delta (p_n)\right)\,\delta \left(\sum_1^n
    p_i\right)\nt\\
  &\qquad \times I_n (m^2; p_1, \cdots, p_n) \,.
    \label{IX-FTT-symmetric}
\end{align}
The Legendre transformation gives
\begin{align}
  \Gamma_{I, t=T}^T [\varphi]
  &= F_{t=T}^T [\sigma ] + \int_p \varphi (p) \left( \sigma (-p) + m^2
    \delta (p)\right) - m^2 \varphi (0)\nt\\
  &= - m^2 \varphi (0) + \sum_{n=2}^\infty \frac{1}{n!} \int_{p_1,
    \cdots, p_n} \varphi (p_1) \cdots \varphi (p_n)\,\delta
    \left(\sum_1^n p_i\right) \, c_n (p_1, \cdots, p_n)\,,
\end{align}
where
\begin{subequations}
\begin{align}
  c_2 (p, -p)
  &= - \frac{1}{f_2 + \F (m^2; p)}\,,\\
  c_3 (p_1, p_2, p_3)
  &= \prod_{i=1}^3 c_2 (p_i, -p_i) \cdot I_3 (m^2; p_1, p_2, p_3)\,,\\
  c_4 (p_1, \cdots, p_4)
  &= \prod_{i=1}^4 c_2 (p_i, -p_i) \cdot \left[
    I_4 (m^2; p_1, p_2, p_3, p_4) + I_4 (m^2; p_1, p_3, p_2,
    p_4)\right.\nt\\
  &\qquad + I_4 (m^2; p_1, p_2, p_4, p_3) \nt\\
  &\quad + I_3 (m^2; p_1, p_2, p_3+p_4) c_2 (p_1+p_2, p_3+p_4) I_3
    (m^2; p_3, p_4, p_1+p_2)\nt\\
  &\left.\quad + (\textrm{t-, u-channels}) \right]\,,
\end{align}
\end{subequations}
and so on.  As in the broken symmetry case, we can show that
$c_2 (p, -p)$ has a tachyon pole, lying at large $p^2$ if $f_2 \gg 1$.

\section{Effective action of the naive continuum
  limit\label{sec-effective action}}

Let us summarize the problems we have found with  the naive continuum
limit:
\begin{enumerate}
\item $F_{t=T}^T (\sigma)$ is convex only for $\sigma > \sigma_{\min}$.
\item $G_{t=T}^T (\varphi)$ is concave only for $\varphi <
  \varphi_{\max}$.
\item $\Gamma_{I, t=T}^T [\varphi]$ suffers from a tachyon pole, an
  indication that its second order differential is not negative
  definite.
\end{enumerate}
What we have computed so far have a dimensionless IR cutoff of order
$1$.  In this and next sections we continue studying the naive
continuum limit by computing its effective action and potential.  We
restore physical dimensions, and bring down the IR cutoff $\Lambda$
all the way to zero.  In Appendix \ref{Appendix-dimension} we have
summarized the rule (or recipe) for restoring physical
dimensions.\footnote{The 1PI Wilson action for finite $\Lambda$ in
  large $N$ has been worked out in \cite{Blaizot:2005xy}.}

In the following we find it convenient to introduce a new notation
\begin{equation}
  F (f_1, f_2) [\sigma] \equiv
  \left(f_1 - \frac{1}{4} \int_p f(p) \right) \sigma (0) +
  \frac{1}{2} \int_p \sigma (p) \sigma (-p) \left( \F (p) + f_2
  \right) + I [\sigma]\,.
\end{equation}
The Legendre transform of the effective action is given by
\begin{equation}
  F_\eff (f_1 \mu^2, f_2) [\sigma]
\equiv \lim_{\Lambda \to 0+} F \left( f_1 \frac{\mu^2}{\Lambda^2} ,
  f_2 + \frac{1}{(4 \pi)^2} \ln \frac{\mu}{\Lambda} \right)
[\sigma_\Lambda]\,,
\end{equation}
where
\begin{equation}
\sigma_\Lambda (p) \equiv \Lambda^2 \sigma (p \Lambda)\,.
\end{equation}
Analogously, the interaction part of the effective action is given by
\begin{equation}
  \Gamma_{I, \eff} (f_1 \mu^2, f_2) [\varphi]
  \equiv \lim_{\Lambda \to 0+} \Gamma_I \left( f_1
    \frac{\mu^2}{\Lambda^2}, f_2 + \frac{1}{(4 \pi)^2} \ln
    \frac{\mu}{\Lambda}\right) [\varphi_\Lambda]\,,
\end{equation}
where
\begin{equation}
  \varphi_\Lambda (p) \equiv \Lambda^2 \varphi (p \Lambda)\,,
\end{equation}
and $\Gamma_I (f_1, f_2) [\varphi]$ is the Legendre transform of $F
(f_1, f_2) [\sigma]$.

The order parameter is $f_1$.  If $f_1 < 0$, the O$(N)$ symmetry is
spontaneously broken, and if $f_1 > 0$, it is unbroken.

\subsection*{Case (a) broken phase $f_1 < 0$}

Let us compute
\begin{align}
&F \left( f_1 \frac{\mu^2}{\Lambda^2} ,
  f_2 + \frac{1}{(4 \pi)^2} \ln \frac{\mu}{\Lambda} \right)
[\sigma_\Lambda]\nt\\
  &= f_1 \frac{\mu^2}{\Lambda^2} \sigma_\Lambda (0) + \frac{1}{2}
    \int_p \sigma_\Lambda (p) \sigma_\Lambda (-p) \left( \F (p) + f_2
    + \frac{1}{(4 \pi)^2} \ln \frac{\mu}{\Lambda} \right) + I
    [\sigma_\Lambda]\nt\\
  &= f_1 \mu^2 \sigma (0) + \frac{1}{2} \int_p \Lambda^4 \sigma (p
    \Lambda) \sigma (- p \Lambda) \left( \F (p) + f_2 + \frac{1}{(4
    \pi)^2} \ln \frac{\mu}{\Lambda} \right)\nt\\
  &\quad + \sum_{n=3}^\infty \frac{1}{2n} \int_{p_1, \cdots, p_n}
    \Lambda^{2n} \sigma (p_1 \Lambda) \cdots \sigma (p_n \Lambda)\,
    \delta \left(\sum_1^n p_i\right)\, I_n (p_1  \cdots, p_n )\nt\\
  &=  f_1 \mu^2 \sigma (0) + \frac{1}{2} \int_p \sigma (p) \sigma (-p)
    \left( \F \left(\frac{p}{\Lambda}\right) + f_2 +
    \frac{1}{(4\pi)^2} \ln \frac{\mu}{\Lambda} \right)\nt\\
  &\quad + \sum_{n=3}^\infty \frac{1}{2n} \int_{p_1, \cdots, p_n}
    \sigma (p_1) \cdots \sigma (p_n)\,\delta \left(\sum_1^n
    p_i\right)\, \Lambda^{4-2n} I_n \left( \frac{p_1}{\Lambda},
    \cdots, \frac{p_n}{\Lambda}\right)\,.
\end{align}
It is easy to find
\begin{align}
& \Lambda^{4-2n} I_n  \left( \frac{p_1}{\Lambda},
                 \cdots, \frac{p_n}{\Lambda}\right)\nt\\
  &=  \Lambda^{4-2n} \int_p \frac{1}{p^2 + R(p)}
    \frac{1}{\left(p + \frac{p_1}{\Lambda}\right)^2 + R \left( p +
    \frac{p_1}{\Lambda}\right)}
    \cdots \frac{1}{\left(p + \frac{1}{\Lambda} \sum_1^{n-1} p_i
    \right)^2
    + R \left(p + \frac{1}{\Lambda} \sum_1^{n-1} p_i
    \right) }\nt\\
  &=     \int_p \frac{1}{p^2 + \Lambda^2 R(p/\Lambda)} \frac{1}{(p+p_1)^2 +
    \Lambda^2 R \left(\frac{p+p_1}{\Lambda}\right)} \cdots
    \frac{1}{\left(p + \sum_1^{n-1} p_i\right)^2 + \Lambda^2 R
    \left(\frac{p+\sum_1^{n-1}p}{\Lambda}\right)}\nt\\
  &\overset{\Lambda \to 0+}{\longrightarrow}
    \int_p \frac{1}{p^2 (p+p_1)^2 \cdots \left( p + \sum_1^{n-1}
    p_i\right)^2}
    \equiv I_{n, \eff} (p_1, \cdots, p_n)\,.\label{V-Ineff}
\end{align}
Using the asymptotic expansion (\ref{IV-calF-asymp}), we obtain
\begin{equation}
  \lim_{\Lambda \to 0+} \left(\F \left(\frac{p}{\Lambda}\right) +
    \frac{1}{(4 \pi)^2} \ln \frac{\mu}{\Lambda} \right)
  = - \frac{1}{2(4 \pi)^2} \ln \frac{p^2}{\mu^2} + B\,.
\end{equation}
We thus obtain
\begin{align}
  F_\eff (f_1 \mu^2, f_2) [\sigma]
  &= f_1 \mu^2 \sigma (0) + \frac{1}{2} \int_p \sigma (p) \sigma (-p)
    \left( f_2 + \frac{1}{2 (4\pi)^2} \ln \frac{\mu^2}{p^2}  + B
    \right)\nt\\
  &\quad + \sum_{n=3}^\infty \frac{1}{2n} \int_{p_1, \cdots, p_n}
    \sigma (p_1) \cdots \sigma (p_n)\, \delta \left(\sum_1^n
    p_i\right)\, I_{n, \eff} (p_1, \cdots, p_n)\,.\label{V-Feff}
\end{align}

Hence, the effective action is obtained by the Legendre transformation as
\begin{align}
  &\Gamma_{I, \eff} (f_1 \mu^2, f_2) [\varphi]
  = F_\eff (f_1 \mu^2, f_2) [\sigma] - f_1 \mu^2 \sigma (0) + \int_p \sigma (-p)
    \left(\varphi (p) + f_1 \mu^2 \delta (p)\right)\nt\\
  &= \sum_{n=2}^\infty \frac{1}{n!} \int_{p_1, \cdots, p_n}
    \left( \varphi (p_1) - \varphi_{\min} \delta (p_1)\right) \cdots
    \left(\varphi (p_n) - \varphi_{\min} \delta (p_n)\right)\,\delta
    \left(\sum_1^n p_i\right)\cdot c_n (p_1, \cdots, p_n)\,,
\end{align}
where the expectation value of $\varphi$ is
\begin{equation}
  \langle \varphi (p)\rangle = \varphi_{\min} \,\delta (p) \equiv - f_1 \mu^2 \delta (p) > 0\,,
\end{equation}
and the vertices are given by
\begin{subequations}
\begin{align}
  c_2 (p, -p)
  &= - \frac{1}{f_2 +  \frac{1}{2 (4\pi)^2} \ln \frac{\mu^2}{p^2} +
    B} = \frac{2(4\pi)^2}{\ln \frac{p^2}{\mu^2 e^{2(4\pi)^2 (f_2+B)}}} \,,\\
  c_3 (p_1, p_2, p_3)
  &= \prod_{i=1}^3 c_2 (p_i, -p_i)\cdot I_{3, \eff} (p_1, p_2, p_3)\,,\\
  c_4 (p_1, \cdots, p_4)
  &= \prod_{i=1}^4 c_2 (p_i, -p_i) \cdot \left[  I_{4, \eff} (p_1, p_2,
    p_3, p_4)\right.\nt\\
  &\quad + I_{4, \eff} (p_1, p_2, p_4, p_3) + I_{4, \eff} (p_1, p_3, p_2,
    p_4)\nt\\
  &\quad + I_{3, \eff} (p_1, p_2, p_3+p_4) c_2 (p_1+p_2. p_3+p_4)
    I_{3, \eff} (p_3, p_4, p_1+p_2)\nt\\
  &\left.\quad + (\textrm{t-, u-channels}) \right]\,,
\end{align}
\end{subequations}
and so on.

It is easy to locate the \textbf{tachyon pole} of $c_2 (p, -p)$ at 
\begin{equation}
  p^2 = \mu^2 e^{2 (4 \pi)^2 (f_2 + B)}\,.
\end{equation}
This reproduces the result first obtained in \cite{Coleman:1974jh}.

\subsection*{Case (b) symmetric phase $f_1 > 0$}

We start with a somewhat lengthy determination of the physical squared
mass.  The result will be used also for the calculation of the
effective potential in the next section.

We start from
\begin{equation}
  F_\eff (f_1 \mu^2, f_2) [\sigma]
  = f_1 \mu^2\sigma  (0) + \frac{1}{2}
    \int_p \sigma (p) \sigma (-p) \left( \F
      \left(\frac{p}{\Lambda}\right) + f_2 
    + \frac{1}{(4 \pi)^2} \ln \frac{\mu}{\Lambda} \right) + I
  [\sigma_\Lambda]\,.
\end{equation}
For the constant field, we find
\begin{equation}
  \sigma_\Lambda (p) = \sigma_\Lambda \delta (p)\,.
\end{equation}
Since
\begin{equation}
  \sigma_\Lambda (p) = \Lambda^2 \sigma (p \Lambda) = \Lambda^2 \sigma
  \, \delta (p \Lambda) = \frac{\sigma}{\Lambda^2} \delta (p)\,,
\end{equation}
we obtain
\begin{equation}
  \sigma_\Lambda =  \frac{\sigma}{\Lambda^2} \,.
\end{equation}
Hence, we obtain
\begin{equation}
  F_\eff (f_1 \mu^2, f_2; \sigma)
  = f_1 \mu^2 \sigma + \frac{1}{2} \sigma^2 \left( f_2 +  \frac{1}{(4
      \pi)^2} \ln \frac{\mu}{\Lambda} \right)
  + \Lambda^4 I \left(\frac{\sigma}{\Lambda^2}\right)\,,
  \label{V-Feff-intermediate}
\end{equation}
where the limit $\Lambda \to 0+$ is implied.  We define the squared
physical mass $\mph^2$ by
\begin{equation}
  \frac{d}{d\sigma} F_\eff (f_1 \mu^2, f_2; \sigma)\Big|_{\sigma = -
    \mph^2} = 0\,.
\end{equation}
This gives
\begin{equation}
   f_1 \mu^2 - \mph^2 \left(f_2 +
    \frac{1}{(4\pi)^2} \ln \frac{\mu}{\Lambda}\right) + \Lambda^2  I'
  \left( - \frac{\mph^2}{\Lambda^2}\right) = 0\,.
\end{equation}
Using the asymptotic behavior, obtained in Appendix \ref{Appendix-asymp},
\begin{equation}
  I' (\sigma) \overset{- \sigma \gg 1}{\longrightarrow} 
- \frac{1}{2} \frac{1}{(4 \pi)^2} \sigma \left( \ln (-\sigma) + A
\right) + \frac{1}{4} \int_p f(p)\,,
\end{equation}
we obtain
\begin{equation}
  f_1 \mu^2 - \mph^2 \left( f_2 + \frac{1}{(4 \pi)^2} \ln
    \frac{\mu}{\Lambda}\right)
  + \Lambda^2 \left[  \frac{1}{2} \frac{1}{(4 \pi)^2}
  \frac{\mph^2}{\Lambda^2}
  \left( \ln \frac{\mph^2}{\Lambda^2} + A \right) + \frac{1}{4} \int_p
  f(p) \right] = 0\,.
\end{equation}
Taking $\Lambda \to 0+$, we obtain
\begin{equation}
  f_1 \mu^2 - \mph^2 \left[ f_2 - \frac{1}{2} \frac{1}{(4 \pi)^2}
    \left( \ln \frac{\mph^2}{\mu^2} + A \right) \right] = 0\,.
\end{equation}
We can rewrite this as
\begin{equation}
  \frac{\mph^2}{\Lambda_L^2} \ln \frac{\mph^2}{\Lambda_L^2} = -
  \frac{2 (4\pi)^2 f_1 \mu^2}{\Lambda_L^2} < 0\,,\label{V-equation-mph2}
\end{equation}
where we define
\begin{equation}
  \Lambda_L \equiv \mu \, e^{(4 \pi)^2 f_2 - \frac{A}{2}}\,.\label{V-Landau}
\end{equation}
$\Lambda_L \gg \mu$ if $(4 \pi)^2 f_2 \gg 1$.
Eq.~(\ref{V-equation-mph2}) can be solved only if
\begin{equation}
  2 (4 \pi)^2 f_1 \mu^2  < \frac{1}{e} \Lambda_L^2\,.
\end{equation}
The solution is then given by the lower branch of the Lambert $W$
function \cite{Corless:1996zz, wiki:Lambert_W_function}, satisfying
$e^{W_{-1} (x)} W_{-1} (x) = x$, as
\begin{equation}
 \mph^2 =  \Lambda_L^2 \exp \left[ W_{-1}
    \left(  - \frac{2 (4 \pi)^2   f_1 \mu^2}{\Lambda_L^2}
    \right)\right]\,.
  \label{V-mph2}
\end{equation}
Since $W_{-1} (x) < -1$ for $- \frac{1}{e} < x < 0$, we obtain
\begin{equation}
  \mph^2 < \frac{1}{e} \Lambda_L^2\,.
\end{equation}
For $- x \ll 1$, we can expand
\begin{equation}
  W_{-1} (x) = \ln (-x) - \ln (- \ln (-x)) + \cdots\,.
\end{equation}
Hence, for $2 (4\pi)^2 f_1 \mu^2 \ll \Lambda_L^2$, we find
\begin{equation}
  \frac{\mph^2}{\mu^2} \simeq e^{2 (4\pi)^2 f_2-A} \frac{2 (4 \pi)^2
    f_1 e^{A-2(4\pi)^2 f_2}}{- \ln \left( 2 (4 \pi)^2
      f_1 e^{A-2(4\pi)^2 f_2}\right)} = \frac{2(4\pi)^2 f_1}{
    2(4\pi)^2 f_2 - A - \ln 2(4 \pi)^2 f_1}\,.
\end{equation}
So much for the squared physical mass.

Now, using (\ref{IX-FTT-symmetric}), we compute
\begin{align}
&  F_\eff (f_1 \mu^2, f_2) [\sigma]
  = \lim_{\Lambda \to 0+} F \left(f_1 \frac{\mu^2}{\Lambda^2}, f_2 +
    \frac{1}{(4 \pi)^2} \ln \frac{\mu}{\Lambda}\right) [\sigma_\Lambda]\nt\\
  &= \lim_{\Lambda \to 0+} \lb \frac{1}{2} \int_p \left(
    \sigma_\Lambda (p) + \frac{\mph^2}{\Lambda^2}  \delta
  (p)\right)\left( \sigma_\Lambda (-p) + \frac{\mph^2}{\Lambda^2} \delta (p)\right)
  \left( f_2 + \frac{1}{(4 \pi)^2} \ln \frac{\mu}{\Lambda} \right)\right.\nt\\
  &\left.\qquad\qquad + I \left(\frac{\mph^2}{\Lambda^2}\right) \left[
    \sigma_\Lambda + \frac{\mph^2}{\Lambda} \delta \right] \rb\nt\\
  &= \lim_{\Lambda \to 0+} \lb \frac{1}{2} \int_p \left[ f_2 + \F
    \left( \frac{\mph^2}{\Lambda^2},
    \frac{p}{\Lambda}\right) + \frac{1}{(4 \pi)^2} \ln \frac{\mu}{\Lambda}
    \right] \left(\sigma (p) + 
    \mph^2 \delta (p) \right)\left(\sigma (-p) +
    \mph^2 \delta (p) \right)\right.\nt\\
  &\qquad\qquad + \sum_{n=3}^\infty \frac{1}{2n} \int_{p_1, \cdots, p_n}
    \left(\sigma (p_1 ) +  \mph^2 \delta  (p_1 ) \right)
    \cdots  \left(\sigma (p_n ) +    \mph^2 \delta
    (p_n) \right)\, \delta \left( \sum_1^n p_i \right)\nt\\
  &\left.\qquad\qquad\qquad \times \Lambda^{-2(n-2)} I_n \left(
    \frac{\mph^2}{\Lambda^2}; \frac{p_1}{\Lambda}, \cdots, 
    \frac{p_n}{\Lambda}\right)\rb\,.
\end{align}
Let us calculate the $\Lambda \to 0+$ limits.  We first consider
\begin{align}
&   \F \left(\frac{\mph^2}{\Lambda^2},
      \frac{p}{\Lambda}\right)  +
                 \frac{1}{(4 \pi)^2} \ln \frac{\mu}{\Lambda}
                 \nt\\
  &= \frac{1}{2} \int_q \left[ h \left(\frac{\mph^2}{\Lambda^2},
    q\right) h\left(\frac{\mph^2}{\Lambda^2},
    q+\frac{p}{\Lambda}\right) -  h (q)^2 \right]
 + \frac{1}{(4 \pi)^2} \ln \frac{\mu}{\Lambda}\nt  \\
  &= \frac{1}{2} \int_q \left[ h \left(\frac{\mph^2}{\Lambda^2},
    q\right) h\left(\frac{\mph^2}{\Lambda^2},
    q+\frac{p}{\Lambda}\right) -  h\left(\frac{\mph^2}{\Lambda^2},
    q\right)^2  \right]   + \F_{\frac{\mph^2}{\Lambda^2}}
    + \frac{1}{(4 \pi)^2} \ln \frac{\mu}{\Lambda}\,.
\end{align}
In Appendix \ref{Appendix-asymp} we derive
\begin{equation}
  \F_{m^2} \equiv \frac{1}{2} \int_q \left( h(m^2, q)^2 -
    h(q)^2\right) \overset{m^2 \to +\infty}{\longrightarrow}
  - \frac{1}{2 (4 \pi)^2} \ln m^2 + C\,,
\end{equation}
where $C$ is a constant dependent on the cutoff function $R$.
Since
\begin{align}
  &\frac{1}{2} \int_q \left( h \left(\frac{\mph^2}{\Lambda^2},
    q\right) h\left(\frac{\mph^2}{\Lambda^2},
    q+\frac{p}{\Lambda}\right) -  h\left(\frac{\mph^2}{\Lambda^2},
    q\right)^2  \right)\nt\\
  &= \frac{1}{2} \int_q  \left( \frac{1}{q^2 + \mph^2 + \Lambda^2
    R (q/\Lambda)} \frac{1}{(q+p)^2 + \mph^2 + \Lambda^2 R
    \left(\frac{q+p}{\Lambda}\right)}\right.\nt\\
  &\left.\qquad - \frac{1}{\left(q^2 + \mph^2 + \Lambda^2
    R (q/\Lambda)\right)^2} \right)\nt\\
  &\overset{\Lambda \to 0+}{\longrightarrow}
    \frac{1}{2} \int_q \left( \frac{1}{q^2 + \mph^2} \frac{1}{(q+p)^2 +
    \mph^2}   - \frac{1}{\left(q^2+\mph^2\right)^2} \right)\,,
\end{align}
we obtain
\begin{align}
&  I_{2, \eff} (\mph^2; p, -p) \equiv   \lim_{\Lambda \to 0+} \left( \F
                             \left(\frac{\mph^2}{\Lambda^2}, 
      \frac{p}{\Lambda}\right)  +
                 \frac{1}{(4 \pi)^2} \ln \frac{\mu}{\Lambda}
                 \right) \nt\\
  &\quad=  \frac{1}{2} \int_q \left( \frac{1}{q^2 + \mph^2} \frac{1}{(q+p)^2 +
    \mph^2}   - \frac{1}{\left(q^2+\mph^2\right)^2} \right)
    - \frac{1}{2(4\pi)^2} \ln \frac{\mph^2}{\mu^2} + C \,.
\end{align}
Note that this has a finite massless limit:
\begin{equation}
  I_{2, \eff} (\mph^2; p, -p) \overset{\mph^2 \to 0}{\longrightarrow}
  = \lim_{\Lambda \to 0+} \left( \F \left(\frac{p}{\Lambda}\right) +
    \frac{1}{(4\pi)^2} \ln \frac{\mu}{\Lambda}\right) = - \frac{1}{
    (4\pi)^2} \ln \frac{p}{\mu} + B\,.\label{V-I2eff}
\end{equation}

We next consider, for $n \ge 3$,
\begin{align}
&  \Lambda^{-2(n-2)} I_n \left( \frac{\mph^2}{\Lambda^2};
                 \frac{p_1}{\Lambda}, \cdots, \frac{p_n}{\Lambda}\right)\nt\\
  &= \Lambda^{-2(n-2)} \int_p \frac{1}{h \left(
    \frac{\mph^2}{\Lambda^2}, p\right)} \cdots \frac{1}{h \left(
    \frac{\mph^2}{\Lambda^2}, p + \sum_1^{n-1}
    \frac{p_i}{\Lambda}\right)} \nt\\
  &=  \Lambda^{4-2n} \int_p \frac{1}{h \left(
    \frac{\mph^2}{\Lambda^2}, \frac{p}{\Lambda}\right)} \cdots \frac{1}{h \left(
    \frac{\mph^2}{\Lambda^2}, \frac{p + \sum_1^{n-1}
    p_i}{\Lambda}\right)} \nt\\
  &= \int_p \frac{1}{p^2 + \mph^2 + \Lambda^2 R (p)} \cdots
    \frac{1}{\left( p + \sum_1^{n-1} p_i \right)^2 + \mph^2 +
    \Lambda^2 R \left( p + \sum_1^{n-1} p_i \right)}\nt\\
  &\overset{\Lambda \to 0+}{\longrightarrow}
    \int_p \frac{1}{p^2 + \mph^2} \cdots \frac{1}{\left(p +
    \sum_1^{n-1} p_i \right)^2 + \mph^2} \equiv I_{n, \eff} (\mph^2;
    p_1, \cdots, p_n)\,.\label{V-Ineff-massive}
\end{align}

We thus obtain
\begin{align}
  F_\eff (f_1 \mu^2, f_2) [\sigma]
  &= \frac{1}{2} \int_p \left(\sigma (p) + \mph^2 \delta
    (p)\right)\left( \sigma (-p) + \mph^2 \delta (p)\right) \left(
    f_2 + I_{2, \eff} (\mph^2; p, -p) \right)\nt\\
  &\quad + \sum_{n=3}^\infty \frac{1}{2n} \int_{p_1, \cdots, p_n}
    \int_{p_1, \cdots, p_n} \left( \sigma (p_1) + \mph^2 \delta
    (p_1)\right) \cdots \left( \sigma (p_n) + \mph^2 \delta
    (p_n)\right)\nt\\
  &\qquad\times \delta \left(\sum_1^n p_i\right) \, I_{n, \eff} (\mph^2; p_1, \cdots, p_n)\,.
\end{align}
This gives
\begin{align}
  \Gamma_{I, \eff} (f_1 \mu^2, f_2) [\varphi]
  &= - \mph^2 \varphi (0) + F_\eff (f_1 \mu^2, f_2) [\sigma] + \int_p \varphi (p)
  \left( \sigma (-p) + \mph^2 \delta (p)\right)\nt\\
  &= - \mph^2 \varphi (0) +  \sum_{n=2}^\infty \frac{1}{n!} \int_{p_1, \cdots, p_n}
  \varphi (p_1) \cdots \varphi (p_n) \, \delta \left(\sum_1^n
    p_i\right)\, c_n (p_1, \cdots, p_n)\,,
\end{align}
where
\begin{subequations}
\begin{align}
  c_2 (p, -p)
  &= - \frac{1}{f_2 + I_{2, \eff} (\mph^2; p, -p)}\,,\\
  c_3 (p_1, p_2, p_3)
  &= \prod_{i=1}^3 c_2 (p_i, -p_i) \cdot I_{3, \eff} (p_1, p_2,
    p_3)\,,
\end{align}
\end{subequations}
and so on.

As in the broken phase, $I_{2, \eff} (\mph^2; p, -p)$ has a tachyon
pole.  For $p^2 \gg \mph^2$, we can ignore the mass, and we find
\begin{equation}
  I_{2, \eff} (\mph^2; p, -p) \overset{p^2 \gg \mph^2}{\longrightarrow}
  - \frac{1}{2(4\pi)^2} \ln \frac{p^2}{\mu^2} + B\,.
\end{equation}
Hence, the tachyon pole is approximately the same as in the broken phase:
\begin{equation}
  p^2 = \mu^2 e^{2(4\pi)^2 (f_2+B)} = \Lambda_L^2 e^{2(4\pi)^2 B + A}\,.
\end{equation}

\section{Effective potential of the naive continuum
  limit\label{sec-effective potential}}

In this section we wish to compute the effective potential.  
We could  derive the effective potential from the effective action by
taking the constant field limit, but we find it easier to go back to the
effective potential obtained in (\ref{V-Feff-intermediate}):
\[
  F_\eff (f_1 \mu^2, f_2; \sigma) = f_1 \mu^2 \sigma + \frac{1}{2}
  \sigma^2 \left(f_2 + \frac{1}{(4 \pi)^2} \ln
      \frac{\mu}{\Lambda}\right) + \Lambda^4 I
    \left(\frac{\sigma}{\Lambda^2}\right)\,.\eqno{(\ref{V-Feff-intermediate})}
\]
To take the limit $\Lambda \to 0+$, we recall the asymptotic behavior
(Appendix \ref{Appendix-asymp})
\begin{equation}
   I (\sigma) \overset{ - \sigma \gg 1}{\longrightarrow} - \frac{1}{4}
   \frac{1}{(4 \pi)^2} \sigma^2 \left( \ln (-\sigma) + A - \frac{1}{2} \right)\,.
\end{equation}
We then obtain
\begin{align}
   F_\eff (f_1 \mu^2, f_2; \sigma)
   &= f_1 \mu^2 \sigma + \frac{1}{2} \sigma^2 \left( f_2 + \frac{1}{2(4
     \pi)^2} \ln \frac{\mu^2}{(- \sigma) e^{A-\frac{1}{2}}}\right)\nt\\
  &= f_1 \mu^2 \sigma + \frac{1}{2(4\pi)^2} \frac{\sigma^2}{2} 
    \left( - \ln \frac{- \sigma}{\Lambda_L^2} + \frac{1}{2} \right)\,,
    \label{VI-Feff}
\end{align}
where $\Lambda_L$ is defined by (\ref{V-Landau}).

 \subsection*{Case (a) broken phase $f_1 < 0$}
 
From (\ref{VI-Feff})  we obtain
\begin{equation}
     \varphi = - \frac{d}{d\sigma} F_\eff (f_1 \mu^2, f_2; \sigma)
= - f_1 \mu^2 + \frac{1}{2(4 \pi)^2} \sigma \ln \frac{(-
  \sigma)}{\Lambda_L^2}\,,\label{VI-varphi}
\end{equation}
and
\begin{equation}
  \varphi_{\min} = - f_1 \mu^2 > 0\,.
\end{equation}
To express $\sigma$ in terms of $\varphi$, we rewrite
(\ref{VI-varphi}) as
\begin{equation}
  - 2 (4 \pi)^2 \frac{\varphi-\varphi_{\min}}{\Lambda_L^2}
  = \frac{-\sigma}{\Lambda_L^2} \ln \frac{-\sigma}{\Lambda_L^2}\,.
\end{equation}
This has the same form as Eq.~(\ref{V-equation-mph2}), and it is
solved by the lower branch of the Lambert $W$ function as 
\begin{equation}
   \frac{-\sigma}{\Lambda_L^2}
   = \exp \left[ W_{-1} \left(  - 2 (4 \pi)^2 
       \frac{\varphi-\varphi_{\min}}{\Lambda_L^2}\right)\right]\,.
   \label{VI-sigma-varphi}
\end{equation}
This is well defined (i.e., real) for
\begin{equation}
   0 < 2 (4 \pi)^2
   \frac{\varphi-\varphi_{\min}}{\Lambda_L^2} < \frac{1}{e}\,.
\end{equation}
Hence, the field has a maximum $\varphi_{\max}$ given by
 \begin{equation}
   \varphi_{\max} - \varphi_{\min}
   = \frac{1}{e} \frac{1}{2 (4 \pi)^2} \Lambda_L^2\,.
 \end{equation}
 Note that at $\varphi = \varphi_{\max}$, we find
 \begin{equation}
\sigma =  \sigma_{\min} = - \frac{1}{e} \Lambda_L^2\,,
 \end{equation}
which corresponds to the inflection point:
 \begin{equation}
   F_\eff'' (\sigma_{\min}) = 0\,.
 \end{equation}
We find
 \begin{equation}
   F_\eff'' (\sigma) \lb\begin{array}{c@{\quad}l}
                  > 0& (\sigma_{\min} < \sigma)\,,\\
                  < 0& (\sigma < \sigma_{\min})\,.
                \end{array}\right.
\end{equation}

To obtain the 1PI potential as a function of $\varphi$, we can
substitute (\ref{VI-sigma-varphi}) into the expression for
\begin{align}
  G_\eff (f_1 \mu^2, f_2; \varphi)
  &= F_\eff (f_1 \mu^2, f_2; \sigma) - f_1 \mu^2 \sigma + \sigma
    \left(\varphi-\varphi_{\min}\right) \nt\\
  &=  \frac{1}{2} \sigma^2 \frac{1}{2(4\pi)^2}
    \left( - \ln \frac{- \sigma}{\Lambda_L^2} + \frac{1}{2} \right)
    + \sigma    \left(\varphi-\varphi_{\min}\right) \nt\\
  &= \frac{1}{2} \Lambda_L^4
    \left(\frac{-\sigma}{\Lambda_L^2}\right)^2
    \frac{1}{2(4\pi)^2} \left( - \ln \frac{-\sigma}{\Lambda_L^2} +
    \frac{1}{2} \right) - \Lambda_L^2 \frac{-\sigma}{\Lambda_L^2}
    (\varphi - \varphi_{\min})\nt\\
  &=  \frac{1}{4 (4 \pi)^2} \Lambda_L^4 \frac{-\sigma}{\Lambda_L^2}
    \left[-  \frac{-\sigma}{\Lambda_L^2} \ln
    \frac{-\sigma}{\Lambda_L^2} + \frac{1}{2}
    \frac{-\sigma}{\Lambda_L^2} - \frac{4 (4\pi)^2
    (\varphi-\varphi_{\min})}{\Lambda_L^2} \right]\nt\\
  &=  \frac{\Lambda_L^4}{4(4\pi)^2} \exp \left[ W_{-1} \left(- 2
    (4\pi)^2 \frac{\varphi-\varphi_{\min}}{\Lambda_L^2}\right)\right]\nt\\
&\quad \times    \lb - 2 (4 \pi)^2 \frac{\varphi-\varphi_{\min}}{\Lambda_L^2} +
    \frac{1}{2} \exp \left[ W_{-1} \left(- 2
    (4\pi)^2 \frac{\varphi-\varphi_{\min}}{\Lambda_L^2}\right)
                                                                          \right]\rb\,.
\end{align}
Using
\begin{equation}
  e^{W_{-1} (z)} = \frac{z}{W_{-1} (z)}\,,
\end{equation}
and introducing a dimensionless variable
\begin{equation}
  \xi \equiv \frac{2 (4\pi)^2 (\varphi - \varphi_{\min})}{\Lambda_L^2}
  = \frac{1}{e} \frac{\varphi -
    \varphi_{\min}}{\varphi_{\max}-\varphi_{\min}}\qquad
  \left(0 < \xi < \frac{1}{e}\right) \,,\label{VI-xi}
\end{equation}
we can rewrite the above as
\begin{equation}
\boxed{  G_\eff (f_1 \mu^2, f_2; \varphi)
  = \frac{\Lambda_L^4}{2 (4\pi)^2} 
    \frac{\xi^2}{2}  \left(  \frac{1}{W_{-1} (-\xi)} +  \frac{1}{2 W_{-1}
        (-\xi)^2}\right) \,.}\label{VI-Geff}
\end{equation}
This agrees with the result obtained previously in \cite{Sonoda:2013jia}.

\begin{figure}[h]
  \centering
  \includegraphics[width=0.5\textwidth]{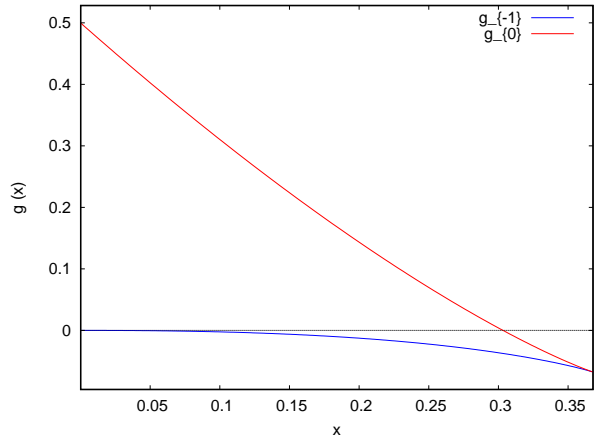}
  \caption{$g_{-1} (x)$ corresponds to $\sigma > \sigma_{\min}$, and
    $g_0 (x)$ corresponds to $\sigma < \sigma_{\min}$}
  \label{fig-gg}
\end{figure}

We plot two functions
\begin{align}
g_{-1} (x) &\equiv x^2 \left(\frac{1}{W_{-1} (-x)} + \frac{1}{2 W_{-1}
             (-x)^2}\right)\,,\\
g_0 (x) &\equiv x^2 \left(\frac{1}{W_{0} (-x)} + \frac{1}{2 W_{0}
          (-x)^2}\right)\,,
\end{align}
for $0 < x < 1/e$ in Fig.~\ref{fig-gg}.  $g_{-1} (x)$, proportional to
our $G_\eff (\varphi)$, is concave, but $g_0 (x)$ is convex and
corresponds to $\sigma < \sigma_{\min}$, where $F_\eff (\sigma)$ is
concave, not convex.

\begin{figure}[h]
  \centering
  \includegraphics[width=0.45\textwidth]{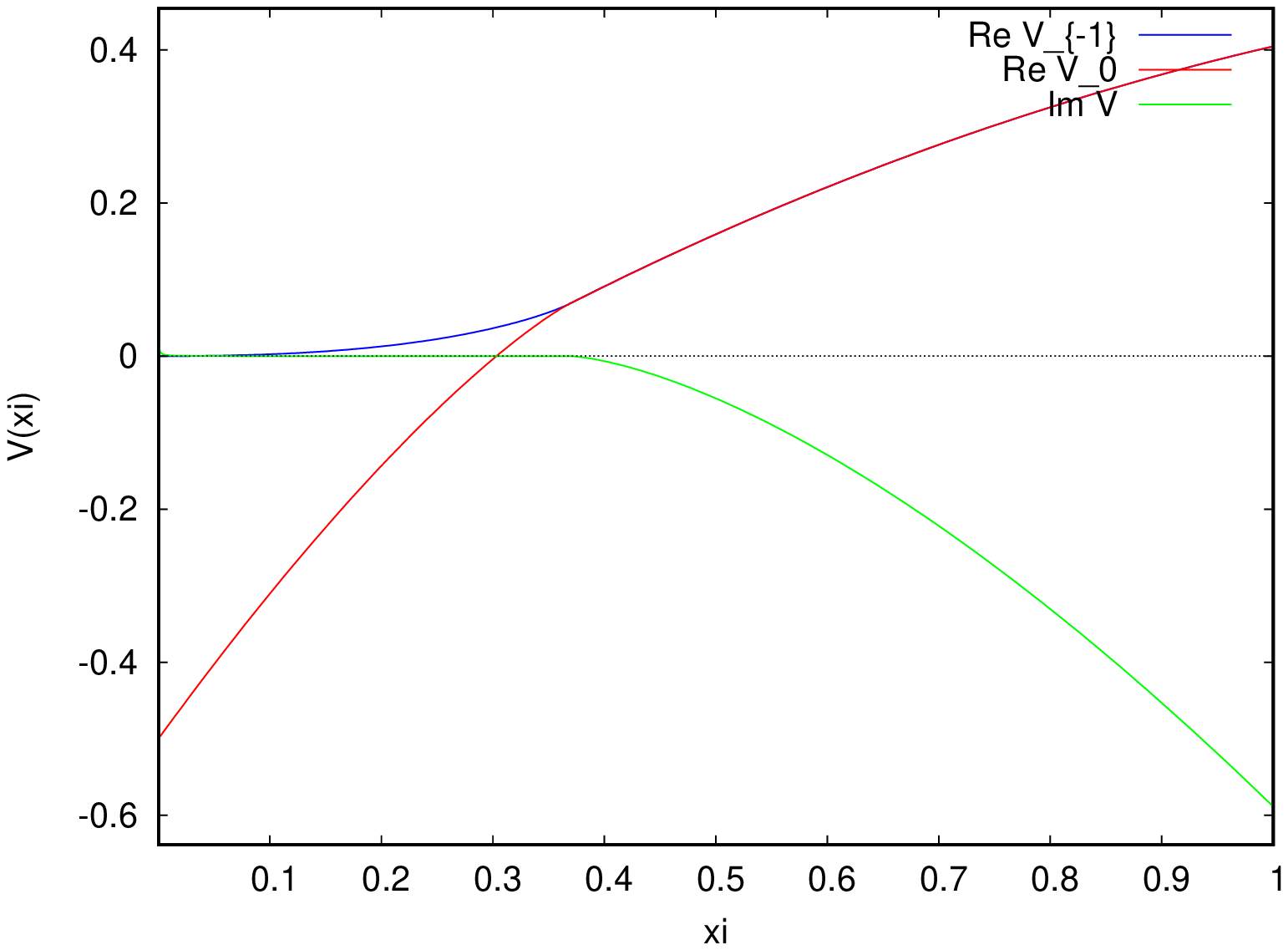}
  \hspace{0.5cm}
  \includegraphics[width=0.45\textwidth]{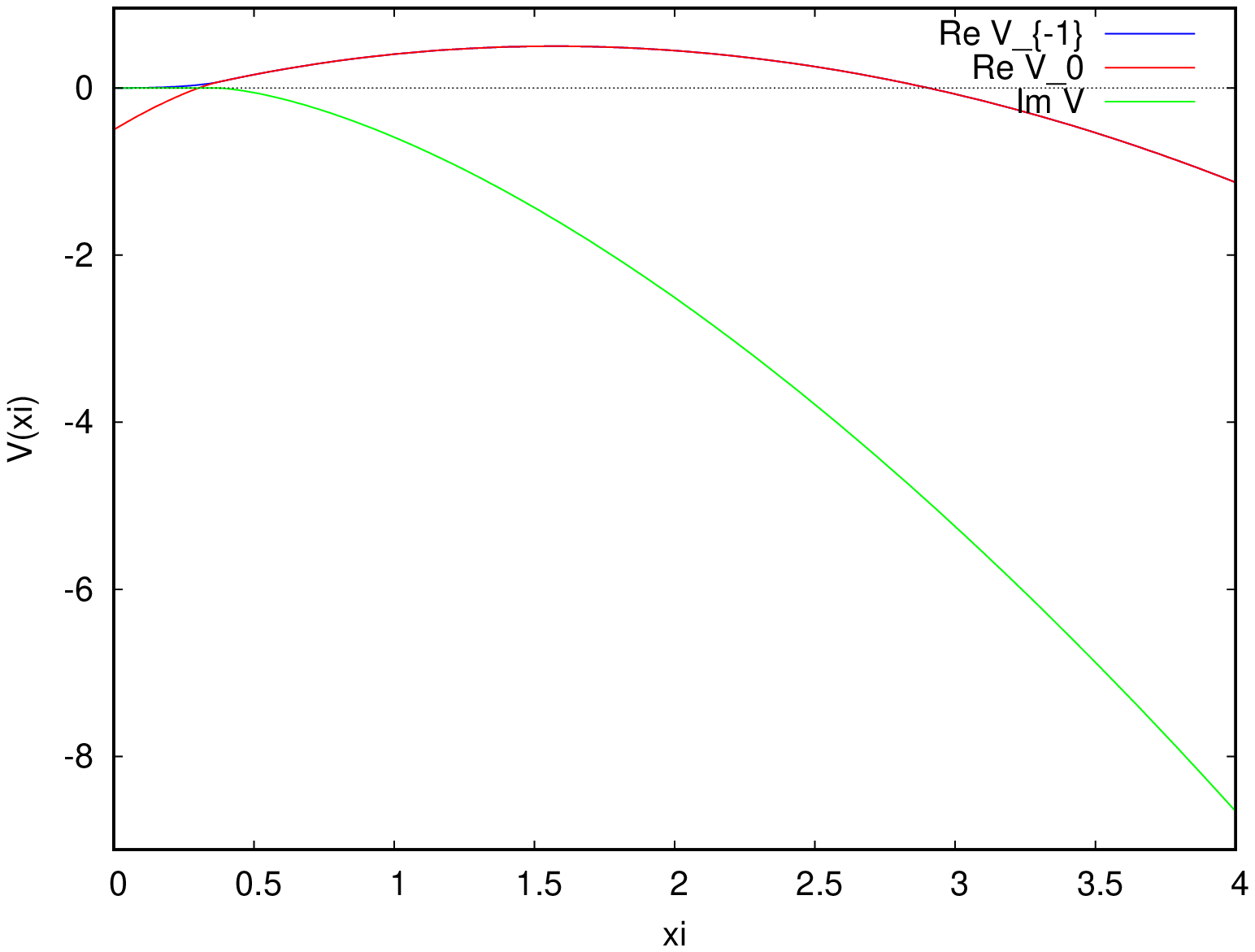}
  \caption{$V_{-1}$ is proportional to $V_\eff$; the right figure
    shows that the real part of $V_{-1}$ is unbounded from below}
  \label{fig-analyticity}
\end{figure}

We have found that $\varphi$ has a maximum value $\varphi_{\max}$
beyond which we seem apparently unable to construct
$G_\eff (\varphi)$.  Let us now speculate that the expression for
$G_\eff (\varphi)$ may be obtained by \textbf{analytic continuation}
for $\xi > \frac{1}{e}$ .  $W_{-1} (z)$ has a branch cut along the
negative real axis $z < - \frac{1}{e}$.  (See Appendix
\ref{Appendix-LambertW} for the Riemann sheets of $W$.)  We choose the
upper side of the branch cut so that the imaginary part vanishes at
$z = - \frac{1}{e}$.\footnote{As is explained in
  Appendix~\ref{Appendix-LambertW}, there is another branch cut
  $- \frac{1}{e} < x < 0$ that connects $W_{-1}$ to $W_1$.  $W_{-1}$
  is real on the upper side of this branch cut.  Hence, by continuity,
  it is natural to choose the upper side of the branch cut
  $x < - \frac{1}{e}$ for the analytic continuation.}  We plot the
real and imaginary parts of
\begin{align}
  V_{-1} (\xi)
  &\equiv  - \xi^2 \left( \frac{1}{W_{-1} (-\xi + i \ep)} + \frac{1}{2 W_{-1}
    (-\xi+i\ep)^2}\right)\,,\\
  V_0 (\xi)
  &\equiv  - \xi^2 \left( \frac{1}{W_{0} (-\xi - i \ep)} + \frac{1}{2 W_{0}
    (-\xi-i\ep)^2}\right)\,,
\end{align}
in Fig.~\ref{fig-analyticity}.  These two are identical for $\xi > \frac{1}{e}$.
The effective potential is given by
\begin{equation}
  V_\eff (\varphi) = - G_\eff (\varphi) = \frac{\Lambda_L^4}{4
    (4\pi)^2} V_{-1} (\xi)\,.
  \end{equation}
$V_\eff (\varphi)$ is convex for $\xi < \frac{1}{e}$, and concave for
$\xi > \frac{1}{e}$.  The imaginary part of $V_\eff (\varphi)$ implies
instability; this must be related to the presence of a tachyon.

In \cite{Coleman:1974jh} it was observed that for $\varphi >
\varphi_{\max}$ the effective potential becomes complex, and it
acquires a negative imaginary part.  They speculate there may be a
true minimum of the effective potential at a very large $\varphi$.
Our $V_\eff (\varphi)$ obtained by analytic continuation has no
such minimum at $\varphi > \varphi_{\max}$.

\subsection*{Case (b) symmetric phase $f_1 > 0$}

We start from
\[
  F_\eff (f_1 \mu^2, f_2; \sigma)
  = f_1 \mu^2 \sigma + \frac{1}{2} \sigma^2 \frac{1}{2 (4 \pi)^2}
\left( - \ln \frac{-\sigma}{\Lambda_L^2} + \frac{1}{2}\right)\,,\eqno{(\ref{VI-Feff})}
\]
and
\begin{equation}
  \varphi = - \frac{d}{d\sigma} F_\eff (\sigma)
  = - f_1 \mu^2 + \sigma  \frac{1}{2(4\pi)^2} \ln \frac{-\sigma}{\Lambda_L^2}\,,
\end{equation}
which gives
\begin{equation}
  \frac{-\sigma}{\Lambda_L^2} = \exp \left[ W_{-1} \left( - 2 (4
      \pi)^2 \frac{\varphi + f_1 \mu^2}{\Lambda_L^2}\right)\right]\,.
\end{equation}
$\mph^2$ is defined so that $\varphi = 0$ at $\sigma = - \mph^2$.
Hence, we obtain
\begin{equation}
  \frac{\mph^2}{\Lambda_L^2} =  \exp \left[ W_{-1}
    \left( - 2 (4 \pi)^2 f_1\frac{\mu^2}{\Lambda_L^2}\right)\right]
  = \exp \left[ W_{-1} \left( - 2 (4\pi)^2 f_1 e^{- 2(4\pi)^2 f_2 + A}\right)\right]\,.
\end{equation}
This agrees with (\ref{V-mph2}).

Replacing $-\varphi_{\min}$ by $f_1 \mu^2$ in (\ref{VI-xi}), we obtain
\begin{equation}
  G_\eff (\varphi)
= \frac{\Lambda_L^4}{4 (4 \pi)^2} \xi^2 \left(\frac{1}{W_{-1} (-\xi)}
                                           + \frac{1}{2 W_{-1} (-\xi)^2}\right)\,,
\end{equation}
where
\begin{equation}
  \xi \equiv \frac{2 (4\pi)^2 \left(\varphi + f_1
      \mu^2\right)}{\Lambda_L^2}\,.
\end{equation}
$f_1$ simply shifts the plot of $G_\eff (\varphi)$ to the left by $f_1 \mu^2$.  At
$\varphi = 0$, the slope of the potential is given by
\begin{equation}
  \frac{d}{d\varphi} G_\eff (\varphi)\Big|_{\varphi = 0}
  = - \mph^2 \,.
\end{equation}
The negative imaginary part of $V_\eff = - G_\eff$ is developed for
$\xi > \frac{1}{e}$, i.e.,
\begin{equation}
  \varphi > - f_1 \mu^2 + \frac{1}{e} \frac{\Lambda_L^2}{2(4\pi)^2}\,.
\end{equation}

\section{With a cutoff\label{sec-cutoff}}

To summarize so far, there are problems with the effective action and
potential in the naive continuum limit.
\begin{enumerate}
\item The effective action, in particular the two-point vertex of
  $\varphi$, suffers from a tachyon pole.
\item Assuming the analytic continuation for
  $\varphi > \varphi_{\max}$ is valid, the effective potential
  $V_\eff (\varphi)$ develops a negative imaginary part, and its real
  part is unbounded from below.
\end{enumerate}
In this section we start from a bare action of the O$(N)$ linear sigma
model with a physical cutoff $\Lambda_0$ and compute its effective
action and potential.  We confirm the absence of any of the problems
suffered by the naive continuum limit.  In the next section we compare
the cutoff theory with the naive continuum limit, and show how the
problems with the latter arise when the cutoff $\Lambda_0$ is raised
beyond the Landau pole.

We start from 
\begin{equation}
  F_{t=0} [\sigma] = \left(- \frac{1}{4} \int_p f(p) + f_1
  \right)\sigma (0) + \frac{f_2}{2} \int_p \sigma (p) 
  \sigma (-p)\,,\label{VII-initialF}
\end{equation}
where
\begin{equation}
  f_2 > 0
\end{equation}
is the inverse of the bare $\phi^4$ coupling.  Unlike $f_1, f_2$ of
the previous sections, $f_1, f_2$ in this section are bare parameters
defined at the cutoff scale $\Lambda_0$ to be introduced shortly.  Let
us construct the 1PI action corresponding to (\ref{VII-initialF}).
\begin{enumerate}
  \item For
\[
- \frac{1}{4} \int_p f(p) +  f_1 < 0
\]
we define
\begin{equation}
  \varphi_{\min} =  \frac{1}{4} \int_p f(p) - f_1 > 0\,.
\end{equation}
We obtain
\begin{align}
  \Gamma_{I, t=0} [\varphi]
  &= F_{t=0} [\sigma] + \varphi_{\min} \sigma (0) + \int_p \sigma (-p)
    \left( \varphi (p) - \varphi_{\min}\delta (p) \right) \nt\\
  &= - \frac{1}{2 f_2} \int_p  \left(\varphi (p) - \varphi_{\min} \delta
    (p)\right) \left(\varphi (-p) - \varphi_{\min} \delta (-p)\right)\,.
\end{align}
This is the action of the O$(N)$ linear sigma model in the broken phase.
\item For
\[
- \frac{1}{4} \int_p f(p) + f_1 > 0
\]
we define $m^2$ by
\begin{equation}
- \frac{1}{4} \int_p f(p) + f_1 - f_2 m^2 = 0
  \Longrightarrow m^2 = \frac{- \frac{1}{4} \int_p f(p) + f_1}{f_2}\,.
\end{equation}
We obtain
\begin{align}
  \Gamma_{I, t=0} [\varphi]
  &= F_{t=0} [\sigma] - m^2 \varphi (0) + \int_p \left(\sigma (-p) +
    m^2 \delta (p)\right) \varphi (p)\nt\\
  &= - m^2 \varphi (0) - \frac{1}{2 f_2} \int_p \varphi (p) \varphi (-p)\,.
\end{align}
This is the action of the O$(N)$ linear sigma model in the symmetric phase.
\end{enumerate}
Calculating the effective action and potential in the following, we
will find that the order parameter is $f_1$, not
$f_1 - \frac{1}{4} \int_p f(p)$.  The model is in the broken phase if
$f_1 < 0$, in the symmetric phase if $f_1 > 0$.  We compute the
effective action and potential in the two subsections below.

\subsection{Broken phase $f_1 < 0$}

Starting from (\ref{VII-initialF}),  we obtain, for $t > 0$,
\begin{align}
  F_t [\sigma]
  &= \left( - \frac{1}{4} \int_p f(p) + f_1 e^{2t} \right) \sigma
    (0)\nt\\
  &\quad + \frac{1}{2} \int_p \sigma (p) \sigma (-p) \left( f_2 +
    \frac{t}{(4 \pi)^2} + \F (p) - \F (p e^{-t}, - p e^{-t})
    \right)\nt\\
  &\quad + \sum_{n=3}^\infty \frac{1}{n!}
    \int_{p_1, \cdots, p_n} \sigma (p_1) \cdots \sigma (p_n) \, \delta
    \left(\sum_1^n p_i \right)\nt\\
  &\qquad \times \left(
    I_n (p_1, \cdots, p_n) - e^{-2(n-2)t} I_n (p_1 e^{-t}, \cdots,
    p_n e^{-t})\right)\,,
\end{align}
where we have used the general formula (\ref{III-generalFt}).

The Legendre transform of the effective action is obtained as
\begin{equation}
  F_\eff [\sigma] = \lim_{\Lambda \to 0+} F_{t = \ln
    \frac{\Lambda_0}{\Lambda}} [\sigma_\Lambda]\,,
\end{equation}
where
\begin{equation}
  \sigma_\Lambda (p) = \Lambda^2 \sigma (p \Lambda)\,.
\end{equation}
We find
\begin{align}
  F_{t=\ln \frac{\Lambda_0}{\Lambda}} [\sigma_\Lambda]
  &= \left( - \frac{1}{4} \int_p f(p) + f_1
    \left(\frac{\Lambda_0}{\Lambda}\right)^2 \right) \sigma_\Lambda
    (0)\nt\\
  &\quad + \frac{1}{2} \int_p \sigma_\Lambda (p) \sigma_\Lambda (-p)
    \left( f_2 + \frac{1}{(4 \pi)^2} \ln \frac{\Lambda_0}{\mu} +
    \F (p) + \frac{1}{(4 \pi)^2} \ln \frac{\mu}{\Lambda} \right) + I [\sigma_\Lambda]\nt\\
  &\quad - \frac{1}{2} \int_p \sigma_\Lambda (p) \sigma_\Lambda (-p) \F
    \left( p \frac{\Lambda}{\Lambda_0} \right) 
 - \sum_{n=3}^\infty \frac{1}{2n} \int_{p_1, \cdots, p_n}
    \sigma_\Lambda (p_1) \cdots \sigma_\Lambda (p_n)\, \delta \left(
    \sum_1^n p_i\right) \nt\\
  &\qquad  \times
    \left(\frac{\Lambda}{\Lambda_0}\right)^{2(n-2)} I_n \left( p_1
    \frac{\Lambda}{\Lambda_0}, \cdots, p_n \frac{\Lambda}{\Lambda_0}
    \right)\,.
\end{align}
Since
\begin{equation}
  \sigma_\Lambda (p) = \Lambda^2 \sigma (p \Lambda)
  = \frac{\Lambda^2}{\Lambda_0^2} \Lambda_0^2 \sigma \left( p
    \frac{\Lambda}{\Lambda_0} \Lambda_0\right)
  = \frac{\Lambda^2}{\Lambda_0^2} \sigma_{\Lambda_0} \left(p
    \frac{\Lambda}{\Lambda_0}\right)\,,
\end{equation}
we obtain
\begin{align}
  \frac{1}{2} \int_p \sigma_\Lambda (p) \sigma_\Lambda (-p)\, \F
    \left(p \frac{\Lambda}{\Lambda_0}\right)
  &= \frac{1}{2} \int_p \frac{\Lambda^4}{\Lambda_0^4}
    \sigma_{\Lambda_0} \left(p \frac{\Lambda}{\Lambda_0}\right)
    \sigma_{\Lambda_0}  \left(-p \frac{\Lambda}{\Lambda_0}\right)
    \, \F   \left(p \frac{\Lambda}{\Lambda_0}\right)\nt\\
  &= \frac{1}{2} \int_p \sigma_{\Lambda_0} (p) \sigma_{\Lambda_0} (-p)
    \F (p)\,,
\end{align}
and
\begin{align}
 & \sum_{n=3}^\infty \frac{1}{2n} \int_{p_1, \cdots, p_n}
    \sigma_\Lambda (p_1) \cdots \sigma_\Lambda (p_n)\, \delta \left(
    \sum_1^n p_i\right) \,
    \left(\frac{\Lambda}{\Lambda_0}\right)^{2(n-2)} I_n \left( p_1
    \frac{\Lambda}{\Lambda_0}, \cdots, p_n \frac{\Lambda}{\Lambda_0}
    \right)\nt\\
  &= \sum_{n=3}^\infty \frac{1}{2n} \int_{p_1, \cdots, p_n}
    \sigma_{\Lambda_0} \left(p_1 \frac{\Lambda}{\Lambda_0}\right)
    \cdots \sigma_{\Lambda_0} \left( p_n
    \frac{\Lambda}{\Lambda_0}\right)\,
    \left(\frac{\Lambda}{\Lambda_0}\right)^{4n} \delta \left(
    \sum_1^n p_i \frac{\Lambda}{\Lambda_0}\right)\nt
  \\
  &\qquad \times I_n \left( p_1
    \frac{\Lambda}{\Lambda_0}, \cdots, p_n \frac{\Lambda}{\Lambda_0}
    \right)\nt\\
  &= \sum_{n=3}^\infty \frac{1}{2n} \int_{p_1, \cdots, p_n}
    \sigma_{\Lambda_0} (p_1) \cdots \sigma_{\Lambda_0} (p_n) \, \delta
    \left(\sum_1^n p_i\right)\, I_n (p_1, \cdots, p_n) = I [\sigma_{\Lambda_0}]\,.
\end{align}
Hence, we can write
\begin{align}
  F_{t=\ln \frac{\Lambda_0}{\Lambda}} [\sigma_\Lambda]
  &= \left( - \frac{1}{4} \int_p f(p) + f_1
    \left(\frac{\Lambda_0}{\Lambda}\right)^2 \right) \sigma_\Lambda
    (0)\nt\\
  &\quad + \frac{1}{2} \int_p \sigma_\Lambda (p) \sigma_\Lambda (-p)
    \left( f_2 + \frac{1}{(4 \pi)^2} \ln \frac{\Lambda_0}{\mu}
    + \F (p)  + \frac{1}{(4 \pi)^2} \ln \frac{\mu}{\Lambda}  \right)
    + I [\sigma_\Lambda]\nt\\ 
  &\quad -  \frac{1}{2} \int_p \sigma_{\Lambda_0} (p) \sigma_{\Lambda_0} (-p)
    \, \F (p)  - I [\sigma_{\Lambda_0}]\,.
\end{align}
Note that the first two lines on the right-hand side is the same as in
the naive continuum limit except that $f_2$ is replaced by
$f_2 + \frac{1}{(4 \pi)^2} \ln \frac{\Lambda_0}{\mu}$ here.  Hence, in
the limit $\Lambda \to 0$, we obtain
\begin{align}
  F_\eff [\sigma]
  &= - \varphi_{\min} \sigma (0)\nt\\
  &\quad + \frac{1}{2} \int_p \sigma (p) \sigma (- p
    )\left(\frac{1}{\lambda}  - \frac{1}{(4 \pi)^2} \ln \frac{p}{\mu} +B 
    - \F
    \left(\frac{p}{\Lambda_0}\right)
    \right)\nt\\
  &\quad + \sum_{n=3}^\infty \frac{1}{2n} \int_{p_1, \cdots, p_n}
    \sigma (p_1) \cdots \sigma (p_n )\, \delta \left(
    \sum_1^n p_i\right) \nt\\
  &\qquad \times  \left( I_{n, \eff} (p_1, \cdots, p_n)
    -  \frac{1}{\Lambda_0^{2(n-2)}} I_n \left( 
    \frac{p_1}{\Lambda_0}, \cdots,  \frac{p_n}{\Lambda_0}
    \right)\right)\,,\label{VII-Feff-broken}
\end{align}
where 
\begin{align}
  \varphi_{\min} &\equiv - f_1 \Lambda_0^2 > 0\,,\label{VII-varphimin}\\
  \lambda &\equiv \frac{1}{f_2 + \frac{1}{(4 \pi)^2} \ln
    \frac{\Lambda_0}{\mu}}\,,\label{VII-lambda}
\end{align}
and $I_{n, \eff}$ is given by (\ref{V-Ineff}).  Fixing
$\varphi_{\min}$ and $\lambda$, we obtain (\ref{V-Feff}) if we take
$\Lambda_0 \to \infty$ naively.

The Legendre transform of (\ref{VII-Feff-broken}) gives
\begin{align}
  \Gamma_{\eff, I} [\varphi]
  &= F_\eff [\sigma] + \varphi_{\min} \sigma (0) + \int_p \sigma (-p) \left(\varphi (p) -
    \varphi_{\min} \delta (p) \right)\nt\\
  &= \sum_{n=2}^\infty \frac{1}{n!} \int_{p_1, \cdots, p_n}
    \left( \varphi (p_1) - \varphi_{\min} \delta (p_1)\right) \cdots
    \left( \varphi (p_n) - \varphi_{\min} \delta (p_n)\right)\,\delta
    \left(\sum_1^n p_i\right)\nt\\
  &\quad \times c_{n, \eff} (p_1, \cdots, p_n)\,.
\end{align}
where
\begin{subequations}
\begin{align}
  c_{2, \eff} (p, -p)
  &= - \frac{1}{\frac{1}{\lambda} - \frac{1}{(4 \pi)^2} \ln
    \frac{p}{\mu} + B - \F \left(\frac{p}{\Lambda_0}\right)}\,,\\
  c_{3, \eff} (p_1, p_2, p_3)
  &= \prod_{i=1}^3 c_2 (p_i, -p_i) \cdot \left[ I_{3, \eff} (p_1,
    \cdots. p_n) - \Lambda_0^{-2} I_3
    \left(\frac{p_1}{\Lambda_0}, \cdots, \frac{p_n}{\Lambda_0} \right) \right]\,,
\end{align}
\end{subequations}
and so on.

We now note that $c_{2, \eff} (p, -p)$ has \textbf{no tachyon pole} since
\begin{equation}
  \frac{1}{\lambda} - \frac{1}{(4 \pi)^2} \ln \frac{p}{\mu} + B - \F
  \left(\frac{p}{\Lambda_0}\right) > 0\,.
\end{equation}
(Proof)
In Appendix \ref{Appendix-inequality} we show, for $t > 0$,
\begin{equation}
  \frac{t}{(4 \pi)^2} + \F (p) - \F (p e^{-t}, - p e^{-t}) > 0\,.
\end{equation}
This gives, for $\Lambda < \Lambda_0$,
\begin{equation}
  \frac{1}{(4 \pi)^2} \ln \frac{\Lambda_0}{\Lambda}
  + \F \left(\frac{p}{\Lambda}\right) - \F
  \left(\frac{p}{\Lambda_0}\right) > 0\,.
\end{equation}
We can rewrite this as
\begin{equation}
  \underbrace{\frac{1}{(4 \pi)^2} \ln \frac{\Lambda_0}{\mu}}_{=
    \frac{1}{\lambda} - f_2} \underbrace{- \frac{1}{(4 \pi)^2}
  \ln \frac{\Lambda}{\mu} + \F
  \left(\frac{p}{\Lambda}\right)}_{\overset{\Lambda \to
    0}{\longrightarrow} - \frac{1}{(4\pi)^2} \ln \frac{p}{\mu} + B} - \F
  \left(\frac{p}{\Lambda_0}\right) > 0\,.
\end{equation}
Hence, we obtain
\begin{equation}
  \frac{1}{\lambda} - \frac{1}{(4\pi)^2} \ln \frac{p}{\mu} + B - \F
  \left(\frac{p}{\Lambda_0}\right) > f_2 > 0\,.
\end{equation}
(End of Proof)\\
The asymptotic behavior is given by
\begin{equation}
  c_{2, \eff} (p, -p)
  \lb \begin{array}{c@{\quad}l}
    \overset{ p \ll \Lambda_0}{\longrightarrow}&
    - \frac{1}{\frac{1}{\lambda} - \frac{1}{(4 \pi)^2} \ln
    \frac{p}{\mu} + B}\,,\\
    \overset{ p \gg \Lambda_0}{\longrightarrow}&
    - \frac{1}{\frac{1}{\lambda} - \frac{1}{(4 \pi)^2} \ln
    \frac{\Lambda_0}{\mu}} = - \frac{1}{f_2}\,.
  \end{array}\right.
\end{equation}
For $p_i \ll \Lambda_0$, the vertex functions $c_n (p_1, \cdots, p_n)$
of the naive continuum limit are reproduced.

We next consider the effective potential.  The Legendre transform of
the effective potential is obtained as
\begin{equation}
  F_\eff (\sigma) = \lim_{\Lambda \to 0+} \Lambda^4 F_{t=\ln
    \frac{\Lambda_0}{\Lambda}} \left(\frac{\sigma}{\Lambda^2}\right)\,,
\end{equation}
where
\begin{align}
  \Lambda^4 F_{t=\ln \frac{\Lambda_0}{\Lambda}}
                 \left(\frac{\sigma}{\Lambda^2}\right)
&= \Lambda^4 \Bigg[
    \left( - \frac{1}{4} \int_p f(p) + f_1
    \left(\frac{\Lambda_0}{\Lambda}\right)^2 \right)
    \frac{\sigma}{\Lambda^2}
 + \frac{1}{2} \left(\frac{\sigma}{\Lambda^2}\right)^2 \left(
    f_2 + \frac{1}{(4 \pi)^2} \ln \frac{\Lambda_0}{\Lambda}\right)\nt\\
&\qquad\qquad\qquad + I \left(\frac{\sigma}{\Lambda^2}\right) -
    \frac{\Lambda_0^4}{\Lambda^4} I
    \left(\frac{\sigma}{\Lambda_0^2}\right) \Bigg]\nt\\
  &= \left( - \Lambda^2 \frac{1}{4} \int_p f(p) + f_1 \Lambda_0^2
    \right) \sigma
  + \frac{1}{2} \sigma^2 \left( f_2 + \frac{1}{(4 \pi)^2} \ln
    \frac{\Lambda_0}{\mu} \right)\nt\\
  &\qquad + \Lambda^4 I \left(\frac{\sigma}{\Lambda^2}\right)
    + \frac{1}{2} \sigma^2 \frac{1}{(4\pi)^2} \ln \frac{\mu}{\Lambda} -
    \Lambda_0^4 I \left(\frac{\sigma}{\Lambda_0^2}\right)\,.
\end{align}
Taking the limit $\Lambda \to 0+$, we obtain
\begin{equation}
  F_\eff (\sigma)
  = - \varphi_{\min} \sigma + \frac{1}{\lambda} \frac{1}{2} \sigma^2
     -  \frac{1}{2 (4 \pi)^2}
  \frac{\sigma^2}{2} \left( \ln \frac{-\sigma}{\mu^2} + A -
    \frac{1}{2} \right) - \Lambda_0^4 I
  \left(\frac{\sigma}{\Lambda_0^2}\right)\,,\label{VII-Feff}
\end{equation}
where $\varphi_{\min}, \lambda$ are given by (\ref{VII-varphimin}),
(\ref{VII-lambda}).  This gives
\begin{equation}
  \varphi
  = \varphi_{\min} - \frac{1}{\lambda} \sigma + \frac{1}{2(4\pi)^2} \sigma
  \left( \ln \frac{-\sigma}{\mu^2} + A \right) + \Lambda_0^2 I'
  \left(\frac{\sigma}{\Lambda_0^2}\right)\,.\label{VII-varphi}
\end{equation}

Let us show that $F_\eff (\sigma)$ is convex in $\sigma < 0$, i.e.,
$\varphi$ is a decreasing function of $\sigma < 0$.  We wish to show
\begin{equation}
\frac{d\varphi}{d\sigma} = - f_2  + \frac{1}{2(4\pi)^2} \left( \ln
  \frac{-\sigma}{\Lambda_0^2} + A + 1 \right) + I''
\left(\frac{\sigma}{\Lambda_0^2}\right) < 0\,.
\end{equation}
(Proof)  In Appendix \ref{Appendix-inequality} we show, for $t > 0$,
\begin{equation}
  - \frac{t}{(4 \pi)^2} - I'' (\sigma) + I'' (\sigma e^{-2t}) < 0\,.
\end{equation}
Substituting $\frac{\sigma}{\Lambda^2}$ for $\sigma$, and choosing $e^{-t} =
\frac{\Lambda}{\Lambda_0}$, we obtain
\begin{equation}
  - \frac{1}{(4 \pi)^2} \ln \frac{\Lambda_0}{\Lambda}
  - I'' \left(\frac{\sigma}{\Lambda^2}\right) + I''
  \left(\frac{\sigma}{\Lambda_0^2}\right) < 0\,.
\end{equation}
In the limit $\Lambda \to 0+$, we obtain
\begin{equation}
  \frac{1}{2(4\pi)^2} \left(\ln \frac{-\sigma}{\Lambda_0^2} + A + 1
  \right) + I'' \left(\frac{\sigma}{\Lambda_0^2}\right) < 0\,.
\end{equation}
Since $f_2 > 0$, we obtain the desired inequality.\\
(End of Proof)\\
Hence, $\varphi$ is well defined in the entire $\sigma < 0$.  At
$\sigma = 0$, we obtain $\varphi = \varphi_{\min}$, and
$\varphi \to +\infty$ as $\sigma \to - \infty$.  For
$- \sigma \gg \Lambda_0^2$, we obtain the asymptotic behavior
\begin{align}
  F_\eff (\sigma)
 &\overset{- \sigma \gg \Lambda_0^2}{\longrightarrow} - \varphi_{\min}
   \sigma + f_2 \frac{1}{2} \sigma^2\,,\\
\varphi
  &\overset{- \sigma \gg \Lambda_0^2}{\longrightarrow} \varphi_{\min}
    - f_2 \sigma\,.
\end{align}

The 1PI potential is given by
\begin{align}
  G_\eff (\varphi)
  &= F_\eff (\sigma) + \sigma \varphi\nt\\
  &= - \frac{1}{2 \lambda }  \sigma^2 + \frac{1}{2(4\pi)^2} \frac{\sigma^2}{2} \left(
    \ln \frac{-\sigma}{\mu^2} + A + \frac{1}{2} \right) - \Lambda_0^4 I
    \left(\frac{\sigma}{\Lambda_0^2}\right) + \Lambda_0^2 \sigma I'
    \left(\frac{\sigma}{\Lambda_0^2}\right)\nt\\
  &\overset{- \sigma \gg \Lambda_0^2}{\longrightarrow} - f_2
    \frac{\sigma^2}{2} - \Lambda_0^4 \frac{1}{8} \int_p
    \frac{f(p)}{h(p)}\,.
\end{align}
This implies the asymptotic behavior
\begin{equation}
  G_\eff (\varphi) \overset{\varphi - \varphi_{\min} \gg f_2
    \Lambda_0^2}{\longrightarrow}
  - \frac{1}{2 f_2} \left(\varphi - \varphi_{\min}\right)^2 - \Lambda_0^4 \frac{1}{8} \int_p
  \frac{f(p)}{h(p)}\,.
\end{equation}
The additive constant can be ignored.  \textbf{The effective potential reduces
to the bare potential for a large field $\varphi$.}

\subsection*{Plot of $V_\eff (\varphi)$ for various $\Lambda_0$}

Let us summarize what is needed to make a plot of the effective
potential.  For simplicity we take $\varphi_{\min} = 0$.  We rewrite
(\ref{VII-Feff}) and (\ref{VII-varphi}) as
\begin{align}
  F_\eff (\sigma)
  &= - \frac{1}{2 (4\pi)^2} \frac{\sigma^2}{2}\left(\ln
    \frac{-\sigma}{\Lambda_L^2} - \frac{1}{2}\right) - \Lambda_0^4 I
    \left(\frac{\sigma}{\Lambda_L^2} \frac{\Lambda_L^2}{\Lambda_0^2}\right)\,,\\
  \varphi
  &= \frac{1}{2(4 \pi)^2} \sigma \ln \frac{-\sigma}{\Lambda_L^2} +
    \Lambda_0^2 I' \left(\frac{\sigma}{\Lambda_L^2}
    \frac{\Lambda_L^2}{\Lambda_0^2}\right)\,, 
\end{align}
where we have defined
\begin{equation}
  \Lambda_L^2 \equiv \mu^2 e^{2 (4\pi)^2 \frac{1}{\lambda} - A}\,.\label{VII-Landau}
\end{equation}
This $\Lambda_L$ is the same as (\ref{V-Landau}) except that $f_2$
there is replaced by $\frac{1}{\lambda}$ here.

\begin{figure}[h]
  \centering
  \includegraphics[width=0.8\textwidth]{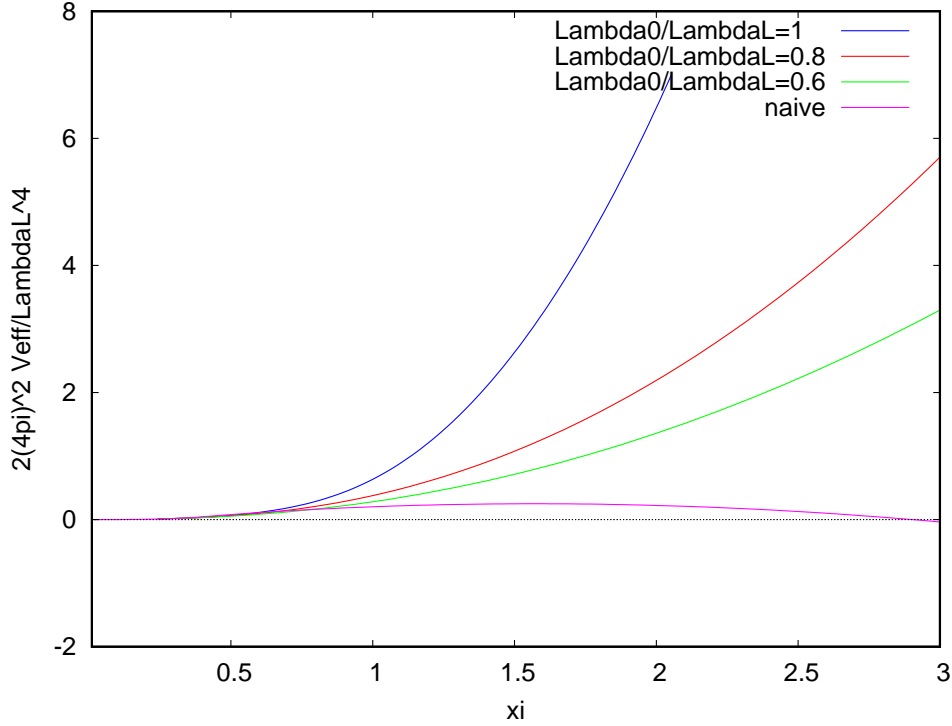}
  \caption{Effective potentials with a fixed $\lambda$ for various
    $\Lambda_0$'s}
  \label{fig-Veff-comparison}
\end{figure}

The definition of $\lambda$, given by (\ref{VII-lambda}), gives
\begin{equation}
  \frac{\Lambda_0^2}{\Lambda_L^2}
  = \exp \left( A - 2 (4 \pi)^2 f_2  \right) < e^A\,,
\end{equation}
since $f_2 > 0$.  Hence, given $\lambda$, the largest cutoff we can
choose is 
\begin{equation}
  \max \Lambda_0 = \Lambda_L \,e^{\frac{A}{2}}
  = \mu \,e^{(4 \pi)^2 \frac{1}{\lambda}}\,.\label{VII-maximum}
\end{equation}
This is the famous Landau pole \cite{Landau:1955}, corresponding to
$f_2=0$.  In Fig.~\ref{fig-Veff-comparison} we plot the effective
potential
\begin{equation}
  V_\eff (\varphi) = - G_\eff (\varphi)
\end{equation}
for three choices of $\Lambda_0$ below the Landau pole:
\[
  \Lambda_0 = (1, 0.8, 0.6) \Lambda_L\,.
\]
(For the cutoff function $R (p) = e^{-p^2}$, we find $A \simeq
0.129$.  Hence, $\Lambda_L$ is below the Landau pole.)  The $x$-axis gives
\begin{equation}
  \xi \equiv \frac{2 (4\pi)^2 \varphi}{\Lambda_L^2}\,.
\end{equation}
(This is the same as (\ref{VI-xi}) except $f_2$ there is replaced by
$\frac{1}{\lambda}$ here.)  We also plot the real part of
$V_\eff (\varphi)$ in the naive limit that corresponds to
$\Lambda_0 \to +\infty$:
\begin{equation}
  V^{\mathrm{naive}}_\eff (\varphi) = - \frac{1}{2(4\pi)^2}
  \Lambda_L^4 \cdot \frac{\xi^2}{2} \left( \frac{1}{W_{-1}
      (-\xi+i\ep)} + \frac{1}{2 W_{-1} (-\xi+i\ep)^2}\right)\,.
\end{equation}
(This is the same as (\ref{VI-Geff})).  As $\Lambda_0$ decreases, the
bare parameter $f_2$ gets larger, and $\frac{1}{f_2}$ smaller.

\subsection{Symmetric phase $f_1 > 0$}

We obtain the physical squared mass $\mph^2$ from
\begin{equation}
  F'_\eff (\sigma = - \mph^2) = 0\,,
\end{equation}
where $F_\eff$ is given by (\ref{VII-Feff}).  This gives
\begin{equation}
  f_1 \Lambda_0^2 - f_2 \mph^2 + \frac{1}{2(4\pi)^2} \mph^2 \left( \ln
    \frac{\mph^2}{\Lambda_0^2} + A \right) - \Lambda_0^2 I' \left( -
    \frac{\mph^2}{\Lambda_0^2}\right) = 0\,.
\end{equation}
For $\mph^2 \ll \Lambda_0^2$, we obtain approximately
\begin{equation}
  f_1 \Lambda_0^2 - f_2 \mph^2 + \frac{1}{2(4\pi)^2} \mph^2 \left( \ln
    \frac{\mph^2}{\Lambda_0^2} + A \right) = 0\,.
\end{equation}
As long as
\begin{equation}
  2 (4\pi)^2 f_1 \Lambda_0^2 < \frac{1}{e} \Lambda_L^2\,,
\end{equation}
where $\Lambda_L$ is given by (\ref{VII-Landau}),
this is solved by
\begin{equation}
  \mph^2 = \Lambda_L^2 \exp \left[ W_{-1} \left(
      - \frac{2 (4\pi)^2 f_1 \Lambda_0^2}{\Lambda_L^2}\right)\right]
  = \frac{2 (4\pi)^2 f_1 \Lambda_0^2 }{- W_{-1} \left( - \frac{2
        (4\pi)^2 f_1 \Lambda_0^2}{\Lambda_L^2}\right)}\,.
\end{equation}
This is the same as (\ref{V-mph2}) except that $f_1 \mu^2$ there is
replaced by $f_1 \Lambda_0^2$ here.

The calculation of the effective action proceeds the same way as for
the broken phase.  We merely state the result.  We obtain
\begin{align}
  F_\eff [\sigma]
  &= \frac{1}{2} \int_p \left(\sigma (p) + \mph^2 \delta (p)\right)
    \left(\sigma (-p) + \mph^2 \delta (p) \right)\nt\\
  &\qquad \times \left[ \frac{1}{\lambda}
 + I_{2,\eff} (\mph^2; p, -p) - \F
    \left(\frac{\mph^2}{\Lambda_0^2}; \frac{p}{\Lambda_0} \right)\right]\nt\\
  &\quad + \sum_{n=3}^\infty \frac{1}{2n} \int_{p_1, \cdots, p_n}
    \left( \sigma (p_1) + \mph^2 \delta (p_1)\right) \cdots \left(
    \sigma (p_n) + \mph^2 \delta (p_n)\right) \, \delta \left(\sum_1^n p_i\right)\nt\\
  &\qquad\times\left[ I_{n,\eff} (\mph^2;
    p_1, \cdots, p_n) - \Lambda_0^{- 2 (n-2)} I_n \left(
    \frac{\mph^2}{\Lambda_0^2}; \frac{p_1}{\Lambda_0}, \cdots,
    \frac{p_n}{\Lambda_0}\right) \right]\,,
\end{align}
where $I_{2, \eff}$ and $I_{n \ge 3,\eff}$ are defined by
(\ref{V-I2eff}) and (\ref{V-Ineff}), respectively.

The Legendre transformation gives
\begin{align}
  \Gamma_{\eff, I} [\varphi]
  &= - \mph^2 \varphi (0) + F_\eff [\sigma] + \int_p \left(\sigma (-p)
    + \mph^2 \delta (p)\right)\varphi (p)\nt\\
  &= - \mph^2 \varphi (0) + \sum_{n=2}^\infty \frac{1}{n!} \int_{p_1,
    \cdots, p_n} \varphi (p_1) \cdots \varphi (p_n)\,\delta
    \left(\sum_1^n p_i \right) \times c_{n, \eff} \left( \mph^2; p_1, \cdots, p_n\right)\,,
\end{align}
where
\begin{subequations}
\begin{align}
  c_{2, \eff} (\mph^2; p, -p)
  &= - \frac{1}{\frac{1}{\lambda} +
    I_{2,\eff} (\mph^2; p, -p) - \F \left(\frac{\mph^2}{\Lambda_0^2};
    \frac{p}{\Lambda_0}\right)}\,,\\
  c_{3, \eff} (\mph^2; p_1, p_2, p_3)
  &= \prod_{i=1}^3 c_{2, \eff} (\mph^2; p_i, -p_i) \nt\\
  &\quad \times \left[
    I_{3, \eff} (\mph^2; p_1, p_2, p_3) - \Lambda_0^{-2} I_3
    \left(\frac{\mph^2}{\Lambda_0^2}; \frac{p_1}{\Lambda_0},
    \frac{p_2}{\Lambda_0}, \frac{p_3}{\Lambda_0}\right)\right]\,,
\end{align}
\end{subequations}
and so on.

Let us show $c_{2, \eff} (\mph^2; p, -p)$ has no tachyon pole:
\begin{equation}
  \frac{1}{(4\pi)^2} \ln \frac{\Lambda_0}{\mu} +
    I_2^\cont (\mph^2; p, -p) - \F \left(\frac{\mph^2}{\Lambda_0^2};
      \frac{p}{\Lambda_0}\right) > 0\,.
\end{equation}
(Proof) In Appendix \ref{Appendix-inequality} we show, for $t > 0$,
\begin{equation}
  \F (m^2 e^{2t}, p e^t) + \frac{1}{(4 \pi)^2} t > \F (m^2, p)\,.
\end{equation}
Hence, for $\Lambda < \Lambda_0$, we obtain
\begin{equation}
  \F \left( \frac{m^2}{\Lambda^2}, \frac{p}{\Lambda}\right) +
  \frac{1}{(4 \pi)^2} \ln \frac{\Lambda_0}{\Lambda}  - \F
  \left(\frac{m^2}{\Lambda_0^2}, \frac{p}{\Lambda_0}\right) > 0\,.
\end{equation}
Taking the limit $\Lambda \to 0+$, and using
\[
  I_{2,\eff} (m^2; p, -p) \equiv \lim_{\Lambda \to 0+} \left( \F
    \left(\frac{m^2}{\Lambda^2}, \frac{p}{\Lambda}\right) +
    \frac{1}{(4 \pi)^2} \ln \frac{\mu}{\Lambda} \right)
\]
we obtain the desired inequality.\\
(End of Proof)

Thus, we have seen that the effective action is well defined, free
from tachyon poles.  The 1PI potential $G_\eff (\varphi)$ is the same
as in the broken phase except for a shift of $\varphi$ by
$f_1 \Lambda_0^2$.  The slope at $\varphi = 0$ gives $- \mph^2$.  All
is well.

\section{Comparison with the naive continuum limit\label{sec-comparison}}

In the previous section we have considered the O$(N)$ linear sigma
model with a finite cutoff $\Lambda_0$, and calculated the effective
action and effective potential in the large $N$ limit.  The model is
well defined: the effective action has no tachyon, and the effective
potential is real and bounded from below.  In this section we would
like to see how it goes wrong when we raise the cutoff beyond the
Landau pole (\ref{VII-maximum}).

We recall how the maximum is derived.  The renormalized coupling
$\lambda$ is given by
\begin{equation}
  \frac{1}{\lambda} = f_2 + \frac{1}{(4 \pi)^2} \ln
  \frac{\Lambda_0}{\mu}
\end{equation}
in the cutoff theory, where $\frac{1}{f_2}$ is the bare $\phi^4$
coupling at the cutoff scale $\Lambda_0$.  Since $f_2 > 0$, we obtain
\begin{equation}
  \frac{1}{\lambda} > \frac{1}{(4 \pi)^2} \ln
  \frac{\Lambda_0}{\mu} \Longrightarrow
  \Lambda_0 < \mu\, \exp \left( \frac{(4 \pi)^2}{\lambda}\right)\,.
\end{equation}
In the naive continuum limit, however, we fix $\lambda$ and take
$\Lambda_0 \to +\infty$, ignoring the above inequality.

Let us take a theory with $\lambda > 0$ and a cutoff $\Lambda_0$ beyond
the Landau pole:
\begin{equation}
  \Lambda_0 > \mu \exp \left( \frac{(4 \pi)^2}{\lambda}\right)\,.\label{VIII-beyond}
\end{equation}
This implies
\begin{equation}
  f_2 = \frac{1}{\lambda} - \frac{1}{(4 \pi)^2} \ln
  \frac{\Lambda_0}{\mu} < 0\,.
\end{equation}
At the cutoff scale $\Lambda_0$ we obtain a bare $\phi^4$ potential
unbounded from below.  This must be the cause of a tachyon pole and
the instability of the effective potential.  This reasoning is not
new; it was first given in \cite{Coleman:1974jh}.

\subsection{Tachyon pole}

For simplicity we consider only the critical case $f_1 = 0$.  The
effective action is given by
\begin{equation}
  \Gamma_{I, \eff} [\varphi]
  = \sum_{n=2}^\infty \frac{1}{n!} \int_{p_1, \cdots, p_n}
    \varphi (p_1) \cdots  \varphi (p_n)\, \delta
    \left(\sum_1^n p_i\right)\,  c_{n, \eff} (p_1, \cdots, p_n)\,,
\end{equation}
where
\begin{subequations}
\begin{align}
  c_{2, \eff} (p, -p)
  &= - \frac{1}{\frac{1}{\lambda} - \frac{1}{(4 \pi)^2} \ln
    \frac{p}{\mu} + B - \F \left(\frac{p}{\Lambda_0}\right)}\,,\\
  c_{3, \eff} (p_1, p_2, p_3)
  &= \prod_{i=1}^3 c_2 (p_i, -p_i) \cdot \left[ I_3^\cont (p_1,
    \cdots. p_n) - \Lambda_0^{-2(n-2)} I_n
    \left(\frac{p_1}{\Lambda_0}, \cdots, \frac{p_n}{\Lambda_0} \right) \right]\,,
\end{align}
\end{subequations}
and so on.  

We wish to show that $c_{2, \eff} (p, -p)$ has a tachyon pole if
$\Lambda_0$ satisfies (\ref{VIII-beyond}).  We have
\begin{align}
  &\frac{1}{\lambda} - \frac{1}{(4 \pi)^2} \ln \frac{p}{\mu} + B - \F
    \left(\frac{p}{\Lambda_0}\right)\\
  &= f_2 - \frac{1}{(4 \pi)^2} \ln \frac{p}{\Lambda_0} + B - \F
    \left(\frac{p}{\Lambda_0}\right)\,,
\end{align}
where $f_2 < 0$.  The first inequality in Appendix
\ref{Appendix-inequality} gives
\begin{equation}
0 \le   - \frac{1}{(4 \pi)^2} \ln \frac{p}{\Lambda_0} + B - \F
  \left(\frac{p}{\Lambda_0}\right) \longrightarrow
    \lb \begin{array}{c@{\quad}l}
          + \infty& (p \to 0)\\
          0& (p \to +\infty)
             \end{array}\right.\,.
\end{equation}
Hence, $c_{2, \eff} (p, -p)$ must have a tachyon pole.

As $\Lambda_0 \to +\infty$, the pole approaches
$p = \mu e^{(4 \pi)^2 \left(\frac{1}{\lambda} + B\right)}$.
Let us find the tachyon pole for $\Lambda_0$ a little beyond the limit:
\begin{equation}
f_2=  \frac{1}{\lambda} - \frac{1}{(4 \pi)^2} \ln \frac{\Lambda_0}{\mu} =
  - \ep\,,
\end{equation}
where $- \ep$ is a small negative constant.  The pole satisfies
\begin{equation}
  - \frac{1}{(4 \pi)^2} \ln \frac{p}{\Lambda_0} + B - \F
  \left(\frac{p}{\Lambda_0}\right) = \ep \ll 1\,.
\end{equation}
For $p \gg \Lambda_0$, we find (Appendix \ref{Appendix-asymp})
\begin{equation}
  \F \left(\frac{p}{\Lambda_0}\right) = - \frac{1}{(4 \pi)^2} \ln
  \frac{p}{\Lambda_0} + B -  \frac{\Lambda_0^2}{2 p^2} \int_q f(q)\,.
\end{equation}
Hence, we obtain
\begin{equation}
  \frac{\Lambda_0^2}{2 p^2} \int_q f(q) = \ep
  \Longrightarrow p^2 = \frac{\Lambda_0^2}{2 \ep} \int_q f(q) \gg \Lambda_0^2\,.
\end{equation}
As $\Lambda_0$ increases, this should approach the limiting value $p = \mu
e^{(4\pi)^2 \left(\frac{1}{\lambda} + B\right)}$.

Incidentally, if the cutoff is a little smaller than the Landau pole
\begin{equation}
f_2 =  \frac{1}{\lambda} - \frac{1}{(4\pi)^2} \ln \frac{\Lambda_0}{\mu} =
  \ep > 0\,,
\end{equation}
there is a bound state pole at
\begin{equation}
-  p^2 = \frac{\Lambda_0^2}{2\ep} \int_q f(q) \gg \Lambda_0^2\,.
\end{equation}
This turns over to a tachyon as the cutoff crosses the limit.  The
transition of a bound state to a tachyon has been observed previously
in a model of a charged harmonic oscillator in the radiation field
\cite{Sonoda:2022nnx} and also in a model of harmonic oscillators with
a configuration interaction that mimics the Lee model
\cite{Sonoda:2013dma}.

\subsection{Instability}

The 1PI potential is given by
\begin{align}
  G_\eff (\varphi)
  &= F_\eff (\sigma) + \sigma \varphi\nt\\
  &= - \frac{1}{2} f_2 \sigma^2 + \frac{1}{4(4\pi)^2}
    \frac{\sigma^2}{2} + \frac{1}{2(4\pi)^2} \frac{\sigma^2}{2} \left(
    \ln \frac{-\sigma}{\Lambda_0^2} + A \right) - \Lambda_0^4 I
    \left(\frac{\sigma}{\Lambda_0^2}\right) + \Lambda_0^2 \sigma I'
    \left(\frac{\sigma}{\Lambda_0^2}\right)\,,
\end{align}
where
\begin{align}
  \varphi
  &= - f_2 \sigma + \frac{1}{2(4\pi)^2} \sigma \left( \ln
    \frac{-\sigma}{\Lambda_0^2} + A \right) + \Lambda_0^2 I'
    \left(\frac{\sigma}{\Lambda_0^2}\right)\nt\\
  &= - \left(f_2 + \frac{1}{(4 \pi)^2} \ln
    \frac{\Lambda_0}{\mu}\right) \sigma + \frac{1}{2 (4\pi)^2} \sigma
    \left( \ln \frac{-\sigma}{\mu^2} + A \right) + \Lambda_0^2 I'
    \left(\frac{\sigma}{\Lambda_0^2}\right)\,.
\end{align}
We wish to show that there exists $\sigma_{\min}$ where
\begin{equation}
  \frac{d\varphi}{d\sigma}\Big|_{\sigma = \sigma_{\min}} = 0\,.
\end{equation}
$\sigma_{\min}$ corresponds to $\varphi_{\max}$ beyond which the
effective potential is likely to develop a negative imaginary part.  We find
\begin{equation}
  \frac{d\varphi}{d\sigma}
  = - f_2 + \frac{1}{2(4\pi)^2} \left( \ln \frac{-\sigma}{\Lambda_0^2}
    + A + 1 \right) + I'' \left(\frac{\sigma}{\Lambda_0^2}\right)\,,
\end{equation}
where $- f_2 > 0$.  Since
\begin{equation}
  \frac{1}{2(4\pi)^2} \left( \ln \frac{-\sigma}{\Lambda_0^2} + A + 1
  \right) + I'' \left(\frac{\sigma}{\Lambda_0^2}\right)
  \longrightarrow \lb\begin{array}{c@{\quad}l}
                       - \infty& (- \sigma \to 0+)\\
                       0 & (- \sigma \to +\infty)
                     \end{array}\right.\,,
\end{equation}
there must be such $\sigma_{\min}$.

Let us find $\sigma_{\min}$ for a very small $- f_2 = \ep > 0$.  We expect
$- \sigma_{\min} \gg \Lambda_0^2$.  Using the asymptotic behavior
derived in Appendix \ref{Appendix-asymp}
\begin{equation}
  I'' (\sigma) \overset{- \sigma \gg 1}{\longrightarrow}
  - \frac{1}{2} \frac{1}{(4 \pi)^2} \left( \ln (-\sigma) + A + 1
  \right) + \frac{1}{6 \sigma^3} \int_q \frac{f(q)}{h(q)^2}\,,
\end{equation}
we obtain
\begin{equation}
     \ep + \frac{1}{6} \frac{\Lambda_0^6}{\sigma_{\min}^3} \int_q
  \frac{f(q)}{h(q)^2} = 0\,.
\end{equation}
Hence,
\begin{equation}
  (-\sigma_{\min})^3 = \frac{1}{6} \frac{\Lambda_0^6}{\ep}\int_q
  \frac{f(q)}{h(q)^2} \Longrightarrow - \sigma_{\min} = \Lambda_0^2
  \left(\frac{1}{6 \ep} \int_q
    \frac{f(q)}{h(q)^2}\right)^{\frac{1}{3}} \gg \Lambda_0^2\,.
\end{equation}
We expect
\begin{equation}
  - \sigma_{\min} \overset{\Lambda_0 \to +\infty}{\longrightarrow}
  \frac{1}{e} \Lambda_L^2 = \frac{1}{e} \mu^2
  e^{\frac{2(4\pi)^2}{\lambda} - A}\,.
\end{equation}

\section{Concluding remarks\label{sec-conclusions}}

We have studied the large $N$ limit of the O$(N)$ linear sigma model
in four dimensions, and computed both the effective action and the
effective potential by solving the exact renormalization group
equations. 

For $N=1$, the absence of a non-trivial continuum limit has recently
been given a rigorous mathematical proof
\cite{10.4007/annals.2021.194.1.3}.  The absence is expected of any
$N$, including the large $N$ limit.  This implies that, given a finite
renormalized coupling $\lambda$, we cannot raise the cutoff beyond a
certain limit.  In the large $N$ limit, the maximum cutoff is the
Landau pole
\[
  \max \left(\Lambda_0 \right) = \mu e^{\frac{(4
      \pi)^2}{\lambda}}\,,\eqno{(\ref{VII-maximum})}
\]
where $\mu$ is the renormalization scale $\lambda$ is defined at.  In
the naive continuum limit, where $\Lambda_0$ is taken to infinity
while $\lambda$ is kept fixed, we expect problems.  The effective
potential, continued analytically for a large field, is found
unbounded from below with a negative imaginary part.  We hope to
justify this analytic continuation in a future publication.

Though the introduction of a cutoff $\Lambda_0$ helps to set the
theory on a firm ground, it has its own problems, such as the
existence of a bound state beyond the cutoff scale as shown in
Sec.~\ref{sec-comparison}.  The nature of the bound state is cutoff
dependent.

There is a lesson to learn from our calculations.  If the effective
potential of a model shows a sign of instability, suspect it is the
fault of a naive continuum limit.  The model may not be complete; it
may be a part of a larger and more complete theory.  We think it would
be interesting to apply our method to fermionic theories in large $N$.

\appendix

\section{Transition to the dimensionful convention\label{Appendix-dimension}}

We can move back and forth between the dimensionless convention, where
the momentum cutoff is $1$, and the dimensionful convention, where the
momentum cutoff is $\Lambda$, by using a set of very simple rules.  We
do not specify the space dimension and leave it as $D$.

Given a solution $\Gamma_{I,t} [\varphi]$ of the ERG equation
(\ref{II-ERG-N}), we define
\begin{equation}
  \Gamma_{I, \Lambda} [\varphi]
  \equiv \Gamma_{I,t} [\varphi_\Lambda]\,,
\end{equation}
where
\begin{align}
  \Lambda &= \mu \, e^{-t}\,,\\
  \varphi_\Lambda (p) &= \Lambda^2 \varphi (p \Lambda)\,.
\end{align}
Note that
\begin{equation}
  \varphi (p) = \frac{1}{2N} \int_q \Phi^I (q+p) \Phi^I (-q)
\end{equation}
has mass the dimension $-2$ for any $D$, and
$\Phi^I (p)$ has the mass dimension $- \frac{D+2}{2}$.

We have introduced $\mu$ as a physical renormalization scale,
corresponding to $t=0$.  $\Gamma_{I,\Lambda}$ satisfies the ERG
differential equation
\begin{equation}
  - \Lambda \frac{\partial}{\partial \Lambda} \Gamma_{I, \Lambda}
  [\varphi]
  = \frac{1}{2} \int_p \Lambda \frac{\partial R_\Lambda (p)}{\partial \Lambda}\,
  \G_{\Lambda; p,-p} [\varphi]\,,
\end{equation}
where
\begin{equation}
  R_\Lambda (p) \equiv \Lambda^2 R (p/\Lambda)\,,
\end{equation}
and
\begin{equation}
  \G_{\Lambda; p,-q} [\varphi] \equiv \Lambda^{-(D+2)} \G_{t; p/\Lambda, -q/\Lambda}
  [\varphi_\Lambda]
\end{equation}
satisfies
\begin{equation}
\int_q \G_{\Lambda; p,-q} [\varphi] \left[ \left( q^2 + R_\Lambda
  (q)\right) \delta (q-r) - \frac{\delta \Gamma_{I,\Lambda}
  [\varphi]}{\delta \varphi (q-r)} \right] = \delta (p-r)\,.
\end{equation}
In the limit $\Lambda \to 0+$, we obtain the 1PI generating functional
of the correlation functions:
\begin{equation}
  \Gamma_{I, \eff} [\varphi] = \lim_{\Lambda \to 0+}
  \Gamma_{I,\Lambda} [\varphi]\,.
\end{equation}

Similarly, the Legendre transform of $\Gamma_{I, \Lambda} [\varphi]$
is given by
\begin{equation}
  F_\Lambda [\sigma] \equiv F_t [\sigma_\Lambda]\,,
\end{equation}
where
\begin{equation}
  \sigma_\Lambda (p) = \Lambda^{D-2} \sigma (p \Lambda)\,.
\end{equation}
$\Gamma_{I, \Lambda} [\varphi]$ and $F_\Lambda [\sigma]$ are related
by
\begin{equation}
  \Gamma_{I,\Lambda} [\varphi] = F_\Lambda [\sigma] + \int_p \sigma
  (p) \varphi (-p)\,,
\end{equation}
and
\begin{equation}
  \sigma (p) = \frac{\delta \Gamma_{I,\Lambda}[\varphi]}{\delta
    \varphi (-p)},\quad
  \varphi (p) = - \frac{\delta F_\Lambda [\sigma]}{\delta \sigma
    (-p)}\,.
\end{equation}

For the 1PI potential, we define
\begin{equation}
  G_\Lambda (\varphi) \equiv \Lambda^D G_t (\varphi_\Lambda)\,,
\end{equation}
where
\begin{equation}
  \varphi_\Lambda \equiv \Lambda^{2-D} \varphi\,.
\end{equation}
The mass dimensions of $\varphi$ and $G_\Lambda$ are $D-2$ and $D$,
respectively.  The ERG equation
\begin{equation}
    \partial_t G_t (\varphi)
  = \left( D  - (D-2) \varphi \,\partial_\varphi \right) G_t
    (\varphi) + \frac{1}{2} \int_q f(q) \frac{\partial_\varphi G_t
      (\varphi)}{1 - h(q) \partial_\varphi G_t (\varphi)}
\end{equation}
gives
\begin{equation}
  - \Lambda \partial_\Lambda G_\Lambda (\varphi)
  = \frac{1}{2} \int_q \frac{\Lambda \partial_\Lambda R_\Lambda
    (q)}{q^2 + R_\Lambda (q)} \frac{\partial_\varphi G_\Lambda
    (\varphi)}{q^2 + R_\Lambda (q) - \partial_\varphi G_\Lambda (\varphi)}\,.
\end{equation}
In the limit $\Lambda \to 0+$, we obtain the effective potential as
\begin{equation}
  V_\eff (\varphi) = - G_\eff (\varphi) = - \lim_{\Lambda \to 0+}
  G_\Lambda (\varphi)\,.
\end{equation}

Similarly, the Legendre transform of $G_\Lambda (\varphi)$ is given by
\begin{equation}
  F_\Lambda (\sigma)
  \equiv \Lambda^D F_t \left( \sigma_\Lambda \right)\,,
\end{equation}
where
\begin{equation}
  \sigma_\Lambda \equiv \Lambda^{-2} \sigma\,.
\end{equation}
$\sigma$ has the mass dimension $2$ independent of $D$.  $G_\Lambda$
and $F_\Lambda$ are related by the Legendre transformation
\begin{equation}
  G_\Lambda (\varphi) = F_\Lambda (\sigma) + \sigma \varphi\,,
\end{equation}
where
\begin{equation}
  \varphi = - \frac{d}{d\sigma} F_\Lambda (\sigma),\quad
  \sigma = \frac{d}{d\varphi} G_\Lambda (\varphi)\,.
\end{equation}
The ERG equation
\begin{equation}
  \partial_t F_t (\sigma)
  = \left( D - 2 \sigma \,\partial_\sigma \right) F_t (\sigma) +
  \frac{1}{2} \int_q f(q) \frac{\sigma}{1 - \sigma h (q)}
\end{equation}
gives
\begin{equation}
  - \Lambda \partial_\Lambda F_\Lambda (\sigma)
  = \frac{1}{2} \int_q \frac{\Lambda \partial_\Lambda R_\Lambda
    (q)}{q^2 + R_\Lambda (q)} \frac{\sigma}{q^2 + R_\Lambda (q) -
    \sigma}\,.\
\end{equation}
The Legendre transform of $G_\eff (\varphi)$ is obtained as
\begin{equation}
  F_\eff (\sigma) = \lim_{\Lambda \to 0+} F_\Lambda (\sigma)\,.
\end{equation}

\section{ $I [\sigma]$ and $I (\sigma)$\label{Appendix-I}}

In this Appendix, the dimensionality of space is fixed as $D=4$.  

\subsection{Functional $I [\sigma]$}

We define
\begin{equation}
  I [\sigma] \equiv \sum_{n=3}^\infty \frac{1}{2n} \int_{p_1, \cdots,
    p_n} \sigma (p_1) \cdots \sigma (p_n)\,\delta \left(\sum_1^n
    p_i\right)\,
  I_n (p_1, \cdots, p_n)\,,
\end{equation}
where
\begin{equation}
  I_{n \ge 3} (p_1, \cdots, p_n) \equiv \int_p h(p) h(p+p_1) \cdots h
  \left( p + \sum_1^{n-1} p_i \right)\,,
\end{equation}
and the high momentum propagator is defined by
\begin{equation}
  h(p) \equiv \frac{1}{p^2 + R (p)}\,.
\end{equation}
The cutoff function $R(p)$ is $1$ at $p=0$ and approaches $0$ rapidly
at large $p$.  For numerical calculations, we use $R (p) = e^{- p^2}$.

The functional $I[\sigma]$ satisfies
\begin{align}
&  \int_p (p \cdot \partial_p + 2 ) \sigma (p) \cdot \frac{\delta I
                 [\sigma]}{\delta \sigma (p)} \nt\\
  &=  \frac{1}{2} \int_p (p \cdot \partial_p - 2) R (p) \cdot
    \left( \G_{p,-p} [\sigma] - h (p) \delta (0) - h (p)^2 \sigma (0)
    - h(p)^2 \int_q \sigma (q) h (p+q) \sigma (-q) \right)\nt\\
  &= - \frac{1}{2} \int_p  f (p) \sum_{n=3}^\infty \int_{q_1,
    \cdots, q_n}\, \delta \left(\sum_1^n q_i\right)\nt\\
  &\quad \times 
    \sigma (q_1) h(p+q_1) \sigma (q_2) h (p+q_1+q_2) \sigma (q_3)
    \cdots \sigma (q_{n-1}) h\left( p + \sum_1^{n-1} q_i\right) \sigma
    (q_n)\,,\label{Appendix-I-diffeq-functionalI}
\end{align}
where
\begin{equation}
  f(p) \equiv (2-p\cdot \partial_p ) R (p) \cdot h (p)^2 > 0\,.
\end{equation}

We wish to show
\begin{equation}
   I [\sigma] - I [- m^2 \delta] - \int_p \frac{\delta I [\sigma]}{\delta
    \sigma (p)}\Big|_{\sigma = - m^2 \delta} \left(\sigma (p) + m^2 \delta (p)\right)\nt\\
  = I (m^2) [\sigma + m^2 \delta]\,,
\end{equation}
where the right-hand side is defined by
\begin{align}
  &I (m^2) [\sigma+m^2\delta]
  \equiv \frac{1}{2} \int_{p,q} \left(\F (m^2; p) - \F (p) \right)\left( \sigma (p) + m^2
    \delta (p)\right) \left(\sigma (-p) + m^2 \delta (p)\right)\nt
  \\
  &\quad + \sum_{n=3}^\infty \frac{1}{2n} \int_{p_1, \cdots, p_n}
    \delta \left(\sum_1^n p_i\right)\, I_n (m^2; p_1, \cdots, p_n)
\,   \left(\sigma (p_1) + m^2 \delta (p_1)\right)
    \cdots \left(\sigma (p_n) + m^2 \delta (p_n)\right)\,,
\end{align}
and
\begin{subequations}
  \begin{align}
    h(m^2, p)
    &\equiv \frac{1}{p^2+m^2+R(p)}\,,\\
  \F (p)
  &\equiv \frac{1}{2} \int_q h(q) \left( h(q+p) - h(q)\right)\,,\\
  \F (m^2; p)
  &\equiv \frac{1}{2} \int_q \left( h(m^2, q) h(m^2, q+p) -
    h(q)^2 \right)\,,\\
  I_{n\ge 3} (m^2; p_1, \cdots, p_n)
  &\equiv \int_p h(m^2, p) h(m^2, p+p_1) \cdots h \left(m^2, p+
    \sum_1^{n-1} p_i\right)\,.
\end{align}
\end{subequations}
Equivalently, we wish to show
\begin{equation}
\boxed{    I [\sigma - m^2 \delta] - I [- m^2 \delta] - \int_p \frac{\delta I [\sigma]}{\delta
    \sigma (p)}\Big|_{\sigma = - m^2 \delta} \sigma (p) 
  = I (m^2) [\sigma] \,.}\label{Appendix-I-functionalIshift}
\end{equation}

Substituting
\begin{equation}
  \sigma (p) \longrightarrow \sigma (p) - m^2 \delta (p) \label{Appendix-I-substitution}
\end{equation}
into Eq.~(\ref{Appendix-I-diffeq-functionalI}), we obtain
\begin{align}
&  \left( - 2 m^2 \partial_{m^2} + \int_p (p \cdot \partial_p + 2 )
                 \sigma (p) \cdot \frac{\delta}{\delta \sigma (p)}
                 \right) I [\sigma - m^2 \delta]\nt\\ 
  &= -  \frac{1}{2} \int_p (2-p \cdot \partial_p ) R (p) \cdot
    \Big[ \G_{p,-p} [\sigma-m^2\delta] - h (p) \delta (0) - h (p)^2
    \left(\sigma (0)-m^2 \delta (0)\right)\nt\\
  &\quad    - h(p)^2 \int_q \left(\sigma (q) - m^2 \delta (q)\right) h (p+q)
    \left(\sigma (-q) - m^2 \delta (q)\right) \Big]\,,\label{Appendix-I-diffeq-Ishift}
\end{align}
where $\G_{p, -q} [\sigma-m^2\delta]$ is determined by
\begin{equation}
  \int_q \G_{p, -q} [\sigma-m^2\delta] \left[ (q^2+m^2 + R(q)) \delta (q-r) -
    \sigma (-q+r) \right] = \delta (p-r)\,.
\end{equation}
(This is obtained from Eq.~(\ref{III-def-G}) by the substitution (\ref{Appendix-I-substitution}).)
We can expand
\begin{align}
  &\G_{p, -q} [\sigma-m^2\delta]
    = h(m^2, p) \delta (p-q) + h (m^2, p) \Big[ \sigma (-p+q) \nt\\
  &\quad+ \int_{p_1, p_2} \sigma
    (p_1) h (m^2, p+p_1) \sigma (p_2) \delta (p_1+p_2+p-q) 
    + \int_{p_1, p_2, p_3} \sigma (p_1) h(m^2, p+p_1) \nt\\
  &\quad \times \sigma
    (p_2) h(m^2, p+p_1+p_2) \sigma (p_3) \delta (p_1+p_2+p_3+p-q) +
    \cdots \Big] h (m^2, q)\,.
\end{align}

Expanding (\ref{Appendix-I-diffeq-Ishift}) up to first order in $\sigma$, we obtain
\begin{align}
&  - 2 m^2 \partial_{m^2} I [-m^2\delta]\nt\\
&  = - \frac{1}{2} \int_p (2-p\cdot \partial_p) R (p) \cdot \left[
    h(m^2, p) \delta (0) - h (p) \delta (0) + h(p)^2 m^2 \delta (0) -
    h(p)^3 (m^2)^2 \delta (0) \right]\,,
\end{align}
and
\begin{align}
  & \int_p \sigma (p) (- 2 m^2 \partial_{m^2}) \frac{\delta}{\delta
    \sigma (p)} I [\sigma - m^2 \delta]\Big|_{\sigma = 0}
    + \int_p (p \cdot \partial_p + 2) \sigma (p) \cdot
    \frac{\delta}{\delta \sigma (p)} I [\sigma -
    m^2\delta]\Big|_{\sigma =0}\nt\\
  &= - \frac{1}{2} \int_p (2 - p \cdot \partial_p) R (p) \cdot
    \left[ h(m^2, p)^2 - h(p)^2  + 2 h(p)^3 m^2 \right] \sigma (0)\,.
\end{align}
Hence, we obtain
\begin{align}
  & \left( - 2 m^2 \partial_{m^2} + \int_p (p \cdot \partial_p + 2)
    \sigma (p) \cdot \frac{\delta}{\delta \sigma (p)} \right)
    \left[ I [\sigma - m^2\delta] - I [-m^2\delta] - \int_q \sigma (q)
    \frac{\delta I[\sigma - m^2\delta]}{\delta \sigma
    (q)}\Big|_{\sigma = 0} \right]\nt\\
  &= - \frac{1}{2} \int_p (2 - p \cdot \partial_p) R (p) \cdot
    \left( \G_{p, -p} [\sigma - m^2\delta] - h(m^2, p) \delta (0) -
    h(p)^2 \int_q \sigma (q) h(q+p) \sigma (-q) \right)\,.
    \label{Appendix-I-diffeq1}
\end{align}

We now derive the differential equation satisfied by
$I (m^2) [\sigma]$, and compare the result with
(\ref{Appendix-I-diffeq1}).  Since
\begin{subequations}
\begin{align}
&  p \cdot \partial_p \F (p)
  = \frac{1}{2} \int_q (p \cdot \partial_p + q \cdot \partial_q + 4)
    \left[ h(q) \left( h(q+p) - h(q)\right) \right]\nt\\
  &\quad= \int_q f(q) \left( h (q+p) - h(q)\right) = \int_q f(q) h(q+p) -
    \frac{1}{(4 \pi)^2}\,,\\
  &\left(p \cdot \partial_p + 2 m^2 \partial_{m^2} \right) \F (m^2,
  p)\nt\\
  &\quad = \frac{1}{2} \int_q \left( p \cdot \partial_p + q \cdot \partial_q
    + 2 m^2 \partial_{m^2} + 4 \right) \left( h(m^2, q) h(m^2, q+p) -
    h(q)^2 \right)\nt\\
  &\quad= \int_q  f(m^2, q) h(m^2, q+p) - \frac{1}{(4 \pi)^2}\,,
\end{align}
\end{subequations}
we obtain
\begin{equation}
  (p \cdot \partial_p + 2 m^2 \partial_{m^2} ) \left(\F (m^2, p) - \F
    (p) \right) = \int_q \left( f(m^2, q) h (m^2, q+p) - f(q) h (q+p)
  \right)\,.
\end{equation}
Using
\begin{align}
&  \left(\sum_1^n p_i \cdot \partial_{p_i} + 2 m^2 \partial_{m^2} +
                                                             2(n-2) \right) I_n (m^2; p_1, \cdots, p_n)\nt\\
  &\quad= \int_p \left( p \cdot \partial_p + \sum_1^n p_i \cdot
    \partial_{p_i} + 2 m^2 \partial_{m^2} + 2n \right) 
   \left[ h(m^2, p) h(m^2, p+p_1)  h\left(m^2, p +
    \sum_1^{n-1}p_i\right)\right]\nt\\
  &\quad= \int_p \Big[ f(m^2, p) h(m^2, p+p_1) \cdots h\left(m^2, p+\sum_1^{n-1}
    p_i\right)\nt\\
  &\qquad\quad + h(m^2, p) f(m^2, p+p_1) \cdots \ h\left(m^2, p+\sum_1^{n-1}
    p_i\right) + \cdots \Big]\,,
\end{align}
we obtain
\begin{align}
  & \left( - 2 m^2 \partial_{m^2} + \int_p (p \cdot \partial_p + 2)
    \sigma (p) \cdot \frac{\delta}{\delta \sigma (p)} \right) I (m^2)
    [\sigma]\nt\\
  &= - \frac{1}{2} \int_p (2-p \cdot \partial_p) R (p) \cdot
    \left( \G_{p, -p} [\sigma - m^2 \delta] - h(m^2, p) \delta (0) -
    h(p)^2 \int_q \sigma (q) h (q+p) \sigma (-q) \right)\,.
    \label{Appendix-I-diffeq2}
\end{align}
The right-hand side is the same as that of Eq.~(\ref{Appendix-I-diffeq1}).
The differential equation has a unique analytic solution, and we obtain the
desired equality.

\subsection{Function $I (\sigma)$}

We define
\begin{align}
  I (\sigma)
  &\equiv \frac{1}{2} \int_p \left[ - \ln \left(1 - h (p)
    \sigma\right) - h (p) \sigma  - \frac{1}{2} h(p)^2 \sigma^2 \right]\nt\\
  &= \sum_{n=3}^\infty \frac{1}{2n} I_n \cdot \sigma^n\,,
\end{align}
where
\begin{equation}
  I_{n\ge 3}  \equiv \int_p h(p)^n\,.
\end{equation}

Given an arbitrary $m^2 > 0$, we would like to show
\begin{equation}
  I (\sigma) - I (-m^2) - I' (-m^2) (\sigma + m^2)
  = I(m^2; \sigma+m^2) \,,\label{Appendix-I-functionIshift}
\end{equation}
where
\begin{align}
  I (m^2; \sigma+m^2)
  &\equiv \frac{1}{2} \F_{m^2} (\sigma + m^2)^2 + \frac{1}{2} \int_p
    \left[ - \ln \left(1 - h(m^2, p) (\sigma+m^2)\right) \right.\nt\\
  &\left.\qquad - h(m^2, p)
    (\sigma+m^2) - \frac{1}{2} h(m^2, p)^2 (\sigma+m^2)^2 \right]\nt\\
& = \frac{1}{2} \F_{m^2} (\sigma + m^2)^2 
  + \sum_{n=3}^\infty
    \frac{1}{2n} I_n (m^2) \left(\sigma+m^2\right)^n\,,
\end{align}
and
\begin{align}
  \F_{m^2} &\equiv \frac{1}{2} \int_p \left( h(m^2, p)^2 -
    h(p)^2\right) = \F (m^2; p=0)\,,\\
  I_{n\ge 3} (m^2) &\equiv \int_p h(m^2, p)^n = I_n (m^2; 0, \cdots, 0)\,.
\end{align}
This is a zero momentum version of
Eq.~(\ref{Appendix-I-functionalIshift}).  Instead of showing that both
sides of Eq.~(\ref{Appendix-I-functionIshift}) satisfy the same
differential equation, we prove it directly by computation.
\begin{align}
  & I (\sigma) - I (-m^2) - I' (-m^2) \cdot (\sigma + m^2)\nt\\
  &= \frac{1}{2(4 \pi)^2} \int_0^\infty dp^2\, p^2 \left[ - \ln
    \frac{p^2-\sigma+R}{p^2+R} - \frac{\sigma}{p^2+R} - \frac{1}{2}
    \frac{\sigma^2}{(p^2+R)^2}+ \ln \frac{p^2+m^2+R}{p^2+R}\right.\nt\\
  &\quad\left.  + \frac{-m^2}{p^2+R} +
    \frac{1}{2} \frac{(m^2)^2}{(p^2+R)^2} - (\sigma+m^2)
    \left(\frac{1}{p^2+m^2+R} - \frac{1}{p^2+R} +
    \frac{m^2}{(p^2+R)^2}\right) \right]\nt\\
  &= \frac{1}{2(4\pi)^2} \int_0^\infty dp^2\, p^2 \left[
    - \ln \left(1 - \frac{\sigma+m^2}{p^2+m^2+R}\right) - \frac{\sigma
    + m^2}{p^2+m^2+R} \right.\nt\\
  &\quad\left. - \frac{1}{2} (\sigma+m^2)^2
    \frac{1}{(p^2+m^2+R)^2} + \frac{1}{2} (\sigma+m^2)^2 \left( -
    \frac{1}{(p^2+R)^2} + \frac{1}{(p^2+m^2+R)^2} \right)\right]\nt\\
  &= \frac{1}{2} \int_p \left( - \ln \left(1 - (\sigma+m^2) h(m^2, p) \right) -
    (\sigma+m^2) h(m^2, p) - \frac{1}{2} (\sigma+m^2)^2 h(m^2, p)^2
    \right)\nt\\
&    \quad+ \frac{1}{2} (\sigma+m^2)^2 \cdot \frac{1}{2} \int_p \left(
                h(m^2, p)^2 - h(p)^2 \right)\nt\\
  &= I (m^2; \sigma + m^2)\,.
\end{align}

\section{Asymptotic behaviors\label{Appendix-asymp}}

\subsection{Asymptotic behavior of $I (\sigma)$}

To derive the asymptotic behavior of
\begin{equation}
  I (\sigma) \equiv \frac{1}{2} \int_p \left(
    - \ln \left(1 - \sigma h(p)\right) - \sigma h(p) - \frac{1}{2}
    \sigma^2 h(p)^2 \right)\,,
\end{equation}
we first derive the differential equation it satisfies.
From
\begin{equation}
  \int_p \left( p \cdot \partial_p + 4 \right) \left[
  - \ln \left(1 - h (p)
    \sigma\right) - h (p) \sigma  - \frac{1}{2} h(p)^2 \sigma^2 \right] =
  0\,,
\end{equation}
we obtain
\begin{align}
  2 \sigma I' (\sigma) - 4 I (\sigma)
  &= 2 \sigma^3 \frac{d}{d\sigma} \left( \frac{1}{\sigma^2} I (\sigma)\right)\nt\\
  &= \frac{\sigma}{2} \int_p f(p) \left( \frac{1}{1 - h(p) \sigma} - 1
    - h(p) \sigma \right)\,.\label{Appendix-asymp-diffeq-I}
\end{align}
For $- \sigma \gg 1$, we expand the right-hand side to obtain
\begin{align}
   2 \sigma I' (\sigma) - 4 I (\sigma)
  &\overset{- \sigma \gg 1}{\longrightarrow}  \frac{\sigma}{2} \int_p
    f(p) \left( - \frac{1}{h (p) \sigma} - \frac{1}{h(p)^2 \sigma^2} - 1 - h (p) \sigma \right)\nt\\
  &= - \frac{1}{2} \int_p \frac{f(p)}{h(p)} - \frac{1}{2 \sigma} \int_p
    \frac{f(p)}{h(p)^2} - \frac{\sigma}{2} \int_p
    f(p) - \frac{\sigma^2}{2} \underbrace{\int_p f(p) h(p)}_{=
    \frac{1}{(4 \pi)^2}}\,.
\end{align}
Hence, we obtain
\begin{equation}
  \frac{d}{d\sigma} \left(\frac{1}{\sigma^2} I (\sigma)\right)
  \overset{- \sigma \gg 1}{\longrightarrow}  -
  \frac{1}{4 \sigma} \frac{1}{(4 \pi)^2}  - \frac{1}{4 \sigma^2} \int_p f(p)
  - \frac{1}{4 \sigma^3}
  \int_p \frac{f(p)}{h(p)} - \frac{1}{4 \sigma^4} \int_p
  \frac{f(p)}{h(p)^2}\,.
\end{equation}
This implies the asymptotic behavior
\begin{equation}
  I (\sigma) \overset{-\sigma \gg 1}{\longrightarrow}
  - \frac{1}{2 (4 \pi)^2} \frac{\sigma^2}{2} \left( \ln (-\sigma) + A
    - \frac{1}{2} \right) + \frac{1}{4} \sigma \int_p f (p) +
  \frac{1}{8} \int_p \frac{f(p)}{h(p)} + \frac{1}{12 \sigma} \int_p
  \frac{f(p)}{h(p)^2}\,,
  \label{Appendix-asymp-I}
\end{equation}
where the constant $A$ depends on the choice of a cutoff function $R$.

\begin{figure}[h]
  \centering
  \includegraphics[width=0.45\textwidth]{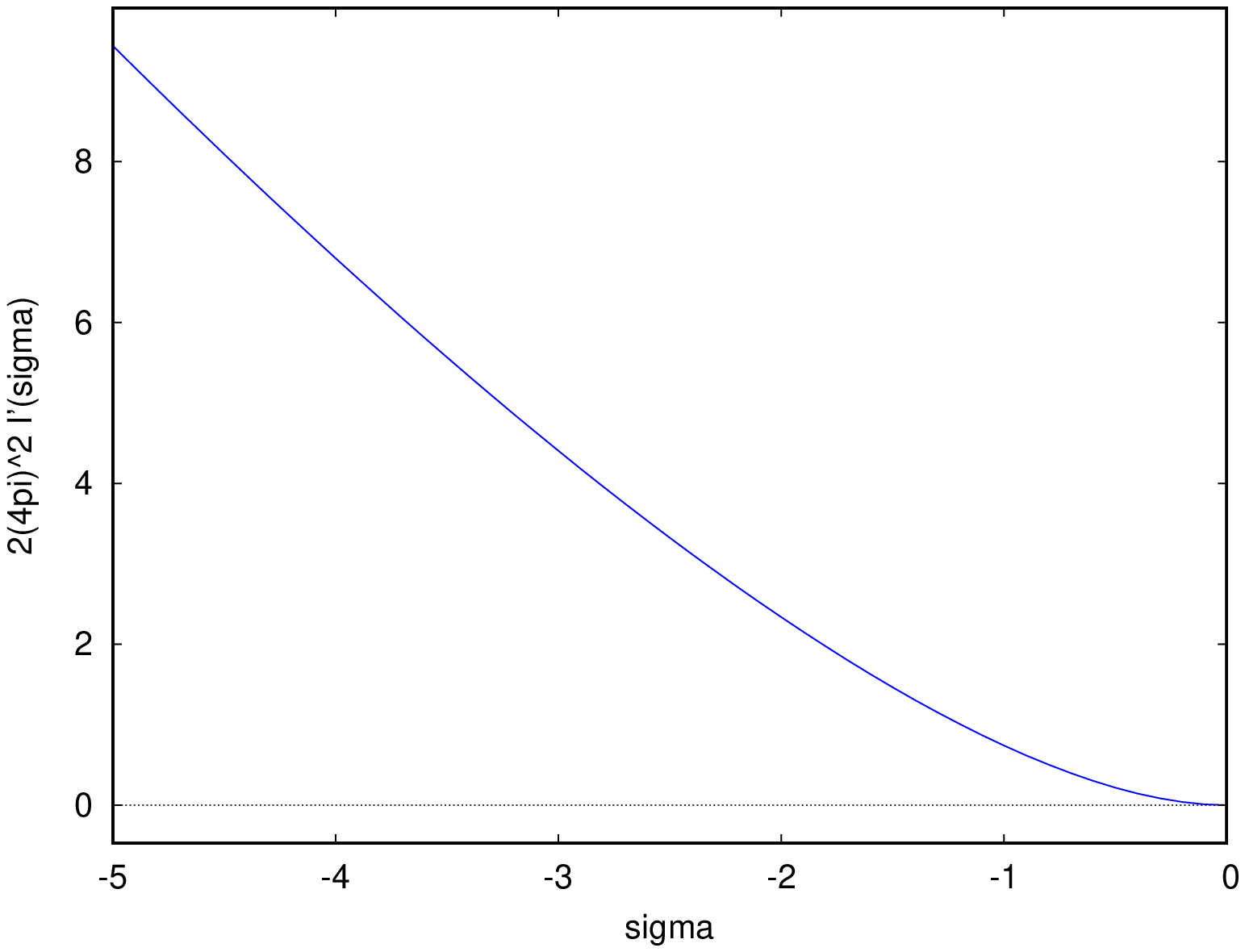}
  \hspace{0.5cm}
  \includegraphics[width=0.45\textwidth]{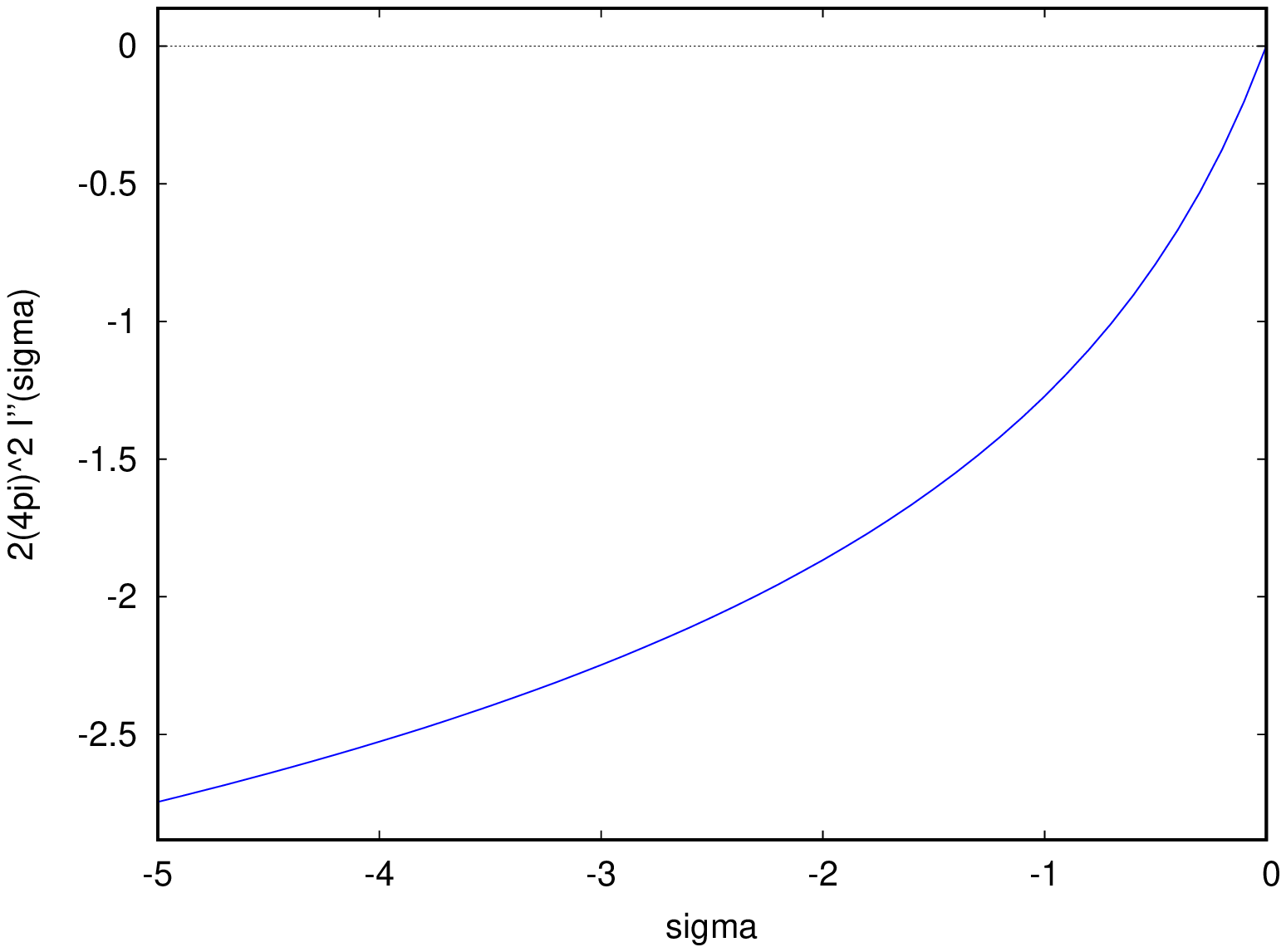}
  \caption{Both $2 (4\pi)^2 I' (\sigma)$ and $2 (4 \pi)^2 I''
    (\sigma)$ are convex in $\sigma < 0$}
  \label{fig-Iprime}
\end{figure}

Differentiating the above further, we obtain
\begin{align}
  I' (\sigma) &= \frac{1}{2} \int_p \left[ \frac{1}{p^2+R(p)-\sigma} -
                h(p) - \sigma h(p)^2 \right]\nt\\
                &\overset{-\sigma \gg 1}{\longrightarrow}
 - \frac{1}{2 (4 \pi)^2} \sigma \left( \ln
  (-\sigma) + A \right) + \frac{1}{4} \int_p f(p) - \frac{1}{12
                  \sigma^2} \int_p \frac{f(p)}{h(p)^2}\,,
  \label{Appendix-asymp-Iprime}\\
  I'' (\sigma) &= \frac{1}{2} \int_p \left[
                 \frac{1}{\left(p^2+R(p)-\sigma\right)^2} - h(p)^2 \right]\nt\\
                 &\overset{-\sigma \gg 1}{\longrightarrow}
   -  \frac{1}{2 (4 \pi)^2} \left( \ln
  (-\sigma) +  A + 1 \right)  + \frac{1}{6 \sigma^3} \int_p
                   \frac{f(p)}{h(p)^2}\,.
                   \label{Appendix-asymp-Iprimeprime}
\end{align}
We plot $I' (\sigma), I'' (\sigma)$ for the choice $R(p) = e^{-p^2}$
in Fig.~\ref{fig-Iprime}.

\subsection{Asymptotic behavior of $\F (p)$}

The function
\begin{equation}
  \F (p) \equiv \frac{1}{2} \int_q h(q) \left( h(q+p) - h(q)\right)
\end{equation}
satisfies
\begin{align}
  p \cdot \partial_p \F (p)
  &= \frac{1}{2} \int_q (p \cdot \partial_p + q \cdot \partial_q + 4)
    \left( h (q) h(q+p) - h(q)^2 \right)\nt\\
  &= \int_q f(q) \left( h(q+p) - h(q)\right)\nt\\
  &\overset{p^2 \to \infty}{\longrightarrow}
    \frac{1}{p^2} \int_q f(q) - \frac{1}{(4 \pi)^2}\,.
\end{align}
This implies
\begin{equation}
  \F (p) \overset{p^2 \to \infty}{\longrightarrow} - \frac{1}{2 (4
    \pi)^2} \ln p^2 + B - \frac{1}{2 p^2} \int_q f(q)\,,
  \label{Appendix-asymp-calF}
\end{equation}
where $B$ is a constant dependent on the cutoff function $R$.  Hence, we obtain
\begin{equation}
  \lim_{\Lambda \to 0+} \left(\F \left(\frac{p}{\Lambda}\right) +
    \frac{1}{(4 \pi)^2} \ln \frac{\mu}{\Lambda} \right)
  = - \frac{1}{2(4 \pi)^2} \ln \frac{p^2}{\mu^2} + B\,.
\end{equation}

\subsection{Asymptotic behavior of $\F_{m^2}$}

The constant
\begin{equation}
  \F_{m^2} \equiv \frac{1}{2} \int_p \left( h(m^2, p)^2 - h(p)^2
  \right)
\end{equation}
satisfies
\begin{align}
  2 m^2 \partial_{m^2} \F_{m^2}
  &= \frac{1}{2} \int_p \left( p \cdot \partial_p + 2 m^2
    \partial_{m^2} + 4 \right) \left( h(m^2, p)^2 - h(p)^2
    \right)\nt\\
  &= \int_p \left( f(m^2, p) h(m^2, p) - f(p) h (p) \right)\nt\\
  &\overset{m^2 \gg 1}{\longrightarrow} - \frac{1}{(4 \pi)^2}\,.
\end{align}
Hence,
\begin{equation}
  \F_{m^2} \overset{m^2 \to +\infty}{\longrightarrow}
  - \frac{1}{2(4\pi)^2} \ln m^2 + C\,,\label{Appendix-asymp-calFm2}
\end{equation}
where $C$ is a constant dependent on the cutoff function $R$.

\section{Riemann sheets of the Lambert $W$ function\label{Appendix-LambertW}}

\begin{figure}[h]
  \centering
  \includegraphics[width=0.8\textwidth]{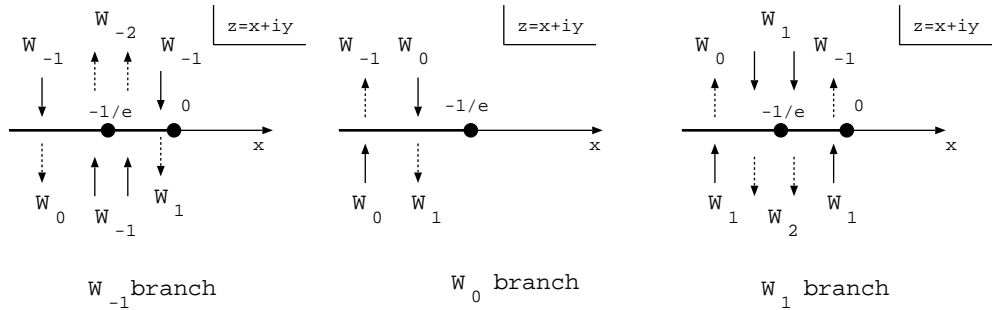}
  \caption{Analytic continuation of the Lambert $W$ function.  The
    figure is based on Fig.~4 of \cite{Corless:1996zz}}
  \label{fig-lambert-sheet}
\end{figure}

The main branch $W (z) = W_0 (z)$ is real on the real axis
$- \frac{1}{e} < z = x < +\infty$.  The lower branch $W_{-1} (z)$ is also
real for $W_{-1} (x+i\ep)$ where $- \frac{1}{e} < x < 0$.  The $W_0$
branch is connected to the $W_{\pm 1}$ branches so that for $x < -
\frac{1}{e}$ 
\begin{subequations}
\begin{align}
  W_0 (x-i\ep) &= W_{-1} (x+i\ep)\,,\\
  W_0 (x+i\ep) &= W_1 (x-i\ep)\,.
\end{align}
\end{subequations}
We also find
\begin{equation}
  W_0 (x+i\ep) =  W_0 (x-i\ep)^*\quad \left(x < - \frac{1}{\ep}\right)\,.
\end{equation}
The $W_{-1}$ branch has more complicated branch cuts.
\begin{subequations}
  \begin{align}
    W_{-1} (x+ i\ep)
    &= W_0 (x-i\ep)\quad \left( x < - \frac{1}{e}\right)\,,\\
    W_{-1} (x+i\ep)
    &= W_1 (x-i\ep) \quad \left( - \frac{1}{e} < x < 0\right)\,,\\
    W_{-1} (x-i\ep)
    &= W_{-2} (x+i\ep)\quad \left( x < 0 \right)\,.
  \end{align}
\end{subequations}
The cut between the $W_{-1}$ and $W_{-2}$ branches lies along the
entire negative real axis $x < 0$.

\begin{figure}[h]
\centering
\includegraphics[width=0.6\textwidth]{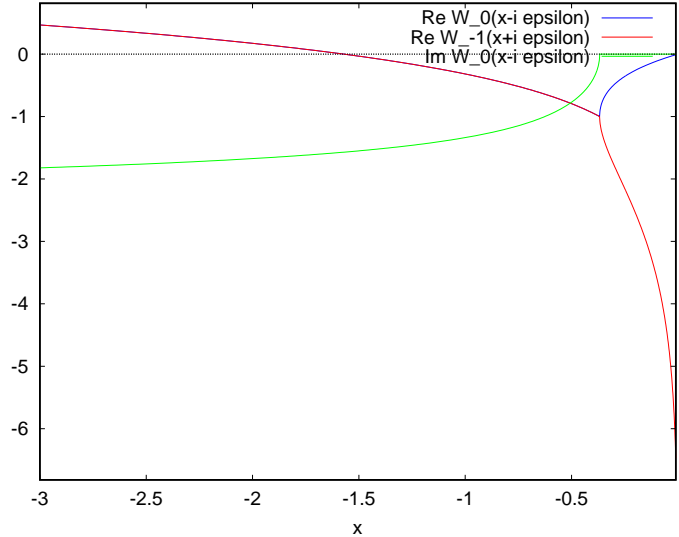}
\caption{$W_0 (x-i \ep), W_{-1} (x+i\ep)$ for $- 3 < x < 0$.  We have
  chosen $\ep = 0.00001$ for the plot}
\label{fig-lambertW}
\end{figure}

We plot $W_0 (x - i \ep), W_{-1} (x+i \ep)$ for $-3 < x < 0$ in
Fig.~\ref{fig-lambertW}.
More details can be found in \cite{Corless:1996zz} and
\cite{wiki:Lambert_W_function}.

\section{Inequalities\label{Appendix-inequality}}

\subsection{First inequality}

We wish to show, for $t > 0$,
\begin{equation}
 S (t, p) \equiv \frac{t}{(4 \pi)^2} + \F (p) - \F (p e^{-t}, -p e^{-t}) > 0\,.
\end{equation}
Note
\begin{equation}
  S (0, p) = 0\,.
\end{equation}
Since
\begin{equation}
  p \cdot \partial_p \F (p) = \int_q f(q) h(q+p) - \frac{1}{(4 \pi)^2}\,,
\end{equation}
we obtain
\begin{equation}
  \partial_t S (t, p) = \frac{1}{(4 \pi)^2} + p \cdot \partial_p \F (p
  e^{-t}, - p e^{-t})
  = \int_q f(q) h (q+p e^{-t}) > 0\,.
\end{equation}
Hence, $S(t, p)$ is an increasing positive function of $t$.
\begin{equation}
  S (t, p) > 0\quad (t > 0)\,.
\end{equation}

For large $t \gg 1$, we find
\begin{equation}
  S (t, p) \simeq \frac{t}{(4 \pi)^2} + \F (p)\,.
\end{equation}
For $p \gg e^t$, we find
\begin{equation}
  S (t, p) \simeq \frac{t}{(4 \pi)^2} - \frac{1}{(4\pi)^2} \ln p +
  \frac{1}{(4\pi)^2} \ln (p e^{-t})
  = 0\,.
\end{equation}

\subsection{Second inequality}

We wish to show, for $\sigma < 0$,
\begin{equation}
  \frac{1}{(4 \pi)^2} + 2 \sigma  \frac{d}{d\sigma} I'' (\sigma) > 0\,.\label{Appendix-inequality-second}
\end{equation}
The asymptotic behavior (\ref{Appendix-asymp-Iprimeprime}) implies
that the left-hand side vanishes as $\sigma \to - \infty$.
Now, differentiating
\begin{equation}
  2 \sigma I' (\sigma) - 4 I(\sigma) = \frac{\sigma}{2} \int_p f(p)
  \left(\frac{1}{1-h(p)\sigma} - 1 - h(p) \sigma\right)\,,
\end{equation}
which is (\ref{Appendix-asymp-diffeq-I}), twice with respect to
$\sigma$, we obtain
\begin{equation}
  2 \sigma I''' (\sigma) + \frac{1}{(4 \pi)^2} = \int_p f(p) \frac{h
    (p)}{(1 - h(p) \sigma)^3} > 0\,.
\end{equation}
Hence, the left-hand side of (\ref{Appendix-inequality-second}) is
increasing, and the inequality holds.

\subsection{Third inequality}

For $t > 0$, we wish to show
\begin{equation}
  \F (m^2 e^{2t}, p e^t) + \frac{1}{(4 \pi)^2} t - \F (m^2,
  p) > 0\,.\label{Appendix-inequality-thrid} 
\end{equation}
To prove this we first note the lhs vanishes at $t=0$.  We then note
\begin{align}
  \partial_t \left(  \F (m^2 e^{2t}, p e^t) + \frac{1}{(4 \pi)^2} t
  \right)
  &= \int_q f(m^2 e^{2t}, q) h (m^2 e^{2t}, q+p e^t)\nt\\
 &= \int_q (2-q\cdot \partial_q) R (q) \cdot h(m^2, q)^2 h(m^2 e^{2t},
 q+p e^t) > 0\,.
\end{align}
This proves the inequality.

\begin{acknowledgments}
  I would like to thank Prof. Bala Sathiapalan of the Institute for
  Mathematical Sciences in Chennai for his help in the early stage of
  the work and his continued support and encouragement.
\end{acknowledgments}

\bibliography{paper}

\end{document}